\documentstyle[]{l-aa}
\input epsf
\def\lsim{\lower.4ex\hbox{$\;\buildrel <\over{\scriptstyle\sim}\;$}}

\def\bib{\bibitem{}}
\newcommand{\xia}{\overline{\xi}}
\newcommand{\gam}{\gamma}
\newcommand{\Na}{\overline{N}}
\newcommand{\inta}{\int_{-i\infty}^{+i\infty}}
\newcommand{\beq}{\begin{equation}}
\newcommand{\eeq}{\end{equation}}
\begin{document}
%
%
\topmargin=2.5 cm
\thesaurus{12.12.1 , 11.19.7 , 11.03.1}
\title{The Multiplicity Function of Galaxies, Clusters and Voids}   
\author{P. Valageas\inst{1} \and R. Schaeffer\inst{1}}  
\institute{Service de Physique Th\'eorique, CEA Saclay, 91191 Gif-sur-Yvette, France} 
\maketitle
\markboth{Valageas \& Schaeffer: The Multiplicity Function of Galaxies, Clusters and Voids}{Valageas \& Schaeffer: The Multiplicity Function of Galaxies, Clusters and Voids}

\begin{abstract}

 	We calculate the multiplicity function of matter condensations with given mass and defined by an arbitrary density threshold which can be a function of the mass of the objects, by directly considering the actual, deeply non-linear density field, which we compare to the popular Press-Schechter approximation. This comparison is made possible owing to an analytic description of the non-linear mass distribution, but also thanks to a modellization of the evolution of the two-body correlation function from its linear to non-linear behaviour based on Peeble's spherical collapse picture. We show the mass function is a function of a unique parameter that contains all the dependence on the mass and radius of the object. We compare this new, still simple, analytic model to a more involved formulation that should be even closer to the results obtained using the standard density threshold algorithms. This gives some hindsight into the ``cloud-in-cloud'' problem, for both Press-Schechter and non-linear prescriptions, and it enables us to derive the mass function of general astrophysical objects (in addition to just-virialized halos) which may be defined by any density threshold, that may even vary with the mass of the object. This is beyond the reach of usual formulations based on the initial gaussian field and gives a clear illustration of the advantages of our approach.

We explain why numerical tests seem to favor both Press-Schechter and non-linear prescriptions even though the two approximations differ by their scaling as a function of mass as well as a function of redshift. 
We argue that numerical simulations will be closer to the non-linear predictions and should be reexamined in the light of our findings. 
The difference between the two is due to the fact that the Press-Schechter prescription assumes that present-day mass fluctuations can be recognized in the early linear universe, and that their number is conserved, while the non-linear approach takes into account their evolution which leads to an increase of the number of highly non-linear objects (very large or small masses). 

This difference is seen to be of the same magnitude as the difference obtained by varying the initial spectrum of density fluctuations. Earlier conclusions drawn about the relevance of the latter using analytical approximations to the mass function must thus be reexamined.

\end{abstract}


\section{Introduction}

There are traditionnally two  ways to characterize the quite irregular distribution of matter (and light) at extragalactic scales. The first is  to count galaxies of given luminosity  to define a {\it luminosity function}. Since Press \& Schechter (1974) and Schechter (1976) it is well accepted that the latter is a low luminosity diverging power-law with a sharp cut-off at the bright end. A related choice is to count very massive clusters (Abell 1958) or somewhat looser associations of galaxies (Turner and Gott 1975). This allows one to determine a {\it multiplicity function}, that is an abundance of objects as a function of mass, or luminosity, or any other parameter used to define the objects. An interesting similarity can be found (Bahcall 1979) in the shape of all these distributions. 

	The same clustering properties may be described by measuring the galaxy two-body correlation function. As may be expected from the existence of dense galaxy clusters, the systematic Lick  survey,  that allowed the first statistically significant studies, showed (Groth and Peebles 1977) this correlation function to be quite large, the three-body correlation being even larger. Obviously, one may then suspect the many-body correlations to be even larger. These correlation functions may be determined by dividing the universe into cells, as the statistics of the counts-in-cells is intimately related to these correlation functions at high order (White 1979). Whether galaxies or the mass contained in the cells is considered, the theoretical evaluation of these counts shows, astonishingly (Schaeffer 1985, Balian and Schaeffer 1989, hereafter BS89) a diverging power-law behaviour for small counts, with an exponential cut-off, quite similar to the one observed for the multiplicity functions. This behaviour, specifically, is related to the very large correlations -and whence fluctuations- of the counts at the deeply non-linear scales.

	With the availability (Peebles 1982) of the first realistic initial spectrum of fluctuations, in the sense that it is indeed related to physical processes in the primordial universe, the Press-Schechter (PS) approximation was revived (Schaeffer and Silk 1985) to usefully supplement numerical simulations (Davis et al 1985). Such an approach, however, was violently questioned (Bardeen et al 1986), as were different other alternative solutions to the problem of defining analytically a mass function, because of the {\it cloud-in-cloud problem}. Indeed, a given density threshold defines an object. But, because the density is fluctuating, this object may contain smaller substructures with larger overdensities surrounded by underdense regions so that the full larger object still is at the required density. When the importance of over and underdense regions is evaluated, as required in the PS approach, using the gaussian initial conditions that are assumed to prevail in the early universe, this problem turns out to be serious. The users of this approximation, however, have argued that this approximation is much better than what it stands for, namely the true count of objects defined by a given overdensity, and provides for the correct qualitative -if not quantitative- multiplicity of condensed objects. Specific tests (Efstathiou et al 1988) against numerical simulations show that the mass function using this approximation and the computed ones follow similar trends, but (Eke et al 1996) in some cases they compare well, and in some cases not at all. The PS approach, nevertheless, remains the most popular way to calculate analytically the mass function (White 1993).

	We however think the situation is quite unsatisfactory. The main reason is that from counts in cells, we see that the matter distribution at scales, say, below 10 Mpc at least, is wildly modified by non-linear evolution. This has been amply checked (Alimi et al. 1990; Maurogordato et al. 1992, Benoist et al. 1996; Bouchet et al. 1993) against the CfA, SSRS or IRAS maps of the galaxy distribution, as well as (in great detail) against numerical simulations (Bouchet et al. 1991; Bouchet and Hernquist 1992; Colombi et al. 1992,1994,1995,1996). The dependence of the counts on scale (according to the $r^{-\gamma}$ power-law behaviour of the correlation function) and on time (according to the $(1+z)^{-(3-\gamma)}$ dependence of the latter) follows the {\it non-linear} predictions. Were the PS formulation exact, the mass multiplicity function would scale as the {\it linear} correlation function extrapolated to the epoch under consideration (and in particular to the present one). Although such a feature would not be impossible, such a behaviour is a priori unlikely, given what we know from the counts-in-cells. Also, some of the detailed numerical tests to see whether not only the result but also whether the assumptions made in deriving the PS formula are correct are far from being successful (White 1993, sect 2.3.6). This calls for a reexamination of the problem of the mass function and its calculation by analytical means, in the light of the experience acquired in understanding the behaviour of the counts-in-cells.

	In this paper, we construct the mass function at a given epoch from the non-linear  density field at the {\it same} epoch, whose statistical properties we know. We present our formulation in Ch.2 . We discuss the various possible definitions of the mass function, how they are related to the hierarchical clustering pictures, their advantages and drawbacks. The distribution of underdense regions (with a low matter density) and of galaxy voids (which implies the proper inclusion of discreteness effects) can be obtained by the same methods. The result will be that, as it was the case for the density fluctuations, the mass multiplicity function can be described by i) a scaling function $H(x)$ that may depend on the initial conditions but does not evolve with time in the fully non-linear regime, and ii) the time-dependent non-linear two-body correlation function. The important many-body correlations are taken into account by the former scaling functions, that depend on a variable $x$ close to the internal velocity dispersion of the object. This function turns out to exhibit a power-law for small $x$ and an exponential fall-off for large $x$. The time evolution of the two-body correlation function can be modelled (Ch.3) by a totally analytic calculation using an extension, based on the spherical collapse picture, of a formulation due to Hamilton et al (1991). In Ch.4 we first compare numerically the predictions obtained using the PS prescription and our new model based on non-linear clustering, which differ significantly. We finally push the theoretical comparison as far as possible within our knowledge of the dynamics of clustering, the difference in the two approaches being reduced to one key assumption on the evolution of the number of mass condensations as the universe goes from the linear to non-linear regimes. In the last section (Ch.5) we present the mass functions of general astrophysical objects (in addition to the usual just-virialized halos) defined by an arbitrary density threshold which can be a function of the mass of the objects.

	As a conclusion, we propose several tests by means of computer simulation that may be performed, as well as the possible applications of the method.

\section{The multiplicity function: non-linear approach}

	Provided one is able to describe the density fluctuations in the non-linear regime, it is possible to deduce the multiplicity function 
 from the known distribution of overdensities above a given threshold $\Delta$ in cells of size $R$. The density field to be considered is, obviously in this case, the one relevant to the epoch at which the mass function is seeked.  
It is done in the same spirit as the very popular PS approach, (rephrased -App.A- in a way that parrallels the present non-linear description), but amounts to count the overdensities directly where they are, in the non-linear regime, rather than trying to identify the future overdensities within the linear fluctuations in the early universe. The properties, outlined by BS89, of the non-linear density field needed to our purpose are reviewed in App.B.

\subsection{Counts-in-cells}

	The knowledge of the two-body correlation function and the theoretically predicted scaling of the counts-in-cells, that has been seen to hold at the present epoch and that we assume to hold also in the past, allows us to compute the probability $P(\delta)$ to have a cell of density $\delta$, for a given partition of the universe in cells of size $R$, at an epoch when the matter distribution is non-linear.

	As seen in App.B, the fraction of matter in cells with overdensity larger than $\Delta$ is, provided the mild constraint  $1+\Delta \gg \xia^{\; -\omega/(1-\omega)}$, that is $x \gg \xia^{\; -1/(1-\omega)}$, is satisfied,
\beq		      			                 
F_m (>\Delta) =  \int_{\Delta}^{\infty}  (1+\delta) P(\delta) d\delta \simeq \int_{x(R)}^{\infty} x h(x) dx 
\eeq
\[ 
\mbox{with} \;\;\;\;\;  x(R) =  \frac{1+\Delta}{\xia(R)}
\]
and we can identify the fraction of matter $\mu(R) dR/R$ in overdense cells truly at scale $R$ with the derivative of $F_m(>\Delta,R) \;$:
\beq
\mu(R) \frac{dR}{R} = - \frac{d}{dR} F_m (>\Delta,R) \; dR = x h(x) \frac{dx}{dR} \; dR   \label{Fmhx}
\eeq
With $M = (1+\Delta) \rho_0 V$, where $V$ is the volume of the elementary cell, we can take as a possible definition for the mass function
\beq
\eta(M) \frac{dM}{M}  = \frac{\rho_0}{M} x h(x) dx  \;\;\;\;\; \mbox{with} \;\;\;\;\; x(M) =  \frac{1+\Delta}{\xia[R(M)]}    \label{etahM}
\eeq
which implies a well-defined scaling of the mass function as a function of the threshold $\Delta$: different choices of $\Delta$ lead to self-similar distributions.

	Bernardeau and Schaeffer (1991) tested successfully against observations some of the consequences of the scaling implied by (\ref{etahM}), by comparing galaxy and cluster multiplicities. This approximation to the multiplicity function as well as several other possibilities with the same scaling had been discussed earlier by BS89. Among all these early attempts, the first one is, in our mind, the only one that should be kept. 
 Our approach in particular has
 the advantage of obeying the normalization condition
\beq	             
\int_0^{\infty} \frac{M}{\rho_0} \eta(M) \frac{dM}{M}  =  \int_0^{\infty} x h(x) dx = 1
\eeq		        
since $M=0$ corresponds to $R=0$ and $\xia = +\infty$, that is $x=0$, whereas $M=+\infty$ corresponds to $x=+\infty \;$: all the matter and not just half of it, as would be the case using linear theory in the PS framework, is within the condensed objects.

 	For a matter correlation function $\xia(R) = (R/R_0)^{-\gamma}$, the mass function is a power-law $\eta(M) \; dM/M \; \propto \; M^{\alpha} \; dM$ with a slope somewhat steeper than $\alpha = -1$ at small masses
\[
M \ll M_* \;\; : \;\;\;\;\; \eta(M) \propto   M^{-1+\omega \gamma/3}  , 
\]
that is  $\alpha = -2+\omega \gamma/3$ .
At $M \gg M_*$, it bends down and decreases exponentially 
\[
M \gg M_* \;\; : \;\;\;\;\; \eta(M)  \propto   M^{-1+(1+\omega_s) \gamma/3} \exp[-(M/M_*)^{\gamma/3}]      
\]
where 
\[
M_* = (1+\Delta)^{-(3/\gamma-1)} \; x_*^{3/\gamma} \; M_0 
\]
and
\[
M_0 = \rho_0 \; 4\pi/3 \; R_0^3  \simeq   3 \; 10^{13} M_{\odot} ,
\]
its precise numerical value
depending on the value of $\Omega$ and the matter correlation length $R_0$.

So, with $x_* \simeq 10$, for $\Delta \simeq 200$, a typical density contrast for clusters, it bends at $M_*
\simeq 4 \; 10^{13} M_{\odot}$, whereas for $\Delta \simeq 5000$, characteristic of a galaxy, the bend indeed is at $M_*
\simeq  5 \; 10^{12} M_{\odot}$, a typical mass for an $L_*$ galaxy. 

\begin{figure}[htb]

\begin{picture}(230,160)(-18,-15)

\epsfxsize=8 cm
\epsfysize=12 cm
\put(-8,-112){\epsfbox{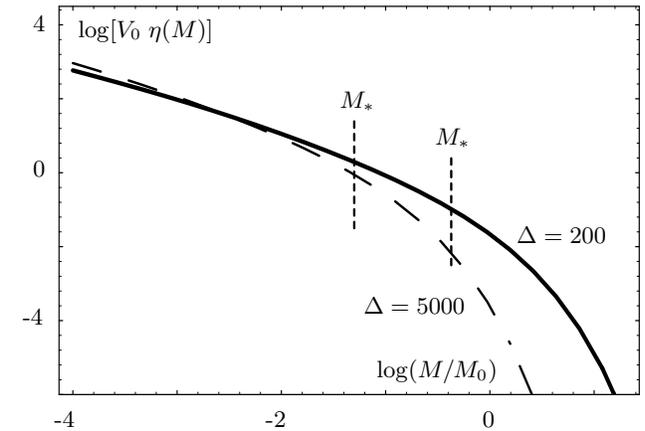}}
\put(-2,-25){-4}
\put(78,-25){-2}
\put(160,-25){0}
\put(-14,14){-4}
\put(-10,70){0}
\put(-10,125){4}
\put(106,96){$M_*$}
\put(142,82){$M_*$}
\put(115,18){$\Delta=5000$}
\put(173,45){$\Delta=200$}
\put(7,123){$\log[V_0 \; \eta(M)]$}
\put(120,-6){$\log(M/M_0)$}

\end{picture}

\caption{Behaviour of the mass function $\eta(M) \; dM/M$. The graph shows $V_0 \; \eta(M)$, where $V_0 = 4\pi/3 \; R_0^3$, with $\omega=1/2 \; , \; x_*=5 \; , \; \gamma=1.8$, in the cases $\Delta=5000$ (dashed line) and $\Delta=200$ (solid line). The former is relevant to galaxies, whereas the latter describes galaxy clusters. The difference in $M_*$ indeed corresponds to the difference between the typical mass of a galaxy and a galaxy cluster.}
\label{figetah}

\end{figure}

The behaviour of the mass function is displayed on Fig.\ref{figetah} for two different values of $\Delta$.

	As explained in App.B, the scaling with $h(x)$ holds provided the correlation function is much larger than unity, in the stable clustering regime.  For values of $\xia \simeq 1$ the non-linear behaviour slowly transforms into the linear gaussian distribution,  but only for overdensities $\Delta$ smaller than unity. For the larger $\Delta$ relevant to our approach since we count real clusters that already exist, and not protoclusters that will potentially become non-linear, we enter the quasi-gaussian regime (Bernardeau 1992, 1994). The gaussian develops an exponential tail that has just the same scaling with $\Delta$ as the tail of $h(x)$ but with a different function $h$. We could interpolate between the two regimes in the way suggested by Colombi et al (1996). However, for $\xia \geq 10$, the difference is minor. When $\xia=1$, we have $x=(1+\Delta)/\xia \gg 1$ and we are already very far in the exponential tail of $h(x)$, so there are practically no more clusters at this scale and the precise form of $h(x)$ is unimportant. For most practical cases we will not go beyond these regimes, and will use the non-linear $h(x)$ throughout. However, for $\xia \leq 10$ and moderate values of $1+\Delta$, such an interpolation would have to be considered.

\subsection{Surroundings of a cell}

	To define overdensities non-linear at scale $R$, another possibility is to consider explicitely the surroundings of each cell. The joint probability $P(\delta,R ;\delta',R+dR )$ (or simply $P(\delta,\delta')$ ), for a cell of scale $R$ to have a density $\delta$ whereas for a slightly larger scale $R+dR$ the density is $\delta'$, is constructed for the non-linear case in App.C.

	The mass fraction in cells with overdensity larger than $\Delta$ in $R$, but smaller than $\Delta$ in $R+dR$, is then
\[
F_m  (>\Delta,R;< \Delta,R+dR) = f_m (>\Delta,< \Delta ; R) dR 
\]
\[
\hspace{1cm} = \int_{\Delta}^{\infty}  d\delta  \int_{-1}^{\Delta} d\delta' \; (1+\delta) \; P(\delta,\delta') 
\]
\beq
\hspace{1cm} = (1+\Delta) \; P(>\Delta,<\Delta)
\eeq
and is of order $dR$. This calls for another possible definition of the mass function
\beq
\eta(M) \frac{dM}{M} =  \frac{\rho_0}{M} \; f_m  (>\Delta, <\Delta ; R) \; dR  \label{FmPsi}
\eeq
which can be usefully compared to the previous definition.

	Clearly, if the latter makes any sense and indeed describes overdensities $\Delta$ with mass $M$, the two definitions should not be very different. The rather technical -but badly needed- calculations used to compare both mass functions are given in App.C. The difference between the two approximations, evaluated in the non-linear case considered here, also gives some information on how the mass function depends on the way in which the objects are defined: directly as overdensities surrounded by underdensities, as considered here, or more implicitely by the number of overdense cell of mass $M$ compared to the number associated to an infinetesimally larger mass as considered above. 

	As shown in App.C, it is quite interesting to note that $\eta(M)$ defined in this section obeys the same scaling relation  as in the previous case. In particular, the trend of the mass function described above, a power-law at small masses and an exponential fall-off at large masses, is exactly the same. Indeed, we have
\beq
\eta(M) \frac{dM}{M}  =  \frac{\rho_0}{M} x H(x) dx   
\eeq
with a function $H(x)$ that differs from $h(x)$. However, as shown  in App.C, $H(x)$ and $h(x)$ have in general the same behaviour $\propto x^{\omega-2}$ at small $x$, with a different normalization but the same value of $\omega$, and $\sim \exp(-x/x_*)$ at large $x$. Explicit models for $h(x)$ show they may not differ by more than 10 per cent, and we have the general bounds (see (\ref{Hhall}) and (\ref{H><})):
\beq
x \geq 0 \; : \; 0 \leq \frac{H(x)}{h(x)} \leq \frac{3}{\gam} \;\; \mbox{and} \;\; x \gg x_* \; : \; 1 \leq \frac{H(x)}{h(x)} \leq \frac{3}{\gam}
\eeq
So, the two definitions for the mass multiplicity functions we have retained are quite similar. The function $h(x)$ is by definition properly normalised, but $H(x)$ is not necessarily: as readily seen in practical implementations of an agorithm that recognizes condensations by means of a density threshold, there is some overcounting (this is the cloud-in-cloud problem) in this case. This is actually in favour of the definition (\ref{etahM}) of section 2.1 which is simpler and has the interesting property that the total mass fraction within the objects is exactly unity by construction.

\subsection{Comparison with the PS approach}

	The same overcounting encountered for the mass fraction $\mu_{><}(\nu)$ defined in this way when overdensities are evaluated in the linear regime with gaussian probabilities (App.A) is much more severe, since in this case, overdensities with a given threshold are counted {\it infinitely} many times at different scales. Indeed, the total mass counted at various scales sums up to infinity. This is due to the fact that for small scales $\Sigma \rightarrow \infty$, hence the density contrast averaged around a given point shows wide fluctuations ($\sim \Sigma$) as the ``smoothing'' scale $R$ varies and it crosses the threshold $\delta_c$ many times, whence a large overcounting. This problem was addressed by Cole (1989), Bond et al.(1991), who calculated at each scale $R$ the mass which lies above $\delta_c$ at a larger scale. The derivative gives then the mass fraction which lies above $\delta_c$ at a scale between $R$ - $R+dR$, and is below this threshold at all larger scales, which solves the ``cloud-in-cloud'' problem. Using a top-hat in $k$ for the window function they were able to show that the mass function obtained in this way is simply the PS result multiplied by the usual factor 2. However, this window function is quite peculiar as in this case the increment of the density at a given scale is statistically independent of the larger scales, which may modify qualitatively the ``cloud-in-cloud'' problem. 

	It is clear that the mass function (\ref{FmPsi}) described above, based on $P(>\delta_c,R;<\delta_c,R+dR)$, overestimates the mass fraction at scale $R$ since one should rather evaluate, as Cole (1989), $P(>\delta_c,R;<\delta_c, \forall R'\geq R+dR)$ which is free of the ``cloud-in-cloud'' problem. At the small mass end, the mass function obtained from (\ref{FmPsi}) is a very bad approximation and it leads to a total mass which is infinite, showing indeed that the ``cloud-in-cloud'' problem is severe at these scales and may lead to a substantial modification of the PS result. On the other hand, we find that for a window function which is a top-hat in $R$ or a gaussian (or any window function which is differentiable in $k$) it tends to the Press-Schechter result {\it without} the factor 2 at the large mass end. This shows that the usual multiplication of this latter mass function by an overall factor of 2 to get the right normalization cannot be justified by the excursion set results, for realistic window functions different from the top-hat in $k$. In fact, taking into account the ``cloud-in-cloud'' problem implies to multiply the PS mass function by a scale-dependent renormalisation factor, smaller or equal to unity for large $\nu$. 

	Even if this geometrical problem were solved, one would still face the problem of the modellization of the dynamics one has to assign to these patches of matter to derive the properties of the non-linear density field obtained later and to get the number of virialized halos. Moreover, it is clear from the previous discussion that the number of large mass virialized objects obtained with the use of the common spherical collapse dynamics would underestimate the results of numerical simulation by $\sim 50\%$, since to match roughly these latter results one needs to multiply the PS mass function by at least the usual factor 2. This casts some serious doubts on the validity of this approach.

	In the non-linear case, the dynamics which governs structure formation, from the initial gaussian universe to the final highly non-linear density field, is partly encapsulated into the scaling function $h(x)$ (or the coefficients $S_n$). At this stage, $h(x)$ cannot be predicted from the initial conditions, and has to be measured in numerical simulations. However, generic predictions can be made from the sole assumption of stable clustering, that is the scale invariance of $\varphi(y)$, see App.B., which can be applied next to any case provided one knows $h(x)$. Note moreover that the measure of $h(x)$ needs only to be done once, if the scale-invariance holds. Then it will apply for any time or scale, in the highly non-linear regime. The advantage of this approach is that all predictions are directly derived from the real non-linear density field, and not from the initial gaussian conditions extrapolated by a highly uncertain dynamical model which leads to some problems (normalization, proper calculation of the mass function). Moreover, we showed in the previous section that, as an extra-bonus, some of these geometrical problems (overcounting) encountered for a gaussian field do not lead to very serious difficulties when one works directly with the actual highly non-linear density field. Indeed, the peaks are well separated as the fraction of volume they occupy tends to 0 and not $1/2$ as in the gaussian case, when the scale tends to 0, see (\ref{FvNL}) and (\ref{Fvgau}). Thus, the same measure of the overcounting due to the cloud-in-cloud problem in the non-linear case shows this problem not to be a serious difficulty when the overdensities are directly counted in the non-linear regime, contrary to the counting based on initial linear gaussian fluctuations. This suggests strongly that the mass functions constructed from (\ref{Fmhx}) or (\ref{FmPsi}) are very close and provide a reliable approximation.

	We can note that in our definitions (\ref{Fmhx}) and (\ref{FmPsi}) for the mass function, what we actually consider is the proportion of matter formed by particules such that each of them is at the center of a sphere of radius $R$ with density contrast $\Delta$. Then we identify this with the mass fraction embedded within virialized halos of scale $R$. Thus, the ``objects'' identified in such a way are not necessarily spherical, although for large $x$ one may expect them to have increasingly spherical shapes. At first glance, an alternative method to count ``actual'' halos would be to start from an unknown distribution of spherical halos $\eta(M,R) \; dM/M \; dR/R$ of size $R$ and mass $M$, with a given density profile, and derive from this the probability distribution of the counts-in-cells. Then, one gets an integral relation between both quantities which one may try to inverse within certain approximations. However, we think that such a procedure cannot give very satisfactory results as it neglects substructures and density fluctuations within halos, which play a major role in the properties of the density field and are necessary if the scaling invariance of the coefficients $S_n$ is to be realised. An obvious way to see this, is to note that within such a picture of a collection of regular halos with a well-defined density profile, the mass which lies above a certain density contrast $\Delta$ tends to zero for very large $\Delta$, whatever small the ``smoothing'' radius $R$ might be. This is clearly in contradiction with (\ref{FmsupD}), which shows that (within the framework of our description of the non-linear density field) for any threshold $\Delta$ one counts nearly all the matter for a sufficiently small smoothing radius. As a consequence, the definitions (\ref{Fmhx}) and (\ref{FmPsi}) seem the most reasonable choices.

\subsection{Underdense regions}

	It is possible to define underdense regions in a way similar to the one we have used to define clusters. Let us define a given density threshold $(1-\Delta)$ below which a region will be considered as underdense. It is more sensible in this case to count the volume occupied by the underdense cells. There are, again, two definitions that may be used to define the number of such regions of a given size $R$. Note that the latter scale is the true scale of the underdense volume.

	The volume fraction occupied by the underdense cells in a partition of the universe into cells of scale $R$ is
\beq
F_v(<-\Delta,R) = \int_{-1}^{-\Delta}  P(\delta) d\delta        
\eeq
For $(1 - \Delta) \ll  \xia$, which for large $\xia$ is true for nearly any value of $(1-\Delta)$ that is smaller than unity, $F_v$ obeys the scaling that results from the considerations of App.B
\beq
F_v(<-\Delta,R)  =  \int_0^{z}  g(z) dz  
\eeq
with
\[
z = (1-\Delta) \; a^{-1/(1-\omega)} \; \xia(R)^{\omega/(1-\omega)}
\]
that is, it does depend on only one scaling variable, and not on the two quantities $\Delta$ and $R$ separately. So the volume fraction of cells truly at scale between $R$ and $R+dR$ is
\[
\left| dF_v \right|  =  - \frac{d}{dR} F_v(<-\Delta,R) \; dR  =  - g(z) dz
\]
and the number density of underdense regions of scale $R$ or volume $V$ is then
\beq
\eta_u(R) \; \frac{dR}{R}  =  - \; \frac{1}{V} \; g(z) \; dz  
\eeq
The normalization properties of the function $g(z)$ then show that the total volume fraction of these low-density regions is equal to unity
\beq     
F_v  =  \int_0^{\infty}   V \; \eta_u(R) \; \frac{dR}{R}  \; = \int_0^{\infty} g(z) \; dz \; =  1   
\eeq
as expected since underdense regions at the non-linear scales occupy nearly all the volume, although all the mass is in the overdense regions.

	For reasons explained in the next section, testing in galaxy catalogues the scaling with $g(z)$ in the underdense regions is difficult because of discreteness effects, but it can be done (Alimi et al. 1990, Maurogordato et al. 1992) and works within the limits of the accuracy of the data extraction from catalogues. This scaling is seen to work beautifully in numerical simulations (Bouchet et al. 1991). There are however conditions of validity (App.B.3.3b). First, at large scales, the $g(z)$ and $h(x)$ scalings merge to rebuild the gaussian behaviour. However, when $(1-\delta) \ll 1$, the dominant contribution to $P(\delta)$ comes from the asymptotic region of $\varphi(y)$ and the results of App.B.3.3b are valid. At small scales, the condition (\ref{cond2}) which reads $(1-\Delta) \ll \xia$ is verified as soon as $\xia \gg 1$. Hence {\em the number density of voids is described by $g(z)$ at all scales}, provided that $(1-\Delta) \ll 1$. However, the scaling functions $g(z)$ describing small scales ($R \ll R_0$) in the highly non-linear regime, and large scales ($R \gg R_0$) in the linear regime are different (as the corresponding functions $h(x)$ and $\varphi(y)$ also vary from the quasi-gaussian regime to the highly non-linear regime).

	At small scales, $z$ is large, and $g(z)$ is a power-law. Whence
\[
(1-\Delta) \gg \xia^{\; -\omega/(1-\omega)} \;\; : \;\;\;\;  \eta_u(R)  \propto  R^{-3+\gamma \omega}      
\]

	At large scales, where $z$ is small, $g(z)$ decreases exponentially and we have, provided $\xia$ is still a power-law 
\[
(1-\Delta) \ll \xia^{\; -\omega/(1-\omega)} \;\; : \;\;\;\;   \eta_u(R)  \propto R^{-3+\gamma (1+\omega)/(2-2\omega)}
\]
\[
\;\;\;\;\;\;\;\;\;\;\;  \times  \exp[-\omega (1-\omega)^{\frac{1-\omega}{\omega}} a^{\frac{1}{\omega}} (1-\Delta)^{-\frac{1-\omega}{\omega}} (R/R_0)^{\gamma}]
\]

\begin{figure}[htb]

\begin{picture}(230,160)(-18,-15)

\epsfxsize=8 cm
\epsfysize=12 cm
\put(-8,-112){\epsfbox{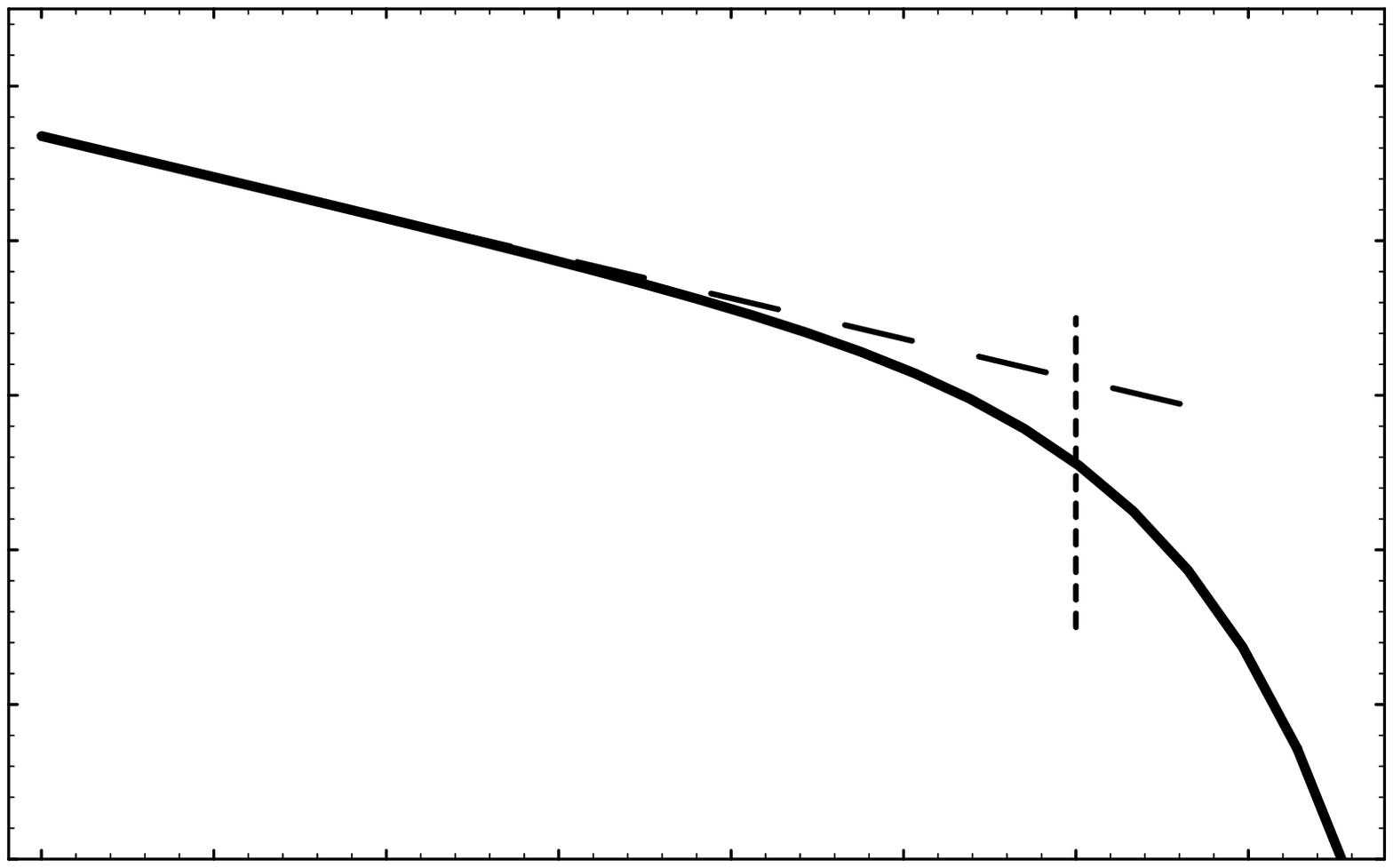}}
\put(-4,-25){-1.5}
\put(55,-25){-1}
\put(105,-25){-0.5}
\put(168,-25){0}
\put(-14,12){-4}
\put(-10,65){0}
\put(-10,119){4}
\put(106,93){$\propto R^{-3+\gamma \omega}$}
\put(165,18){$R_0$}
\put(10,123){$\log[V_0 \; \eta_u(R)]$}
\put(158,-6){$\log(R/R_0)$}

\end{picture}

\caption{Behaviour of the multiplicity function of underdense regions $\eta_u(R) \; dR/R$ predicted in the non-linear approach. The graph shows $V_0 \; \eta_u(R)$ with $V_0 = 4\pi/3 \; R_0^3$ in the case $\omega=1/2 \; , \; \gamma=1.8 \; , \; \Delta=0.9 \; .$}
\label{figetaVoid}

\end{figure}

	Thus the multiplicity function of underdense regions is steadily decreasing, and is a power-law with a cut at a scale close to the correlation length $R \simeq R_0$ where the averaged matter correlation function is unity, as we can see on Fig.\ref{figetaVoid}.

	As for the overdensities, there is another possible definition for these underdense regions, that stems from the consideration of the surroundings of a given cell. The probability for a cell to be underdense at scale $R$, and not at scale $R+dR$, is of order $dR$ and may be identified with the volume fraction occupied by underdense regions of scale $R$:
\[	     		    
P (< -\Delta,R ; > - \Delta, R+dR)  = f_u (< -\Delta, > - \Delta ; R) dR
\]
\beq
\hspace{1.5cm} = \int_{-1}^{-\Delta} d\delta  \int_{-\Delta}^{\infty}  d\delta' \; P(\delta,\delta')  
\eeq
The relation with the previous definition can be inferred from the idendity
\beq
\begin{array}{ll} P (< -\Delta,> - \Delta) = & P(< -\Delta,R ) - P(< -\Delta,R+dR) \\  & + P(>-\Delta,< -\Delta) \end{array}  \label{PVoid}
\eeq
The number density of underdense regions is then
\beq
\eta_u (R) \frac{dR}{R}  =  \frac{1}{V} \; f_u  (< -\Delta, > - \Delta ; R) \; dR
\eeq
that scales with a function $G(z)$ which has the same behaviour, except for a normalization constant, as the function $g(z)$ we introduced previously (see App.C and (\ref{gGg})). Note that because of (\ref{PVoid}) we have $G(z) \geq g(z)$, and $\int_0^{\infty} V \eta_v (R) dR/R \geq 1$. This is natural, since such a prescription does not  solve the ``cloud-in-cloud'' (or ``void-in-void'') problem entirely, and one may count the same volumes several times, but since $G(z)$ and $g(z)$ have the same behaviour and we expect $\int_0^{\infty} \; V \; \eta_v(R) \; dR/R$ to be close to 1 (see App.C and (\ref{gGg})), it is an acceptable solution to this problem.

\subsection{Galaxy voids}

       A continuous density field may be sampled by elementary points with number density $n$. All the above considerations are valid provided the number of elementary points per cell is large. This condition, at the non-linear scales, translates into $(1-\Delta) n V \xia(R)^{-\omega/(1-\omega)} \gg 1$. When this quantity is of order unity, or smaller, discreteness effects become increasingly important. This is most often the case when the galaxy distribution is considered. Indeed, for typical values $\Delta = 0.9 , n = 10^{-2} \; \mbox{Mpc}^{-3} , \omega = 0.5 , N_v$ is smaller than unity at all scales below 8 Mpc, that is at all non-linear scales. Incidentally, note that the counts of the overdensities using the continuous density field concepts that we have presented above, on the other hand, require $(1+\Delta) n V \xia(R)$ to be large. For $\Delta = 9$ and the same galaxy number density, this is the case, provided one is not too demanding, at nearly all interesting scales provided $R > 0.3$ Mpc.

\begin{figure}[htb]

\begin{picture}(230,160)(-18,-15)

\epsfxsize=8 cm
\epsfysize=12 cm
\put(-8,-117){\epsfbox{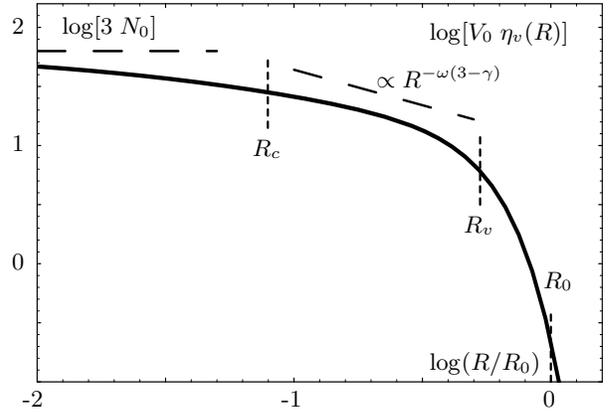}}
\put(0,-25){-2}
\put(98,-25){-1}
\put(197,-25){0}
\put(-4,27){0}
\put(-4,72){1}
\put(-4,116){2}
\put(134,96){$\propto R^{-\omega(3-\gamma)}$}
\put(15,116){$\log[3\;N_0]$}
\put(87,70){$R_c$}
\put(167,41){$R_v$}
\put(197,20){$R_0$}
\put(154,114){$\log[V_0 \; \eta_v(R)]$}
\put(154,-11){$\log(R/R_0)$}

\end{picture}

\caption{Behaviour of the void multiplicity function $\eta_v(R) \; dR/R$. The graph shows $V_0 \; \eta_v(R)$ in the case $\omega=1/2 \; , \; \gamma=1.8 \; , \; N_0 = 21.$}
\label{figetagalVoid}

\end{figure}

	It is possible to extend the above considerations about the void distribution to the discrete case by considering the probability $P(N,R)$, or simply $P(N)$, for counting $N$ galaxies in a cell of scale $R$. A partition of the universe gives a volume fraction of empty cells 
\beq
F_v (R)  =  P(0,R)  = \exp\{ - \varphi[N_c(R)] / \xia(R) \}  
\eeq
with
\[
N_c(R) = n V \xia(R)
\]
The fraction of cells corresponding to voids at scale $R$ and not above is then $- d/dR \; [F_v(R)] \; dR$ and the number of such cells gives the number of voids
\beq
\eta_v(R) \frac{dR}{R}  =  -\frac{1}{V} \; \frac{d}{dR} P(0,R)  \; dR    
\eeq
The alternative definition obtained by considering the edge of an empty cell implies to construct the probability $P( 0,R ; >0,R+dR )$ for having nothing in the cell of size $R$ and at least one point (or one galaxy) in the larger cell of scale $R+dR$. Since
\beq
P( 0,R ; >0,R+dR ) = P(0,R) - P(0,R+dR ) 
\eeq
the two possible definitions are in this case the same: obviously there is no ``void-in-void'' problem if we consider regions truly empty of galaxies. The multiplicity function in this case is also a power-law with an exponential cut, as we can see on Fig.\ref{figetagalVoid}, but with exponents that bear no relation with the continuous case (see Fig.\ref{figetaVoid} ). Moreover, the cut is at $R_v$ and not $R_0$.

We define the quantities:
\[
V_0 = 4\pi/3 \; R_0^3 \; , \; N_0 = n V_0 \; , \; \Na= n V \; , \; N_c = \Na \; \xia 
\]
\[
N_v^{1-\omega} = N_c^{1-\omega}/\xia \; ,  \; R_c = R_0 \; N_0^{-1/(3-\gamma)} 
\]
\[
R_v = R_0 \; N_0^{-1/(3+\gamma \omega /(1-\omega))}
\]
With $n=10^{-2} \; \mbox{Mpc}^{-3} \; , \; R_0=8$ Mpc, we have $N_0 \simeq 21$.

At small scales, where $\varphi(N_c) \simeq N_c \ll 1$, the behaviour is dominated by Poisson statistics
\[
R \ll R_c \; : \;\; \eta_v(R)  \simeq  3 n   
\]

At intermediate scales, where $\varphi(N_c)/\xia \sim N_v^{1-\omega} \ll 1$, it transforms into a power-law:
\[
R_c  \ll  R   \ll  R_v \; : \;\; \eta_v(R) \propto  R^{-\omega(3-\gamma)}    
\]

while at large scales, where $\varphi(N_c)/\xia \sim N_v^{1-\omega} \gg 1$, it decreases as
\[
R \gg R_v \; : \;\; \eta_v(R)  \propto   R^{-\omega(3-\gamma)}  \exp[-(R/R_v)^{3-\omega(3-\gamma)} ] 
\]
The scale $R_v$ may thus be interpreted as the typical size of voids. It depends on the number density $n$ and on the correlation length $R_0$, that explicitely are functions of the magnitude limit of the galaxy sample considered. The value of $R_v$ and its variations according to the magnitude limit of the sample, and the associated scaling of $P(0,R)$, have been extensively discussed by Maurogordato et al. (1992), and follow very closely the theoretical predictions.

\section{Evolution of the correlation function}

	To get the non-linear mass function at any time $t$ once we are given an initial spectrum of the density fluctuations, or alternatively the correlation function of the present universe, we need to evaluate the correlation function at this time $t$. The conservation of pairs of particules gives an exact equation for $\xi(x,a)$ ( Peebles,LSS), from which Hamilton et al (1991) followed by Padmanabhan (1996), argued that $\tilde{\xi}(x,a)$ should depend simply on the linear correlation function $\tilde{\xi}_L(x_L,a)$ evaluated at a different spatial point $x_L$ but at the same time $t$ or expansion factor $a = (1+z)^{-1}$:
\beq
\left\{ \begin{array}{rcl} \tilde{\xi}(x,a) & = & \tilde{F} \left[ \tilde{\xi}_L(x_L,a) \right] \\  \\ x_L^3 & = & \left[ 1+\tilde{\xi}(x,a) \right] \; x^3  \end{array} \right.   \label{Padd}
\eeq
where $x$ and $x_L$ are comoving lengths and 
\beq
\tilde{\xi}(x,a) = \frac{3}{x^3} \int_0^x \xi(y,a) y^2 dy
\eeq
Such a dependence has been checked in numerical simulations by Hamilton et al.(1991), Jain et al.(1995) and Padmanabhan (1996), which showed that the function $\tilde{F}$ had also a small dependence on the form of the initial spectrum of the density fluctuations, that we shall, in a first step, ignore here. However, in all cases $\tilde{F}(\tilde{\xi}_L) \simeq \tilde{\xi}_L$ for small $\tilde{\xi}_L$ and $\tilde{F}(\tilde{\xi}_L) \propto \tilde{\xi}_L^{\;3/2}$ for large $\tilde{\xi}_L$, when $\tilde{\xi} > 200$.
\\

	The reason why the Hamilton et al.(1991) approach works can be traced back to the hierarchical clustering picture (Peebles, LSS) based on the spherical collapse model. In the linear regime, a density contrast $\delta$ within a comoving radius $x_L$ evolves as $\delta \propto a$. As a consequence we have $\xia_L(x_L,a) = \;<\delta^2>\; \propto a^2$. If we note $\Sigma_0(x_L)^2$ the initial correlation function calculated at a mass scale such that $R_M=x_L$ and extrapolated to the present universe by the linear theory, we have:
\beq
\xia_L(x_L,a) = a^2 \; \Sigma_0(x_L)^2  \label{XiL}
\eeq
A density contrast $\delta_L=a\;\Sigma_0(x_L)$ will reach unity at a time $t_{NL}$ given by $a_{NL} = \Sigma_0(x_L)^{-1}$ and the radius of the overdensity at this time is $r_{NL} \simeq \Sigma_0(x_L)^{-1} \; x_L$. Its density at the time of non-linearity is of the order of the universe density at this epoch $n_{NL} \sim n_b(t_{NL})$. In the regime where clustering is stable, the excess of the number of neighbours within the blob over the mean is, as a function of time,
\beq
\xia = \frac{n_{NL}}{n_b(a)} \sim \frac{n_b(a_{NL})}{n_b(a)} = \left( \frac{a}{a_{NL}} \right)^3
\eeq
whence
\beq
\xia \sim  ( a \; \Sigma_0(x_L) )^3
\eeq
In both regimes $\xia$ is a function of $\xia_L = a \; \Sigma_0(x_L)$. The non-linear correlation function $\xia$ is taken at a comoving length $x$ given by the conservation of the number of pairs since there is a similar conservation law (Peebles, LSS) for $\xia$ which relates linear and non-linear scales. In the linear regime, where $\xia \ll 1$, we have $\xia \simeq \xia_L$ and $x \simeq x_L$. In the non-linear regime, $\xia > 200$, we have $\xia \sim \xia_L^{\;3/2}$ and the lenght scales are linked by $x \sim x_L \; a_{NL}/a \sim x_L \; \xia^{\;-1/3}$. Hence we recover the behaviour given by (\ref{Padd}), so we write a similar relation for $\xia$:
\beq
\left\{ \begin{array}{rcl} \xia(x,a) & = & F \left[ \xia_L(x_L,a) \right] \\  \\ x_L^3 & = & \left[ 1+\xia(x,a) \right] \; x^3   \end{array} \right.   \label{Padd1}
\eeq

	If we consider the physical length $r$ we get, using (\ref{XiL}),
\beq
\xia(r,z) = F \left\{ (1+z)^{-2} \Sigma_0^2 \left[ (1+z) r [1+\xia(r,z)]^{1/3} \right] \right\}
\eeq
which is an implicit equation for $\xia(r,z)$. Numerical simulations (Jain et al.1995) show that
\beq
\Omega=1 \; : \;\; \left\{  \begin{array}{rcl}   \xia \ll 1 & : & \xia \simeq \xia_L  \\ \\ \xia \gg 200 & : & \xia \simeq \left(\frac{10}{3\alpha}\right)^3 \; \xia_L^{\;3/2} \end{array} \right.  \label{xiaxiaL1}
\eeq
where $\alpha \simeq 1$, see App.E.\\

	The function $F$ can be obtained from numerical simulations. It may also be modellized in the spirit of the spherical collapse picture detailed above if we know i) the relation  between $\xia$ and $\delta$, which changes from $\xia \sim \delta^2$ in the linear regime to $\xia \sim \delta$ in the non-linear case, and ii) the evolution of $\delta$, which can be traced using the spherical collapse model (App.D) according to the ideas underlying the above qualitative picture. Such a model is presented in App.E, based on the spherical collapse picture. It is calibrated on the numerical results obtained by Jain et al.(1995), and it allows one to get a relation for the non-linear correlation function in the case of an open universe $\Omega<1 \; , \; \Lambda=0$ which is consistent with the behaviour seen by Peacock \& Dodds (1996) in simulations. Note that in this case the simple relation $\xia_L - \xia$ is no longer valid and one gets $\xia = F \left( \xia_L , a \right)$. More precisely, we have:
\beq
\Omega<1 \; : \;\; \left\{  \begin{array}{rcl}   \xia \ll 1 & : & \xia \simeq \xia_L  \\ \\ \xia \gg 200 & : & \xia \simeq \left(\frac{10}{3\alpha\; d(\Omega)}\right)^3 \; \xia_L^{\;3/2} \end{array} \right.  \label{xiaxiaL2}
\eeq
where $d(\Omega) \propto D(t)/a$ (with $d \rightarrow 1$ for $\Omega \rightarrow 1$ and $t \rightarrow 0$), see App.E. From this model for the evolution of $\xia$, and the scaling function $h(x)$ measured in simulations, we can now obtain the mass functions of any astrophysical objects at all times, in the highly non-linear regime. We shall first compare in more detail in the next chapter this approach with prescriptions based on an extrapolation from the initial gaussian field.

\section{Comparison of the linear and non-linear approaches}

\subsection{Counts-in-cells implied by the PS approximation}

The hierarchical picture based on the spherical collapse we developped in the previous chapter to get the evolution of the correlation function can also give the counts-in-cells, in the same spirit as the PS derivation seen in App.A. Indeed, in both cases the fundamental hypothesis is that one can recognize in the early universe the overdensities which will eventually form clusters, or more precisely one assumes that it is possible to follow the evolution of each initial overdensity (or of most of them). The dynamics of these objects is close to the one given by the spherical model and, more importantly, they keep their identity throughout their evolution: there is no redistribution of mass. Of course, a given overdensity will be part of different virialized objects as time goes on, but neighbours remain close to each other, so that mass is conserved by the dynamics. As we recalled in App.A, this picture of gravitational clustering leads to the PS mass function (\ref{etaM}). {\it The PS prescription also implies a prediction of the counts-in-cells in the non-linear regime}, that may be usefully compared to the one seen in numerical simulations, as we will undertake now.

\subsubsection{Density field implied by the PS and stable clustering approximations}

It is possible to define at any time an effective density field that, by the same counting done by PS in the linear regime and done by us in the non-linear regime, would at any time give the number of overdensities implied by the PS approach. If the latter is exact, this effective density field should be the same as the non-linear density field. Obviously, this cannot be the case as all density fluctuations cannot simultaneously follow a spherical dynamics, and the behaviour of underdensities is known to be quite inacurate, but we shall consider overdense regions which are usually assumed to be rather well modelled by such an approach and examine the consequences which can be inferred from such a picture.

Similarly to (\ref{Padd1}) and (\ref{xiaxiaL1}) we get from simply applying the spherical collapse rules that an overdensity $\delta_L$ at a comoving scale $R_M$ transforms into an overdensity $\delta$ at scale $R$:
\beq
\left\{ \begin{array}{rcl} \delta & = & {\cal F}[\delta_L] \\ \\ R_M^3 & = & (1+\delta) \; R^3  \end{array} \right.   \label{deldelL}
\eeq
where we considered the case $\Omega=1$ (no explicit time dependence), and ${\cal F}$ satisfies:
\beq
\left\{ \begin{array}{rcl} |\delta_L| \ll 1 & : & {\cal F}(\delta_L) \simeq \delta_L   \\  \\ \delta_L \gg 1 & : & {\cal F}(\delta_L) \simeq ( \frac{10}{3} )^3 \; \delta_L^3  \end{array} \right.
\eeq
Note that we took $\alpha=1$, as we consider here the evolution given by the usual spherical collapse model, with no kinetic energy at the time of maximum expansion, in order to follow closely the usual PS approximation. Within this framework, the mass fraction above a given density contrast $\delta$ at scale $R$ is:
\beq
 F_m(>\delta,R)  =  F_{m0}(>{\cal F}^{-1}(\delta),R_M)   \label{FmFm0}
\eeq
where $F_{m0}(>\delta_L,R_M)$ is relative to the early universe, that is to the linearly extrapolated values. Of course, this formulation gives back the PS prescription when overdensities are counted directly in the non-linear density field constructed in this way.
\[
F_m(>\Delta_c,R) = F_{m0}(>\delta_c,R_M) \;\;\; \mbox{with} \;\;\; (1+\Delta_c)
 \; R^3= R_M^3
\]
where $\delta_c = 3/20 \; (12 \pi)^{2/3} \simeq 1.68$ and $1+\Delta_c=18 \pi^2$ are the density contrasts at the time of collapse given by the usual spherical model. Then, the mass fraction in the form of just virialized objects (i.e. with a density contrast $\Delta_c$) is:
\[
\mu(M) \frac{dM}{M} = -\frac{\partial}{\partial R}  F_m(>\Delta_c,R)  dR 
\]
\[
\;\;\;\;\;\;\;\;\;\;\;\;\;\;\;\; = -\frac{\partial}{\partial R_M}  F_{m0}(>\delta_c,R_M)  dR_M
\]
which is just the PS mass function. We can write (\ref{FmFm0}) as:
\beq
F_m(>\delta,R) = \int_{{\cal F}^{-1}(\delta)/\Sigma(R_M)}^{\infty} e^{-\nu^2/2} \; \frac{d\nu}{\sqrt{2\pi}} 
\eeq
The mass fraction in cells of scale $R$ with a density contrast $\delta$ to $\delta+d\delta$ is:
\beq
(1+\delta) \; P_{PS}(\delta) \; d\delta = - \; \frac{\partial}{\partial \delta} \; F_m(>\delta,R) \; d\delta
\eeq
defining $P_{PS}(\delta)$ as the approximate probability distribution of the density contrast within cells of size $R$ implied by the PS approach. Thus we obtain for the counts-in-cells:
\[
P_{PS}(\delta) = \frac{1}{\sqrt{2\pi}} \frac{1}{1+\delta} \frac{\partial}{\partial \delta} \left[ \frac{{\cal F}^{-1}(\delta)}{\Sigma(R_M)} \right] \exp\left[-\frac{1}{2} \left( \frac{{\cal F}^{-1}(\delta)}{\Sigma(R_M)} \right)^2 \right]
\]
Naturally, by construction we recover all the mass of the universe $\int (1+\delta) \; P_{PS}(\delta) \; d\delta = 1$ (indeed ${\cal F}^{-1}(-1) = -\infty$). In fact, half of the mass is in overdensities and the other half in underdensities, since overdensities (underdensities) remain overdensities (underdensities) forever (${\cal F}(0)=0$). However, in general this probability distribution is not correctly normalized: $\int P_{PS}(\delta) \; d\delta \neq 1$ (but in the linear regime, $\xia \rightarrow 0$ and $\Sigma \rightarrow 0$, its normalization tends to unity). This is linked to the fact that the Press-Schechter approximation (using the usual spherical dynamics) cannot describe underdensities (which are generally taken into account in an ad-hoc fashion by the factor 2). In the regime $|\delta| \ll 1$, we have ${\cal F}^{-1}(\delta) \simeq \delta$, since $\delta \simeq \delta_L$, and $R_M \simeq R$, so we recover the gaussian:
\beq
|\delta| \ll 1 \; : \;\; P_{PS}(\delta) \simeq \frac{1}{\sqrt{2\pi} \; \Sigma(R)} \; e^{- \delta^2 / (2 \Sigma(R)^2) } \label{PSgaussian}
\eeq
For large overdensities $\delta > \Delta_c$ we saw that ${\cal F}(\delta_L) \simeq ( \frac{10}{3} )^3 \; \delta_L^3$, so ${\cal F}^{-1}(\delta) \simeq  3/10 \; \delta^{1/3}$. We consider the case where the power-spectrum is a power-law: $P(k) \propto k^n$ and $\Sigma(l)^2 \propto l^{-(3+n)}$.  Then we have:
\[
\frac{{\cal F}^{-1}(\delta)}{\Sigma[(1+\delta)^{1/3}R]} \simeq \frac{3}{10} \; \frac{\delta^{(5+n)/6}}{\Sigma(R)}
\]
We can also get the correlation function $\xia(R)$ at the scale $R$ from (\ref{xiaxiaL1}), which leads to:
\[
\Sigma(R) = \frac{3\alpha}{10} \; \xia(R)^{(5+n)/6} \;\;\; \mbox{when} \;\;\; \xia > \Delta_c
\]
Hence we get:
\[
\nu \equiv \frac{\delta_L}{\Sigma(R_M)} = \frac{{\cal F}^{-1}(\delta)}{\Sigma[(1+\delta)^{1/3}R]} \simeq \frac{1}{\alpha} \; \left( \frac{\delta}{\xia} \right)^{(5+n)/6}
\]
and eventually, for $\xia > \Delta_c$ and $\delta > \Delta_c$ :
\[
P_{PS}(\delta) = \frac{1}{\xia^2} \; \frac{1}{\sqrt{2\pi}} \; \frac{5+n}{6 \alpha} \; \left( \frac{\delta}{\xia} \right)^{(n-7)/6} 
\]
\beq
\;\;\;\;\;\;\;\;\;\;\;\;\;\;\;\;\;\;\;\;\;\;\;\;\;\;\; \times  \exp \left[ - \frac{1}{2 \alpha^2} \left( \frac{\delta}{\xia} \right)^{(5+n)/3} \right]
\eeq
We can note that we recover exactly the form (\ref{cond1}) we had for the non-linear prescription, with the same parameter $x$:
\beq
P_{PS}(\delta) = \frac{1}{\xia^2} \; h_{PS}(x) \;\;\; \mbox{with} \;\;\; x=\frac{\delta}{\xia}   \label{PdelhxPS}
\eeq
since in the regime of large overdensities we consider here $(1+\delta) \simeq \delta$. Hence the probability distribution exhibits the same {\it scaling law} with
\beq
x^2 \; h_{PS}(x) = \frac{1}{\sqrt{2\pi}} \; \frac{5+n}{6 \alpha} \; x^{(n+5)/6} \; e^{- x^{(5+n)/3} / (2 \alpha^2)}  \label{hxPS}
\eeq
This is due to the fact that the usual Press-Schechter scaling parameter $\nu$ can be expressed as a function of the sole non-linear scaling parameter $x$:
\beq
\nu = \frac{1}{\alpha} \; x^{(5+n)/6}
\eeq
and the mass fraction in objects within a given range of mass is simply:
\beq
\mu(M) \frac{dM}{M} = \frac{1}{\sqrt{2\pi}} \; \nu \; e^{-\nu^2/2} \; \frac{d\nu}{\nu} = x^2 \; h_{PS}(x) \; \frac{dx}{x}  \label{massfrac}
\eeq 
The existence of the relation $\nu - x$ is made possible thanks to two effects: i) in this range $\xia \simeq \delta_+$ so $\delta$ and $\xia$ follow the same evolution, and ii) the power-spectrum is a power-law. We can notice that these functions $h_{PS}(x)$ satisfy the normalization
\beq
\int_0^{\infty} x \; h_{PS}(x) \; dx = \frac{1}{2} \label{shPS}
\eeq
Naturally, this expresses again that overdensities only contain half of the total mass, as we saw above. The normalization of $h_{PS}(x)$ is ``correct'' in this sense, although (\ref{PdelhxPS}) is valid only for $\delta>\Delta_c$, because all overdensities will eventually reach and grow beyond this threshold. Hence at large times the formulation (\ref{PdelhxPS}) is valid for all overdensities, except a vanishing fraction. 
To cure this problem we multiply arbitrarily
the PS formula by a normalization factor of 2, a procedure that, as discussed earlier, is not justified.
 In this case the mass function is assumed to be given by the function $2 \; h_{PS}(x)$.

\subsubsection{Moments implied by the PS approximation}

Thus, the effective PS density field has been seen in the previous section to scale in the same way as the actual non-linear field (for large overdensities), but to differ quantitatively as is readily shown by the normalization problem. When the PS mass function is multiplied by the usual (but rather arbitrary) factor 2, we can calculate the scale-invariant coefficients $S_p$ implied by the above density field since all the mass is now described by the function $2 \; h_{PS}(x)$. Then we can compare them with the values obtained in numerical simulations to describe the difference between these two density fields. Indeed, using (\ref{Snhx}) where we replace $h(x)$ by $2\;h_{PS}(x)$ we get:
\beq
p \geq 1 \; : \;\;\; S_p = \frac{1}{\alpha \sqrt{2 \pi}} \;\; (2 \alpha^2)^{\frac{6 p+n-1}{2 (5+n)}} \;\;\; \Gamma \left[ \frac{6 p+n-1}{2 (5+n)} \right] 
\eeq
Naturally $S_1=1$ since we normalized the mass function to unity, but generally we get $S_2 \neq 1$. This contradiction (since $\int x^2 \; h(x) \; dx = 1$ by definition of $\xia$) means that one cannot simply multiply the PS mass function by a factor 2 to get an acceptable mass function, as the additional constraint $S_2=1$ has to be fulfilled. Thus the PS density field is not only inexact for the overall density it predicts in the clustered regions, but also for the correlation function it implies (after correction by the factor 2) which is slightly different from the actual non-linear correlation function. This shows very clearly that the PS formula, even with a corrected normalization, cannot describe exactly the mass function of non-linear objects, and does not give the exact value of the parameter $\alpha$. Naturally, it is possible to ``correct'' the evolution of individual overdensities by a parameter $\alpha'$, as we did for $\xia$, such that $S_2=1$, which means that we would get the correct two-body correlation function $\xia$. However, such a refinement is certainly illusory, as such a simple model is probably too crude to be cured significantly by such a small correction. For completeness, table 1 gives the coefficients $S_p$, for $p \leq 5$, obtained in this way, compared with those given by numerical simulations from Colombi et al. (1995). 

\begin{table}
\begin{center}
\caption{Coefficients $S_p$ from the density field implied by the PS prescription and from numerical simulations (in parenthesis), for various indexes $n$ of the power-spectrum. Note that $S_2=1$ by definition. In the PS case, $S_1$ has been readjusted to unity; without this modification, all values of $S_p$ given by this model should be divided by 2.}
\begin{tabular}{ccccc}\hline

$n$ & $S_2$ & $S_3$ & $S_4$ & $S_5$ \\ 
\hline\hline
\\ 

0 & 1.25 & 2.78 & 8.31 & 30.3 \\
 & (1) & (3.54) & (19.9) & (158) \\

-1 & 1.10 & 2.62 & 9.10 & 40.5 \\
 & (1) & (5.62) & (56.2) & (891) \\

-2 & 1.19 & 4.23 & 25.2 & 209 \\
 & (1) & (11.2) & (316) & (12589) \\

 & & & &  \\

\end{tabular}
\end{center}
\label{table1}
\end{table}

We can see that the coefficients $S_p$ we obtain are quite far from those given by numerical simulations. This means that the PS mass function must differ appreciably from the actual one, and that a correction by a factor $\alpha'$ to get $S_2=1$ cannot be sufficient.

\subsubsection{Numerical comparison of the non-linear and the PS scaling functions}

\begin{figure}[htb]

\begin{picture}(230,160)(-18,-15)

\epsfxsize=8.22 cm
\epsfysize=12 cm
\put(-14.5,-117){\epsfbox{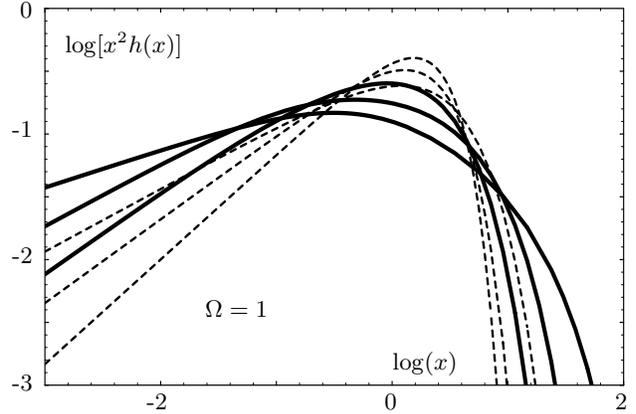}}
\put(38,-25){-2}
\put(128,-25){0}
\put(215,-25){2}
\put(-13,-18){-3}
\put(-13,30){-2}
\put(-13,78){-1}
\put(-10,123){0}
\put(60,10){$\Omega=1$}
\put(7,110){$\log[x^2 h(x)]$}
\put(131,-10){$\log(x)$}

\end{picture}

\caption{Scaling functions $h(x)$ given by numerical simulations (solid line) and the PS prescription (dashed-line), in the cases $n=-2, \; n=-1$ and $n=0$. We multiplied the PS result, eq.(\ref{hxPS}), by the usual factor 2 so that the normalization is the same for both functions. The exponential cutoff of $h(x)$ gets sharper as $n$ increases.}
\label{fighPSNL}

\end{figure}

Indeed, the functions $h_{PS}(x)$ we get from (\ref{hxPS}) have a different shape from those we considered in the context of the non-linear approach. We still have a power-law multiplied by an exponential cutoff, but this cutoff is no longer a simple exponential but the exponential of a power-law. In terms of the function $\varphi(y)$ it means that for $n>-2$ the singularity $y_s$ is repelled to infinity $y_s=-\infty$ while for $n<-2$ it goes to 0: $y_s=0$. However, in the case $n=-2$ we get:
\beq
n=-2 \; : \;\; h_{PS}(x) = \frac{1}{2.18 \sqrt{2\pi}} \; x^{-3/2} \; e^{-x/2.38}
\eeq
which corresponds to $\omega=1/2 \; , \; \omega_s=-1/2$ and $x_*=2.38$. We can note that this value of $x_*$ is much smaller than the value $x_* \simeq 18$ obtained by Colombi et al. (1996) in numerical simulations. In fact, the functions $h(x)$ we get from (\ref{hxPS}) differ from those found by Colombi et al. (1996) by several orders of magnitude after the cutoff, see Fig.\ref{fighPSNL}. We can note that for all these power-spectra, the functions $h(x)$ have a shallower power-law for small $x$, and a smoother cutoff at large $x$. Moreover, if we ``correct'' $h_{PS}(x)$ by a factor $\alpha'>1$ so that $S_2=1$, the exponential cutoff of $h_{PS}(x)$ becomes even sharper, which encreases even more the difference with the actual scaling function $h(x)$.

Fig.\ref{fighPSNL} shows the scaling functions $h(x)$ given by numerical simulations and the PS prescription, for different indexes of the power-spectrum. The difference between the two approaches is of the same order as the difference obtained by varying the initial spectrum of density fluctuations. This implies that conclusions derived for the latter using the PS formulation should be reexamined.

\subsubsection{Discussion of the PS approach}

The important differences which we have seen to exist between the non-linear and the PS based formulations cast some doubts on the latter. Indeed, the fact that the high tail of the functions $h(x)$ we get is not consistent with what is observed, that we do not recover the power-law $h(x) \propto x^{\omega-2}$ for small overdensities in accordance to the condition (\ref{cond1}), because of (\ref{PSgaussian}), that the coefficients $S_p$ are far from those obtained by numerical simulations, and that half the mass remains locked in underdensities, show that {\it there is some redistribution of mass during the process of gravitational clustering}, which prevents one to follow the evolution of a given overdensity which will lose its identity sooner or later. Besides, the spherical collapse model supplemented by virialization itself may be inadequate. Moreover this cannot be cured by a simple change of normalization, since the shape itself of the probability distribution has to be modified, even for large density contrasts. Hence it would seem quite surprising that the PS formulation predicted accurately the results of gravitational clustering. It is actually quite simple to understand why the non-linear distribution (Fig.\ref{fighPSNL}) is {\it systematically broader} than the PS one: assuming that every initial overdensity evolves independently amounts to assume the existence of {\it conservation laws}. Relaxing the latter constraint increases the available phase-space and thus yields {\it broader distribution}. Whence we expect this to hold also for $\Omega < 1$ as well as for $\Lambda \ne 0$. We can note that for $n<-2$ this would imply that $y_s=0$, which is however a bit surprising.

\subsubsection{$\Omega<1 \; , \; \Lambda=0$}

For an open universe we now have an additional time dependence as compared to the previous case $\Omega=1$. Thus, the system (\ref{deldelL}) becomes:
\beq
\left\{ \begin{array}{rcl} \delta & = & {\cal F}[\delta_L,a] \\ \\ R_M^3 & = & (1+\delta) \; R^3  \end{array} \right.  
\eeq
In a fashion similar to the study of the evolution of the correlation function, see (\ref{xiaxiaL2}) and App.E, we obtain the asymptotic behaviour of ${\cal F}$, such that:
\beq
\left\{ \begin{array}{rcl} |\delta_L| \ll 1 & : & {\cal F}(\delta_L,a) \simeq \delta_L   \\  \\ \delta_L \gg 1 & : & {\cal F}(\delta_L,a) \simeq ( \frac{10}{3} )^3 \; d(\Omega)^{-3} \; \delta_L^3  \end{array} \right.
\eeq
Since in the virialized regime $\delta$ and $\xia$ follow the same behaviour the pre-factor $d(\Omega)^{-3}$ disappears in the calculation, and we recover:
\beq
\nu = \frac{1}{\alpha} \; x^{(5+n)/6} \;\;\; \mbox{if} \;\;\; \xia, \; \delta \; > 178 \; d(\Omega)^{-3} 
\eeq
as in the case $\Omega=1$. Since the mass fraction in objects within a given range of mass still verifies by definition the relation (\ref{massfrac}) we get the same scaling function $h_{PS}(x)$ as in the case $\Omega=1$. Hence the remarks we drew in the previous section can be readily extended to the case of an open universe, with $\Lambda=0$. Note however that, contrary to the case $\Omega=1$, all overdensities will not reach and grow beyond the threshold $\delta > \Delta_c$. Hence, although $\int \; x \; h_{PS}(x) \; dx = 1/2$ some overdensities will never be described by the scaling function $h_{PS}(x)$. The latter corresponds to overdensities such that $\delta_L > 10$ and $\delta > \Delta_c(\Omega=1) \; d(\Omega)^{-3}$.

Thus, the scaling function $h_{PS}(x)$ we obtain is independent of $\Omega$, as long as we restrict ourselves to scales ($\xia$) and objects ($\delta$) which collapsed while $\Omega \simeq 1$. This is due to our hypothesis of stable clustering, which implies that structures do not evolve once they are formed. This is consistent with the fact that the dependence on redshift of the correlation function $\xia$ is entirely accounted for by the factor $d(\Omega)$, with a constant parameter $\alpha$ (see Chapter 3 and App.E), as shown by numerical simulations (Peacock and Dodds 1996).

\subsection {Analytical comparison of the PS and the non-linear mass functions}

\subsubsection {Direct comparison of analytical expressions}

 	Although the above considerations on the non-linear density field that would give exactly the PS counts are instructive for understanding this approximatiom, there is a more direct way to compare it to the non-linear mass function, which needs only to assume that the model (Ch.3) for the evolution of the correlation function (which fits well into the picture given by the numerical simulations) is correct.

	Using the correspondance between linear and non-linear correlation functions ((\ref {Padd1}) and (\ref{xiaxiaL1})) only, we can directly express (for $\Omega = 1$ and $z = 0$) the ratio $\nu = \delta_c/\Sigma(M)$ appearing in the PS mass function (\ref {etaM}) as a function of the parameter $x$ of the non-linear formulation. Indeed, by definition of the usual PS formulation we have 
\beq
\delta_c = 3/10 \; \Delta_c^{1/3}
\eeq 
for just-virialized objects, since it assumes $\alpha=1$. Our model for the evolution of the correlation function leads to:
\beq
\Sigma(M) = 3 \alpha/10 \; \Delta_c^{1/3} \; \left( \frac{\xia(R)}{\Delta_c} \right)^{\frac{5+n}{6}}
\eeq 
This gives
\beq
\nu = \frac{1}{\alpha} \left( \frac{\Delta_c}{\xia(R)} \right)^{\frac{5+n}{6}}
\eeq 
Thus we obtain again a relation between the PS scaling parameter $\nu$ and the parameter $x=\Delta_c/\xia$. Since the mass fraction in objects within a given range of mass still verifies (\ref{massfrac}) we see that the usual PS mass function (\ref{etaM}) for just-virialized objects has the same scaling as the non-linear one (\ref{etahM}) with the function $h_{PS}$ defined in the previous section, and a scaling parameter $x = (\alpha \nu)^{6/(5+n)}$.

	This consideration is important as it directly compares the PS prescription, which evaluates the mass function of just-virialized objects (\ref{etaM}), to the same quantity given by the non-linear formulation (\ref{etahM}). Thus it does not rely on the assumption of stable clustering for individual halos, as in the previous sections. Moreover, the fact that both scaling functions $h(x)$ and $H(x)$ are very close (see App.C) shows this comparison is valid although $h(x)$ was originally obtained from the counts-in-cells.

\subsubsection{Broadness of the distribution}

We have seen in the previous section on Fig.\ref{fighPSNL} that the slope of $h(x)$ for small $x$ is smaller than the slope of $h_{PS}(x)$, and that the large $x$ cutoff should be smoother. We have also argued that this should probably be true for any power spectrum or cosmological parameters $\Omega \; , \; \Lambda$. This broadening obviously is translated into the mass functions calculated in the two approaches. It is also readily seen on the figure that the non-linear processes, which are certainly the main cause for the deviations of the actual $h(x)$ from the PS approximation (rather than the corrections due to the cloud-in-cloud problem, as discussed in section 2.3), tend to broaden the scaling function $h(x)$ (or the corresponding mass function). The position of the maximum of $x^2 h(x)$ is always close to $x \sim 1$ because of the definition of $x$, since $\xia \sim \delta_+$ in the virialized regime where $\delta_+$ is the characteristic overdensity of non-linear objects at the considered scale, by definition of $\xia$ (see App.E).

\subsubsection{Earlier discussions of the PS approach}

 Blanchard et al.(1992) compare the actual mass function with the PS approximation. These authors assume that the fraction of mass embedded in non-linear objects of mass larger than $M$, defined in the present case by an actual density contrast $\delta$ larger than a density threshold $\Delta_c$, depends linearly on the initial conditions as:
\beq
\int_{M}^{\infty} \mu(M) \; \frac{dM}{M} = \int_0^{\infty} s(\delta_L,\nu) \; F(\nu) \; d\nu  \label{selec}
\eeq
The integral $\int_{\nu_s}^{\infty} F(\nu) d\nu$ is the fraction of volume, calculated in the {\it linear approximation}, in which each point can be embedded in a sphere such that its mean relative overdensity $\delta = \nu \Sigma(M)$ satisfies $|\nu| >\nu_s$, $\Sigma(M)$ being the r.m.s. fluctuation of $\delta$ at scale $M$ . The kernel $s(\delta,\nu)$ is a selection function which gives the fraction of mass in between $\delta$ and $\delta+d\delta$ that ends up in a non-linear object of mass larger than $M$ for a threshold $\nu$. They do not give  $s$ or $F$, but discuss the mass function according to various assumptions made for these functions. They use for $F$ the usual gaussian expression $e^{-\nu^2}$ obtained from linear theory (but however argue that, because of the cloud-in-cloud problem $F(\nu)$ may be different). Taking for granted that $s$ has a sharp threshold, they show that the non-linear mass function is exactly given by the PS expression (but again, discuss also  other possibilities at the faint mass end). We have seen that the actual mass function differs sensibly from the PS approximation, contrary to the conclusion of Blanchard et al.(1992). This is because they only consider the possible variations of $F(\nu)$: the sharp threshold assumption for $s$, that seemed at first sight innocent, indeed implies a one-to-one correspondance between the initial overdensities and the actual non-linear ones, as can be seen by simple derivation of (\ref {selec}), in the same way as has been originally assumed in the derivation of the PS expression.Their derivation that the non-linear mass function is equal to the PS mass function is thus not a prediction, but a consequence of their sharp threshold assumption.

\subsubsection{Non-linear scaling of the PS approach}

Since the PS approximation appears to be plagued by many problems, one may wonder why we recover the {\it scaling law} (\ref{PdelhxPS}) for large densities. This is simply due to the fact that in such a model overdensities are ``frozen'' after virialization: their radius and their density do not evolve any longer so their density contrast grows as $a^3$. Indeed, the many-body correlation functions $\xia_n$ are mainly sensitive to the rare high-density peaks, and increasingly so with larger $n$. Thus we may write $\xia_n \sim <\delta^n> \sim \sum \eta_i \; \delta_i^n \propto a^{3(n-1)}$ similarly to (\ref{etapetam}), where $\eta_i$ is the fraction of volume occupied by overdensities $\delta_i$ and $\eta_i \propto a^{-3}$ since the radius of these overdensities remains constant. Then the coefficients $S_n = \xia_n/\xia^{\;(n-1)}$ are constants, which implies the scaling (\ref{PdelhxPS}), as we recalled in Chapter 2 and App.B. As a consequence, the fact that overdensities ``freeze'' after they reach the threshold $\Delta_c$ is sufficient to get the correct scaling law for large overdensities. We can note that such an argument may also apply to low density universes $\Omega<1$ where structure formation stops after $\Omega$ gets sufficiently small. In fact, the scale invariance of the coefficients $S_n$, within the highly non-linear regime, could be an even better approximation in this case than for $\Omega=1$ where structure formation never stops. Thus, Padmanabhan et al.(1996) find in numerical simulations that the two point correlation function $\xia$ satisfies the predictions of the stable clustering hypothesis with a better accuracy for an open universe than for a critical universe.

\subsubsection{Away from the deeply non-linear regime}

As we can check on Fig.\ref{fighPSNL}, the non-linear approach gives more numerous large overdensities than the PS prescription. This holds as long as $\xia > \Delta_c$ at the considered scales. When $\xia$ becomes lower than this threshold, it suddenly falls to get close to the linear value $\xia_L$. This decrease in $\xia$ is quite fast, as in the usual sperical collape model it corresponds to $\xia$ decreasing from $\Delta_c \simeq 178$ to $\sim 1$ while $\xia_L$ goes from $\delta_c^2 \simeq 2.8$ to $\sim 1$. Hence, for a given density contrast $\delta$ the parameter $x=\delta/\xia$ suddenly rises, so that the mass function given by the non-linear approach decreases sharply. Moreover, when $\xia \leq 1$ the function $h(x)$ itself changes to the one given by the quasi-gaussian prediction, which is characterized by a stronger exponential cutoff (smaller $x_*$). This enhances even further the sudden falloff of the number of these objects. We can also verify this effect if we perform a comparison between the non-linear and PS approaches in the regime $\xia \sim 1$ and $\delta > \Delta_c$. We now have:
\beq
\nu = \xia^{\;(2+n)/6} \; \frac{3\alpha}{10} \; x^{(5+n)/6} 
\eeq
Thus there is no longer a scaling in $x$, except for $n=-2$. However, even in this case the exponential cutoff is much smaller since now $x_* = 2 (10/3\alpha)^2 \simeq 22$. In fact, for $n=-2$ the quasi-gaussian function $h(x)$ has $x_* \simeq 4$ (see Bernardeau 1994), so we see that the multiplicity function it predicts will suddenly fall when $\xia \sim 1$, even below what the PS prescription gives. This means that in the non-linear picture of structure formation through gravitational clustering, objects appear suddenly at a redshift such that the two-points correlation function taken at their scale becomes larger than unity. Thus the evolution we obtain is much more violent than what is implied by the PS approximation: we get more objects in the recent past, but they suddenly disappear at a given redshift, which defines a well-determined redshift of structure formation. The latter will depend on the astrophysical objects one considers (more precisely on their scale).

\subsection{Mass functions}

As we recall in App.A, the comoving mass function for virialized objects, at a redshift $z$, given by the PS prescription (which we multiply by the usual factor 2), is:
\beq
\eta(M) \frac{dM}{M}  = \sqrt{\frac{2}{\pi}} \frac{\rho_0}{M} \frac{\delta_c(z)}{\Sigma_0} \left| \frac{\mbox{dln} \Sigma_0}{\mbox{dln}M} \right| e^{-\delta_c(z)^2/(2\Sigma_0^2)}  \; \frac{dM}{M} 
\label{ps}
\eeq
where, following the notations of App.D,
\[
\Omega = 1 \; : \;\; \delta_c(z) = \delta_{c0} \; (1+z)
\]
and
\[
\Omega < 1 \; , \; \Lambda=0 \; : \;\; \delta_c(z) =  \frac{3}{2} D(t_0) \left[ 1+\left( \frac{2\pi B_b}{t(z)} \right)^{2/3} \right]    
\]
If the power-spectrum is a power-law: $P(k) \propto k^n$, we can write $\Sigma_0(M) = \sigma_0 (M/M_0)^{-(n+3)/6}$, where $\sigma_0$ is the normalization of $\Sigma_0(M)$ at the mass $M_0$, or at the radius $R_0$. Since $R_0$ is defined by $\xia(R_0,z=0)=1$, we have $\sigma_0 \simeq 1$, as we are still close to the linear regime so that $\xia(R_0) \simeq \Sigma_0^2$. If $\Omega=1$, this leads to:
\[
\eta(M) \frac{dM}{M}  = \sqrt{\frac{2}{\pi}} \frac{\rho_0}{M_0} \frac{n+3}{6}  \frac{\delta_{c0}}{\sigma_0} \left( \frac{M}{M_0} \right)^{-(3-n)/6} (1+z)
\]
\beq
\;\;\;\;\;\;\; \times  \exp \left[ - \; \frac{\delta_{c0}^2}{2\sigma_0^2} \; \left( \frac{M}{M_0} \right)^{(n+3)/3} \; (1+z)^2 \right]  \; \frac{dM}{M} 
\eeq
Hence, we can see that the mass function switches from a power-law to an exponential behaviour at $M \sim M_{*PS}(z)$ with
\[
\Omega = 1 \; : \;\; M_{*PS}(z) = M_0 \; (1+z)^{-6/(n+3)}
\]
and more generally:
\[
\forall \Omega \; : \;\; M_{*PS}(z) = M_0 \; \left[\frac{\delta_c(z)}{\delta_{c0}}\right]^{-6/(n+3)}
\]

On the other hand, the non-linear prescription we presented in chapter 2 leads to the comoving mass function:
\beq
\eta(M) \frac{dM}{M} = \frac{\rho_0}{M} \; x \; h(x) \; dx \;\;\;\; \mbox{with} \;\;\;\; x = \frac{1+\Delta_c(z)}{\xia(R,z)}
\label{nl}
\eeq
where $\Delta_c(z)$ is the density contrast at the time of virialization, at the redshift $z$. If $\Omega=1$, $\Delta_c \simeq 178$ is a constant. If we are in the regime of stable clustering with a power-law for $\xia$, we have:
\[
x = \left[(1+\Delta_c(z)) (1+z)^3 \right]^{1-\gamma/3} \; \left( \frac{M}{M_0} \right)^{\gamma/3} 
\]
We define:
\[
M_*(z) = \left[(1+\Delta_c(z)) (1+z)^3 \right]^{-(3/\gamma-1)} \; x_*^{3/\gamma} \; M_0 
\]
Then, for small masses $ M \ll M_* \;$ :
\[
\eta(M) \propto  \left( \frac{M}{M_0} \right)^{-1+\omega \gamma/3} \left[(1+\Delta_c(z)) (1+z)^3 \right]^{(1-\gamma/3) \omega} 
\]
and for large masses $ M \gg M_* \;$ :
\[
\eta(M)  \propto  \left( \frac{M}{M_0} \right)^{-1+ (1+\omega_s) \frac{\gamma}{3}} \left[(1+\Delta_c) (1+z)^3 \right]^{(1-\frac{\gamma}{3}) (1+\omega_s)} 
\]
\[
\hspace{1.3cm}  \times \exp \left[ - \frac{\left[(1+\Delta_c(z)) (1+z)^3 \right]^{1-\gamma/3}}{x_*} \left( \frac{M}{M_0} \right)^{\gamma/3} \right]  \; 
\]
The mass function bends at $M \sim M_*(z)$.
\\

Hence both prescriptions give similar mass functions, a power-law multiplied by an exponential cutoff, but their redshift dependence is very different: the exponent of the redshift term in the exponential cutoff is 2 for the PS prescription, while it is $(3-\gamma) = 6/(5+n)$ for the other mass function, when $\Omega=1$. Indeed, in the case of stable clustering the power-law of the non-linear correlation function is $\gamma= 3(3+n)/(5+n)$. Thus, for a given mass the number density of halos will decrease much more slowly according to the non-linear prescription than the PS formulation would imply if the index $n$ of the power spectrum verifies $n>-2$. In the particular case $n=-2$, the dependence on redshift is the same for both prescriptions. Moreover, both cutoffs $M_*(z)$ and $M_{*PS}(z)$ show the same evolution with redshift. This is consistent with the results seen in sections 4.1. and 4.2., since we noticed that the mutiplicity functions can be written:
\beq
\eta(M) \frac{dM}{M} = \frac{\rho_0}{M} \; \sqrt{\frac{2}{\pi}} \; \nu \; e^{-\nu^2/2} \; \frac{d\nu}{\nu}
\eeq
for the PS approximation (multiplied by the usual factor 2), while
\beq
\eta(M) \frac{dM}{M} = \frac{\rho_0}{M} \; x^2 \; h(x) \; \frac{dx}{x}
\eeq
for the non-linear prescription, with
\beq
\nu \propto x^{(5+n)/6}       \;\;\;\; \mbox{if} \;\;\; \Omega=1
\eeq
Hence both mass functions must evolve in parallel, and their relative position does not change with time: a given value of $x$ corresponds to a constant value of $\nu$ whatever the redshift. The fact that the evolution with time of the mass functions evaluated for a given mass varies from one formulation to the other, is simply due to the fact that their slopes are different. 
\\

In the case of a low-density universe, the relation $\nu - x$ no longer holds, and the multiplicity functions given by both approximations have a different redshift evolution. Indeed, for small scales where clustering is stable (that is $\xia > 178 d(\Omega)^{-3}$) we have for large times ($\Omega \ll 1$): 
\beq
\nu(M,z) \simeq \frac{3 D(t_0)}{2 \Sigma_0(M)}
\eeq
which is independent of $z$ while:
\beq
x \propto M^{(3+n)/(5+n)} (1+z)^{4/(5+n)}
\eeq
Thus, at large times $(1+z) \rightarrow 0$ and $x \rightarrow 0$ for a fixed mass $M$ and linear parameter $\nu$. This means that the small mass end of the multiplicity function given by the non-linear approach decreases with time relative to the PS prediction. However, in the intermediary regime where $1 \ll \xia \ll 178 d(\Omega)^{-3}$, one can show that within the description presented in App.E, the scale defined by $\delta_L = \alpha^6 \; 3/2 \; D(t)$ verifies: 
\beq
\nu \propto x^{(3+n)/6}
\eeq
Hence both multiplicity functions remain close to each other, and follow a similar evolution at these scales. This is natural, since both mass fractions have the same normalization (we must recover the average comoving density of the universe) and their peak is close to $\nu=1$ and $x=1$ by definition of $\Sigma$ and $\xia$, which correspond to similar objects because of the relation $\xia_L - \xia$.

\subsection{Numerical comparison of the mass functions}

We explained in the previous chapter how we evaluate the average two-body correlation function $\xia$ at any redshift once we are given an initial power spectrum. Then, one has to ``choose'' a function $h(x)$, which can for instance be obtained by a fit to the observed present mass function of clusters. The function $h(x)$ has been obtained (Bouchet et al 1991) in a simulation with CDM initial conditions. In the case the primordial power-spectrum is a power-law, $P(k) \propto k^n$, Colombi et al. (1996) showed that different values of $n$ led to different functions $h(x)$ in the deeply non-linear regime. Hence, we can study the mass functions for various initial conditions using these simulations.

\begin{figure}[htb]

\begin{picture}(230,400)(-18,-15)

\epsfxsize=8 cm
\epsfysize=8.5 cm
\put(-5,180){\epsfbox{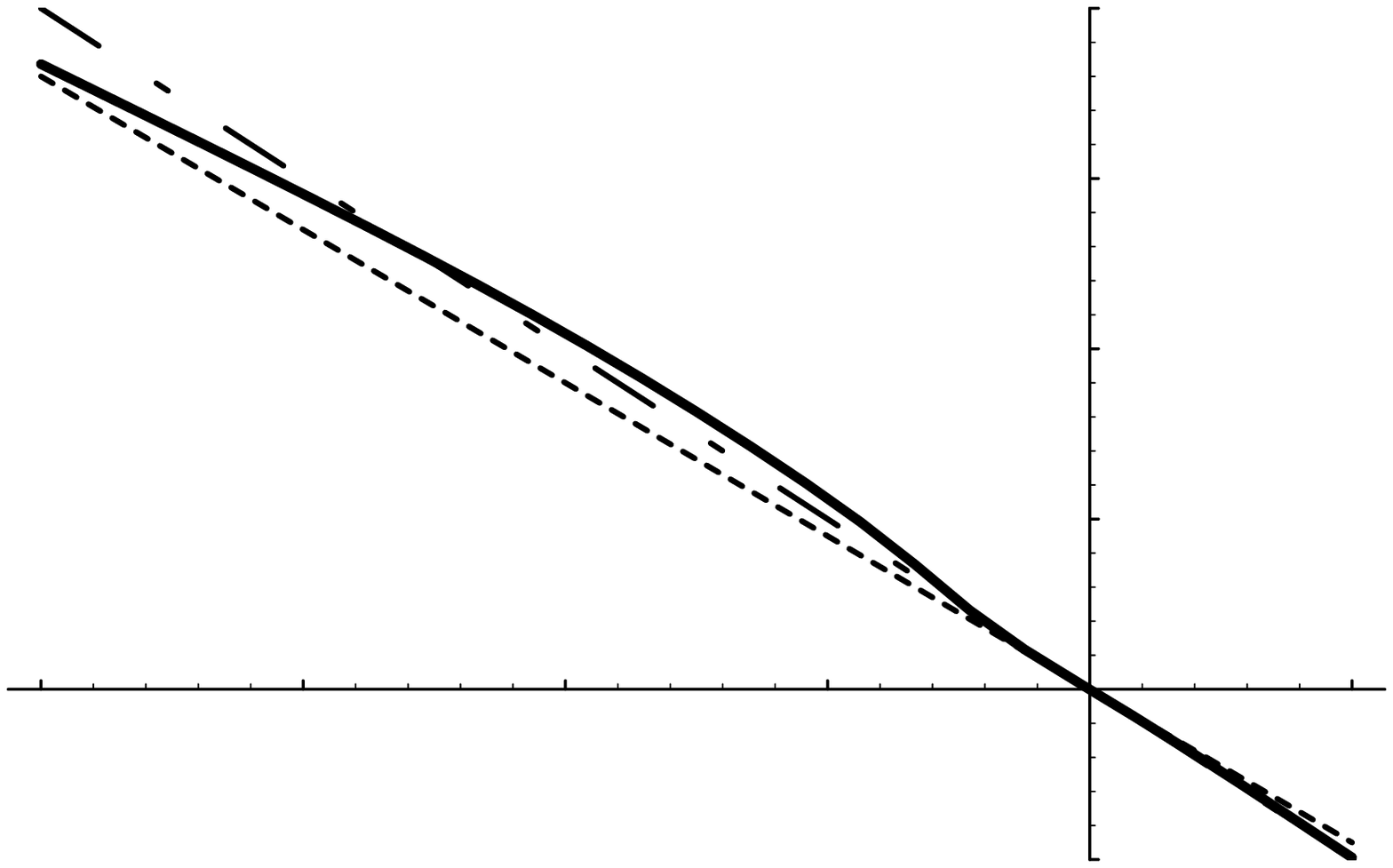}}
\put(-4,261){-2}
\put(83,261){-1}
\put(165,261){0}
\put(163,245){-1}
\put(165,290){1}
\put(165,330){3}
\put(165,360){$\log(\xia)$}
\put(188,273){$\log(R/R_0)$}
\put(20,301){$n=-1$}
\put(20,286){$\Omega=1$}

\epsfxsize=8 cm
\epsfysize=10.5 cm
\put(-8,21){\epsfbox{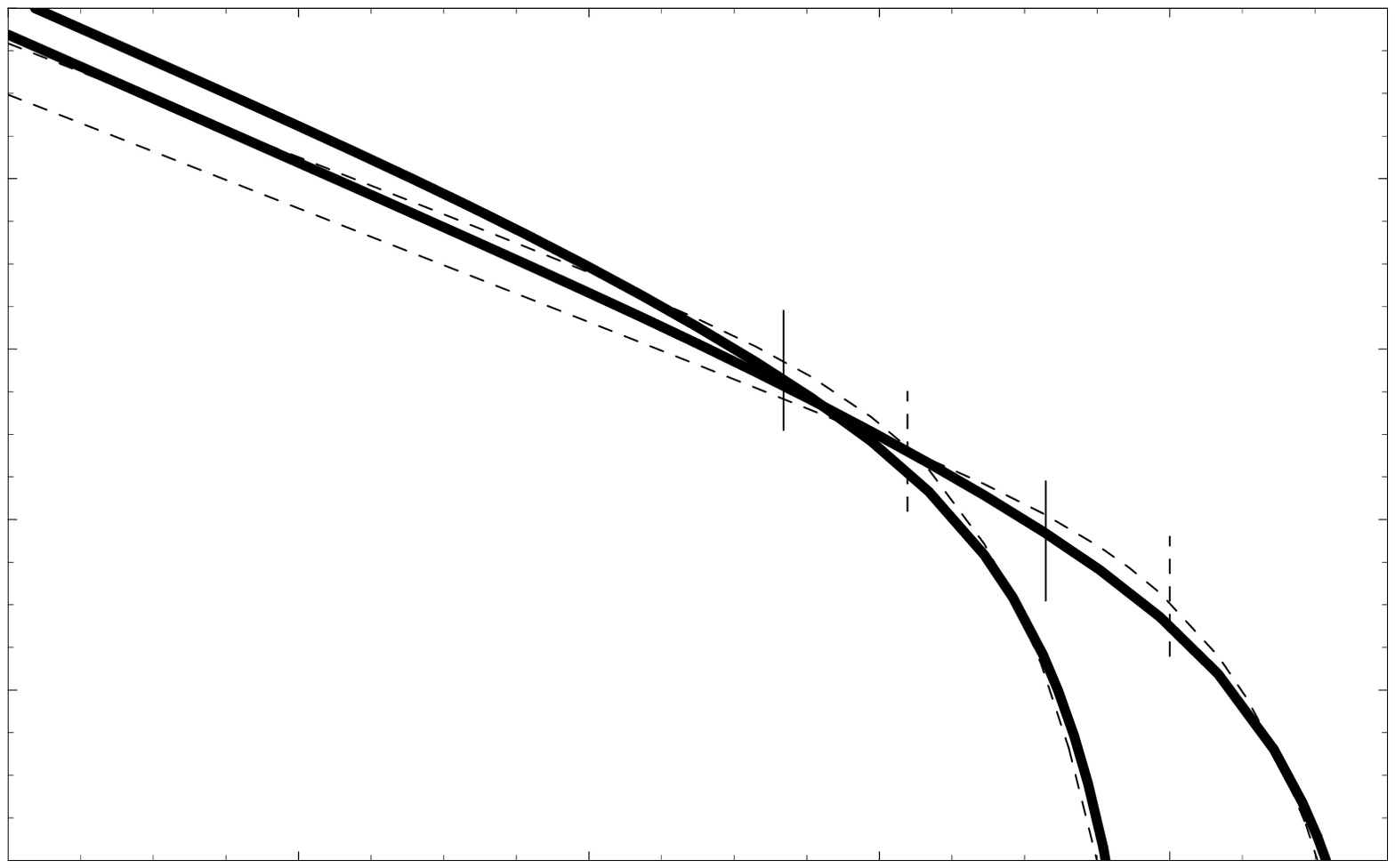}}
\put(-11,132){-2}
\put(-9,184){2}
\put(-9,234){6}
\put(180,162){$M_{*PS}$}
\put(160,170){$M_*$}
\put(136,135){$z=3$}
\put(193,140){$z=0$}
\put(15,145){$n=-1$}
\put(15,130){$\Omega=1$}
\put(160,220){$\log[V_0 \; \eta(M)]$}
\put(115,114){$\log(M/M_0)$}

\epsfxsize=8.22 cm
\epsfysize=10.5 cm
\put(-14.5,-110){\epsfbox{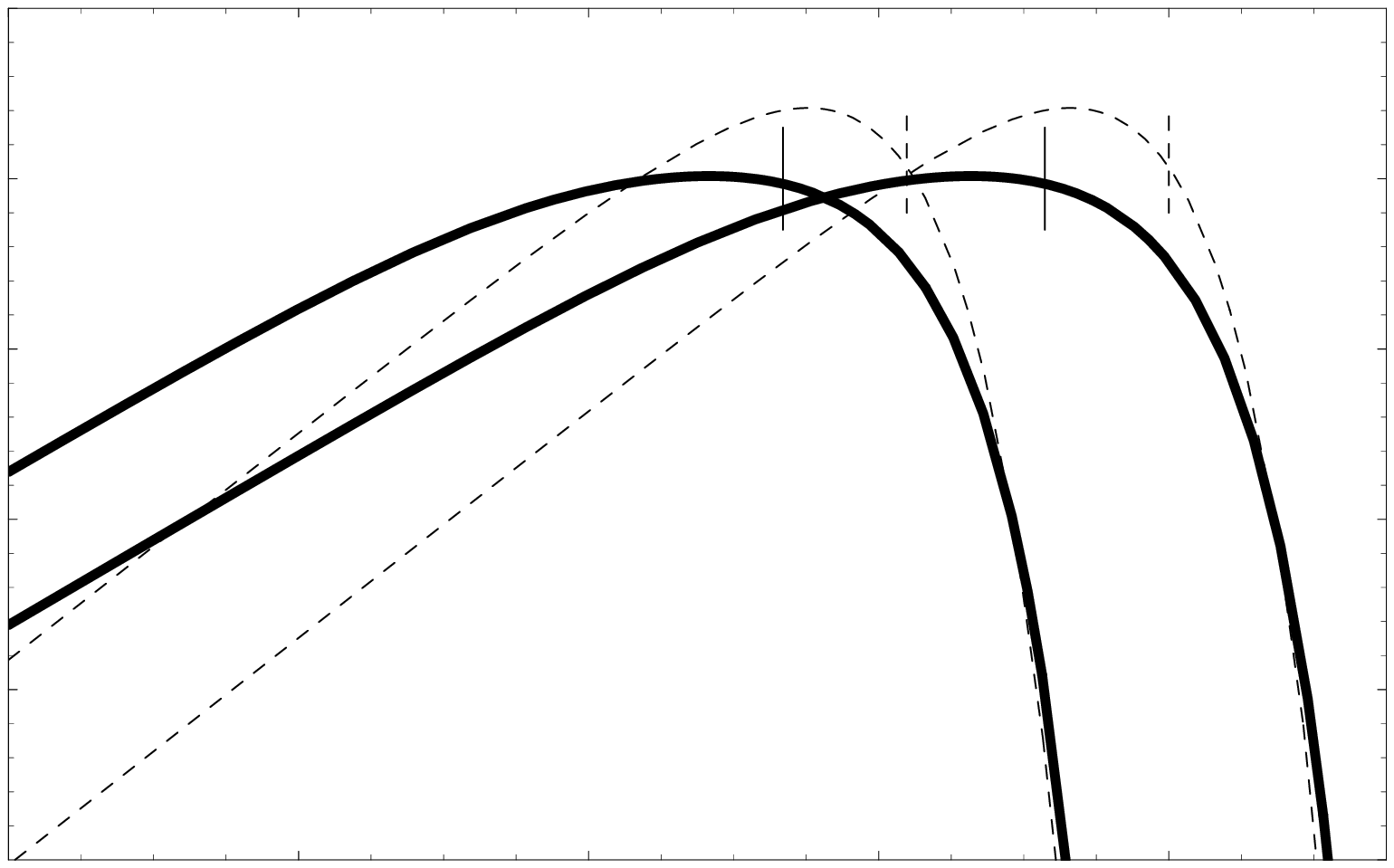}}
\put(40,-28){-6}
\put(88,-28){-4}
\put(134,-28){-2}
\put(183,-28){0}
\put(-11,-22){-3}
\put(-11,27){-2}
\put(-11,77){-1}
\put(180,92){$M_{*PS}$}
\put(160,62){$M_*$}
\put(137,18){$z=3$}
\put(175,35){$z=0$}
\put(60,7){$n=-1$}
\put(60,-8){$\Omega=1$}
\put(7,85){$\log[\mu(M)]$}
\put(115,-15){$\log(M/M_0)$}

\end{picture}

\caption{ {\it Upper figure:} The evolved correlation function $\xia(R,z)$ (solid line), and the linear extrapolation $\Sigma(R)^2$ (dot-dashed line), in the case $\Omega=1 \; , \; P(k) \propto k^{-1} \; (n=-1)$. The short dashed line is the power-law $(R/R_0)^{-\gamma}$ with $\gamma=1.8$.  {\it Intermediate figure:} Comoving mass function $\eta(M) \; dM/M$. The graph shows the quantity $V_0 \; \eta(M)$, for the non-linear prescription (solid line), with $\omega=1/2 \; , \; x_*=5 \; , \; \omega_s=-1.1$, and for the PS prescription (dashed line). The vertical solid lines show the position of $M_*$, while the vertical dashed lines show the position of $M_{*PS}$, for both redshifts. {\it Lower figure:} Mass fraction $\mu(M) \; dM/M$ for the same case.}
\label{figeta1}

\end{figure}

\begin{figure}[htb]

\begin{picture}(230,400)(-18,-15)

\epsfxsize=8 cm
\epsfysize=8.5 cm
\put(-5,180){\epsfbox{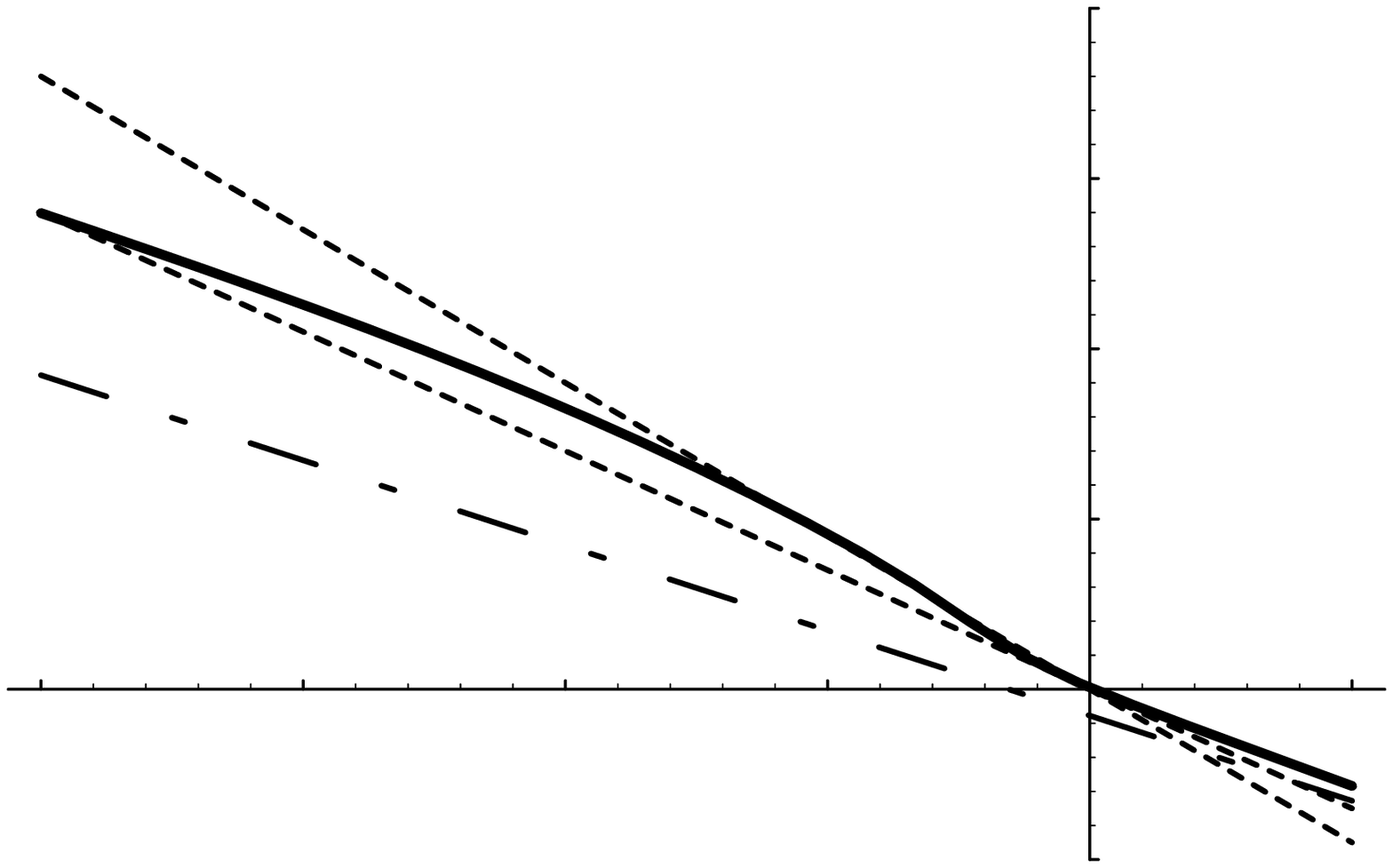}}
\put(-4,261){-2}
\put(83,261){-1}
\put(165,261){0}
\put(163,245){-1}
\put(165,290){1}
\put(165,330){3}
\put(165,360){$\log(\xia)$}
\put(188,273){$\log(R/R_0)$}
\put(100,330){$n=-2$}
\put(100,315){$\Omega=1$}

\epsfxsize=8 cm
\epsfysize=10.5 cm
\put(-8,21){\epsfbox{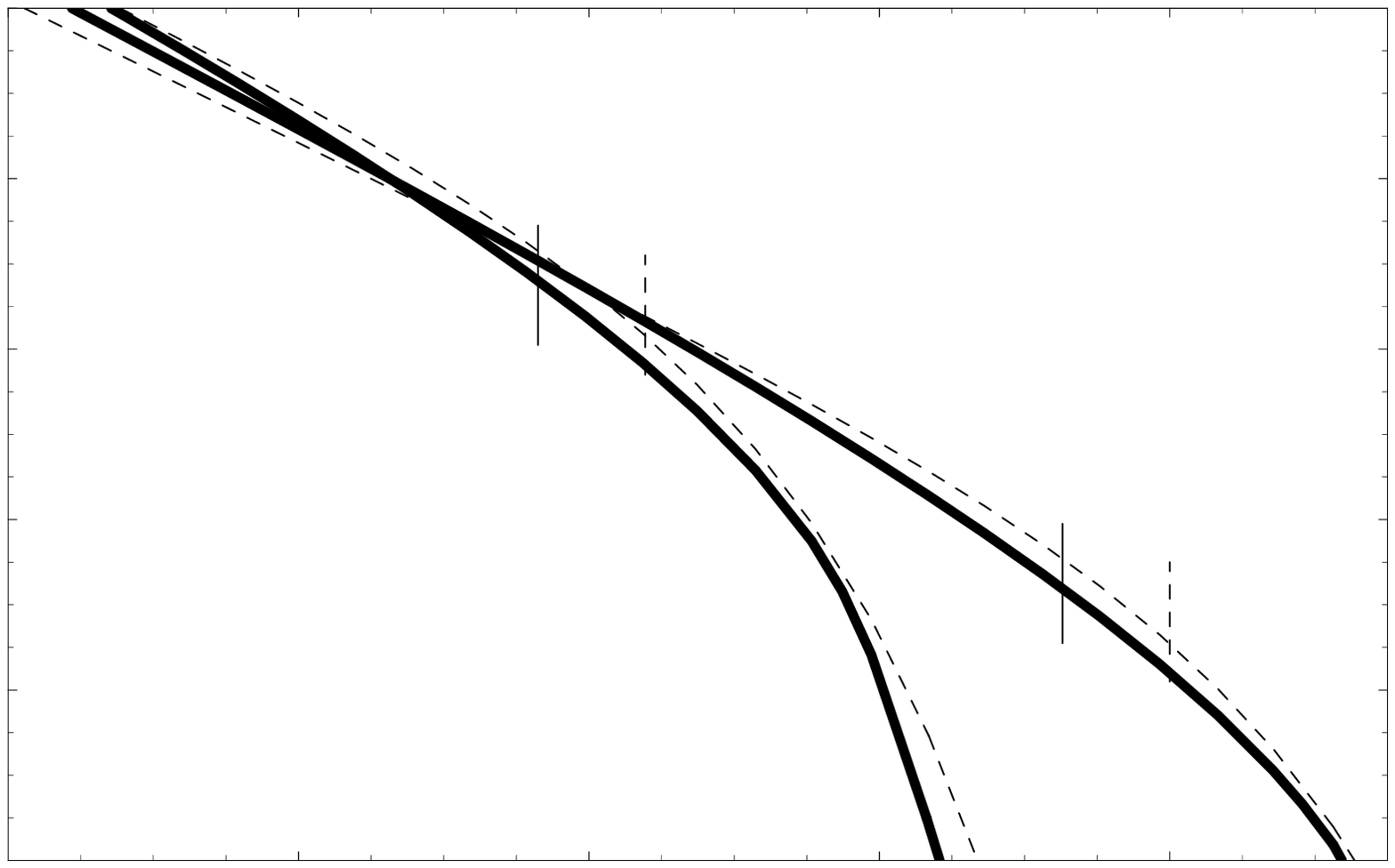}}
\put(-11,132){-2}
\put(-9,184){2}
\put(-9,234){6}
\put(180,160){$M_{*PS}$}
\put(162,165){$M_*$}
\put(105,135){$z=3$}
\put(193,140){$z=0$}
\put(15,145){$n=-2$}
\put(15,130){$\Omega=1$}
\put(160,220){$\log[V_0 \; \eta(M)]$}
\put(88,114){$\log(M/M_0)$}

\epsfxsize=8.22 cm
\epsfysize=10.5 cm
\put(-14.5,-110){\epsfbox{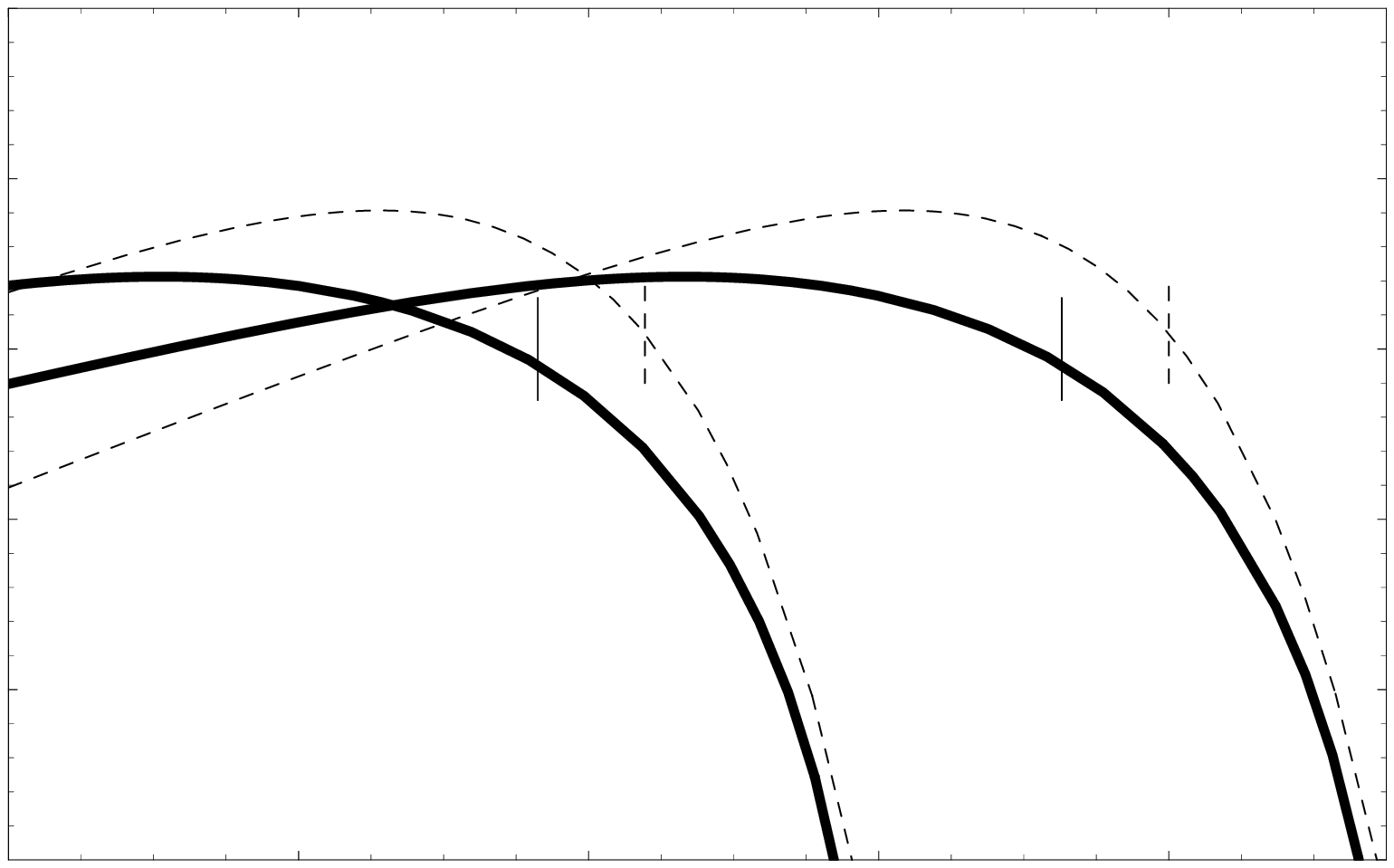}}
\put(40,-28){-6}
\put(88,-28){-4}
\put(134,-28){-2}
\put(183,-28){0}
\put(-11,-22){-3}
\put(-11,27){-2}
\put(-11,77){-1}
\put(180,68){$M_{*PS}$}
\put(160,74){$M_*$}
\put(80,25){$z=3$}
\put(160,30){$z=0$}
\put(30,7){$n=-2$}
\put(30,-8){$\Omega=1$}
\put(7,85){$\log[\mu(M)]$}
\put(145,-15){$\log(M/M_0)$}

\end{picture}

\caption{ {\it Upper figure:} The evolved correlation function $\xia(R,z)$ (solid line), and the linear extrapolation $\Sigma(R)^2$ (dot-dashed line), in the case $\Omega=1 \; , \; P(k) \propto k^{-2} \; (n=-2)$. The short dashed lines are the power-laws $(R/R_0)^{-\gamma}$ with $\gamma=1.8$ and $\gamma=1.4$.  {\it Intermediate figure:} Comoving mass function $\eta(M) \; dM/M$. The graph shows the quantity $V_0 \; \eta(M)$, for the non-linear prescription (solid line), with $\omega=0.3 \; , \; x_*=18 \; , \; \omega_s=-1.6$, and for the PS prescription (dashed line). The vertical solid lines show the position of $M_*$, while the vertical dashed lines show the position of $M_{*PS}$, for both redshifts. {\it Lower figure:} Mass fraction $\mu(M) \; dM/M$ for the same case.}
\label{figeta2}

\end{figure}

\begin{figure}[htb]

\begin{picture}(230,400)(-18,-15)

\epsfxsize=8 cm
\epsfysize=8.5 cm
\put(-5,180){\epsfbox{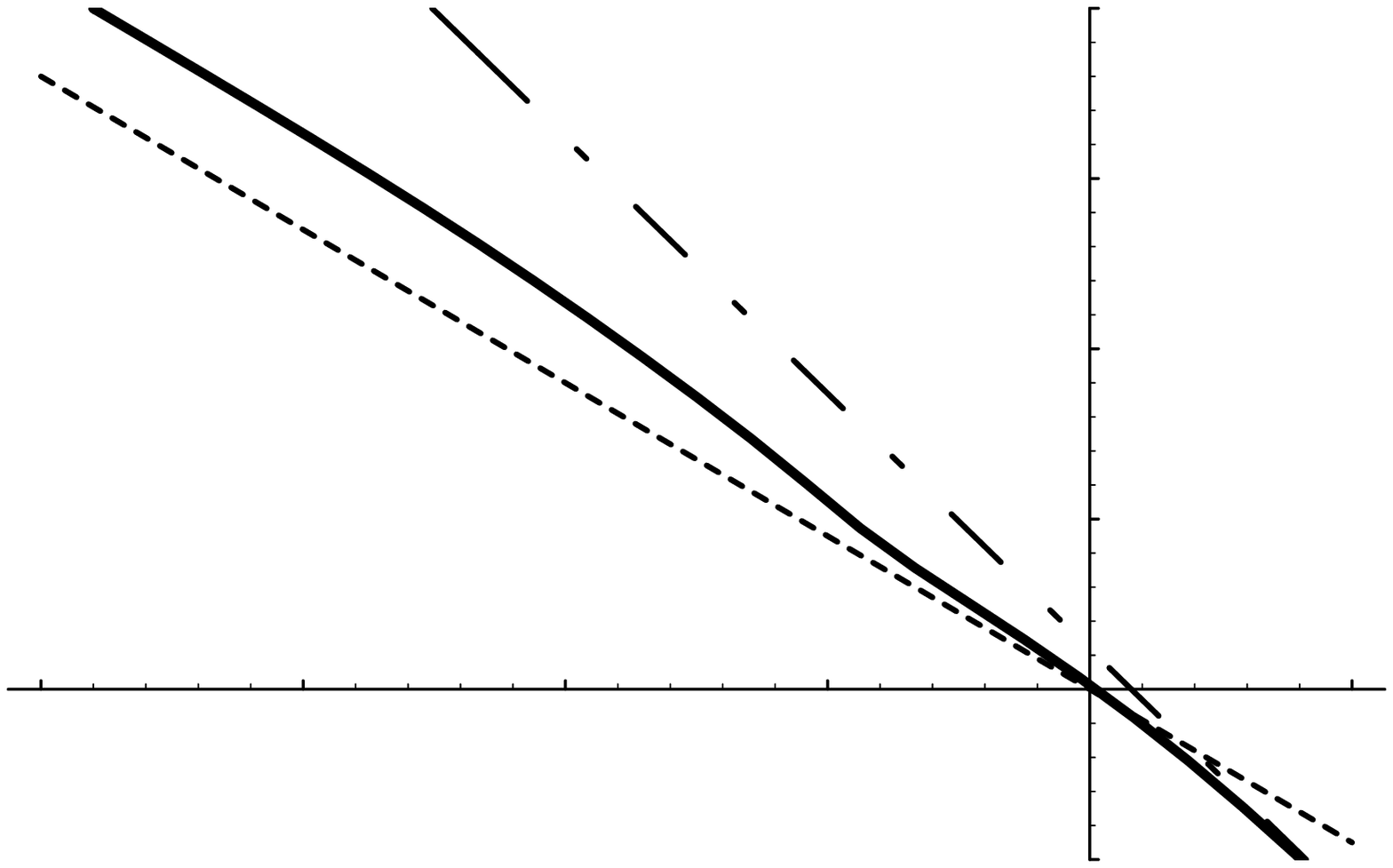}}
\put(-4,261){-2}
\put(83,261){-1}
\put(165,261){0}
\put(163,245){-1}
\put(165,290){1}
\put(165,330){3}
\put(165,360){$\log(\xia)$}
\put(188,273){$\log(R/R_0)$}
\put(20,301){$n=0$}
\put(20,286){$\Omega=1$}

\epsfxsize=8 cm
\epsfysize=10.5 cm
\put(-8,21){\epsfbox{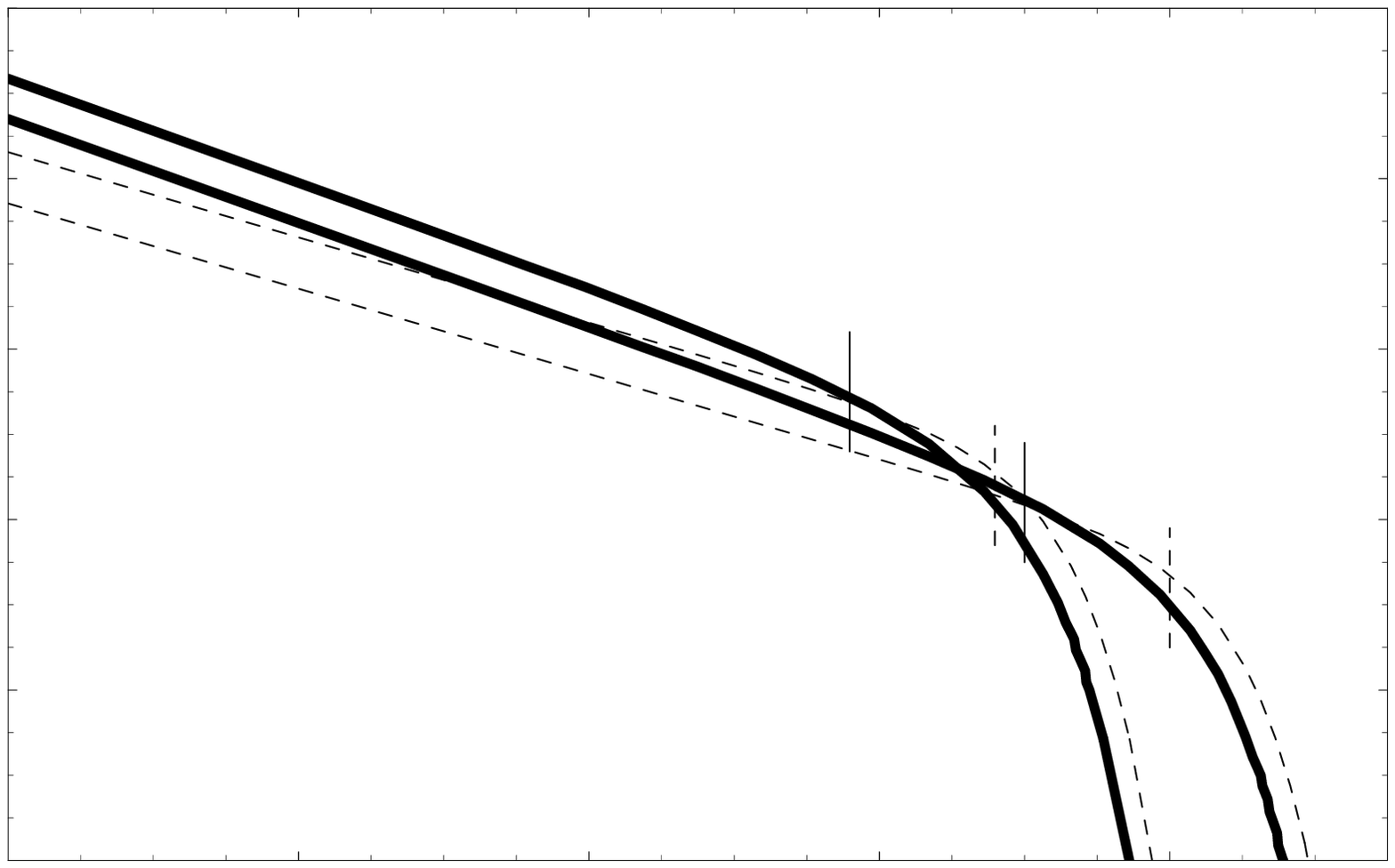}}
\put(-11,132){-2}
\put(-9,184){2}
\put(-9,234){6}
\put(180,162){$M_{*PS}$}
\put(160,177){$M_*$}
\put(142,135){$z=3$}
\put(195,140){$z=0$}
\put(15,145){$n=0$}
\put(15,130){$\Omega=1$}
\put(160,220){$\log[V_0 \; \eta(M)]$}
\put(115,114){$\log(M/M_0)$}

\epsfxsize=8.22 cm
\epsfysize=10.5 cm
\put(-14.5,-110){\epsfbox{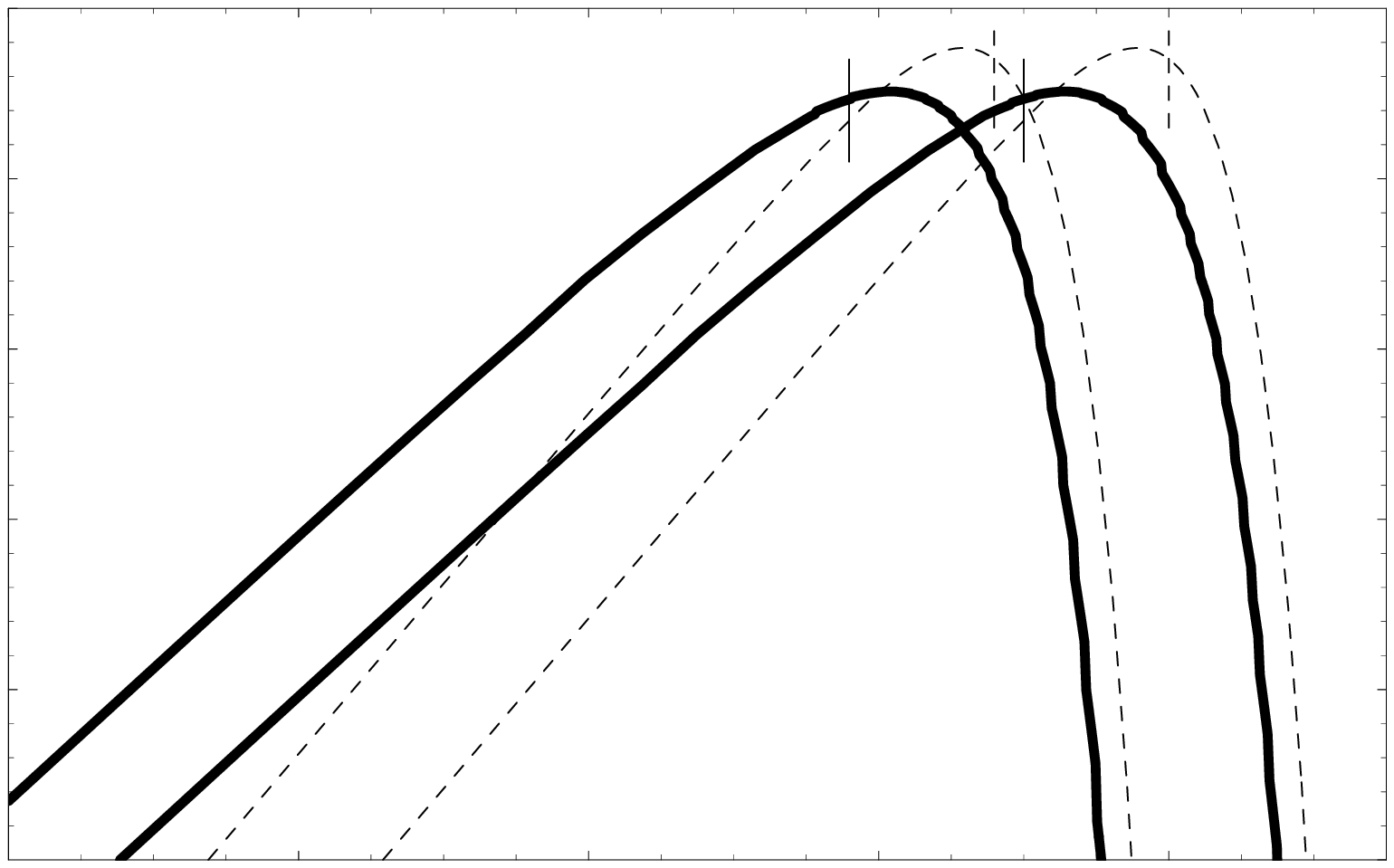}}
\put(40,-28){-6}
\put(88,-28){-4}
\put(134,-28){-2}
\put(183,-28){0}
\put(-11,-22){-3}
\put(-11,27){-2}
\put(-11,77){-1}
\put(186,92){$M_{*PS}$}
\put(118,94){$M_*$}
\put(12,30){$z=3$}
\put(115,35){$z=0$}
\put(70,7){$n=0$}
\put(70,-8){$\Omega=1$}
\put(7,85){$\log[\mu(M)]$}
\put(115,-15){$\log(M/M_0)$}

\end{picture}

\caption{ {\it Upper figure:} The evolved correlation function $\xia(R,z)$ (solid line), and the linear extrapolation $\Sigma(R)^2$ (dot-dashed line), in the case $\Omega=1 \; , \; P(k) \propto k^{0} \; (n=0)$. The short dashed line is the power-law $(R/R_0)^{-\gamma}$ with $\gamma=1.8$.  {\it Intermediate figure:} Comoving mass function $\eta(M) \; dM/M$. The graph shows the quantity $V_0 \; \eta(M)$, for the non-linear prescription (solid line), with $\omega=0.65 \; , \; x_*=2 \; , \; \omega_s=-0.575$, and for the PS prescription (dashed line). The vertical solid lines show the position of $M_*$, while the vertical dashed lines show the position of $M_{*PS}$, for both redshifts. {\it Lower figure:} Mass fraction $\mu(M) \; dM/M$ for the same case.}
\label{figeta0}

\end{figure}

\begin{figure}[htb]

\begin{picture}(230,400)(-18,-15)

\epsfxsize=8 cm
\epsfysize=8.5 cm
\put(-5,180){\epsfbox{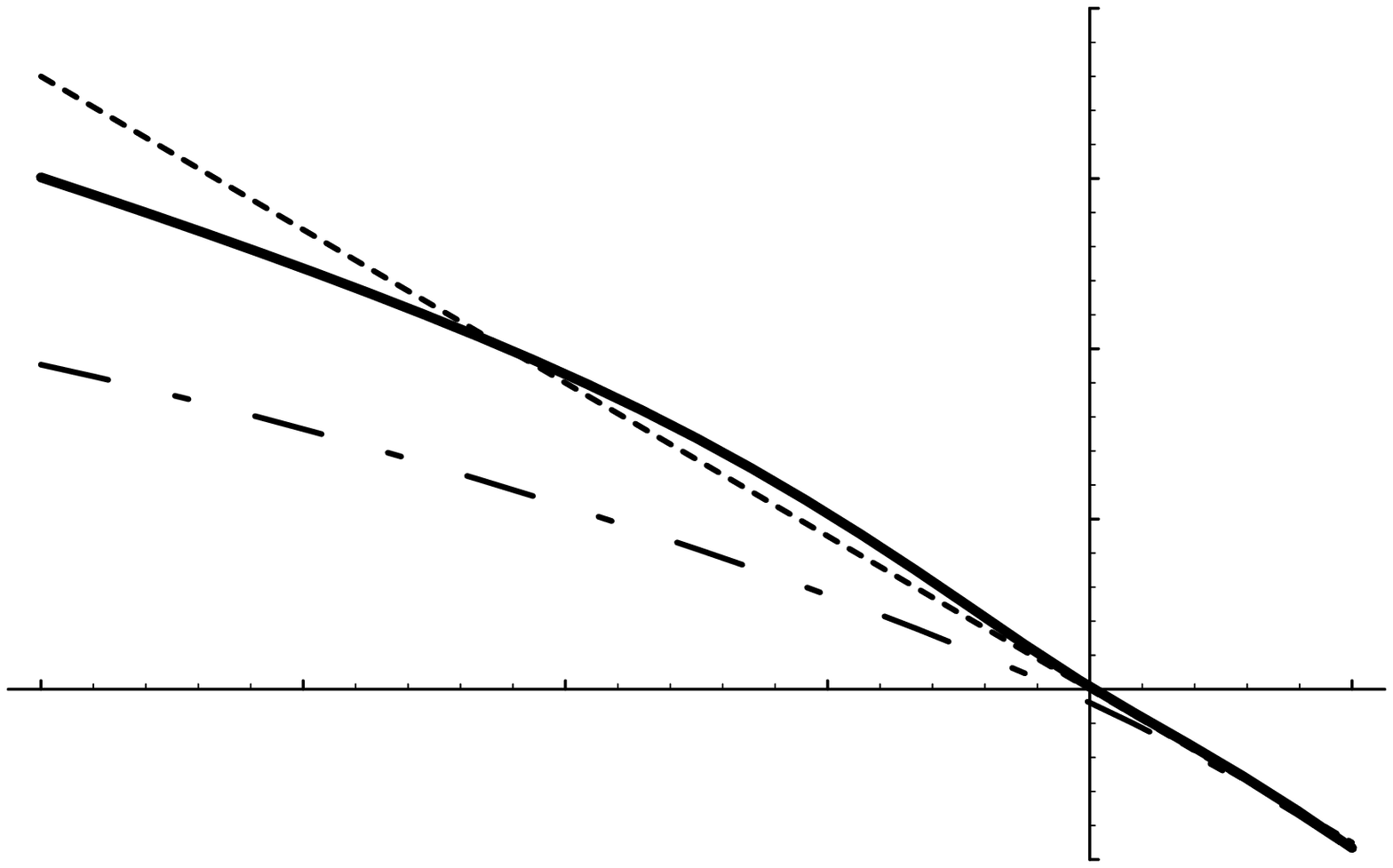}}
\put(-4,261){-2}
\put(83,261){-1}
\put(165,261){0}
\put(163,245){-1}
\put(165,290){1}
\put(165,330){3}
\put(165,360){$\log(\xia)$}
\put(188,273){$\log(R/R_0)$}
\put(100,341){CDM}
\put(100,326){$\Omega=1$}

\epsfxsize=8 cm
\epsfysize=10.5 cm
\put(-8,21){\epsfbox{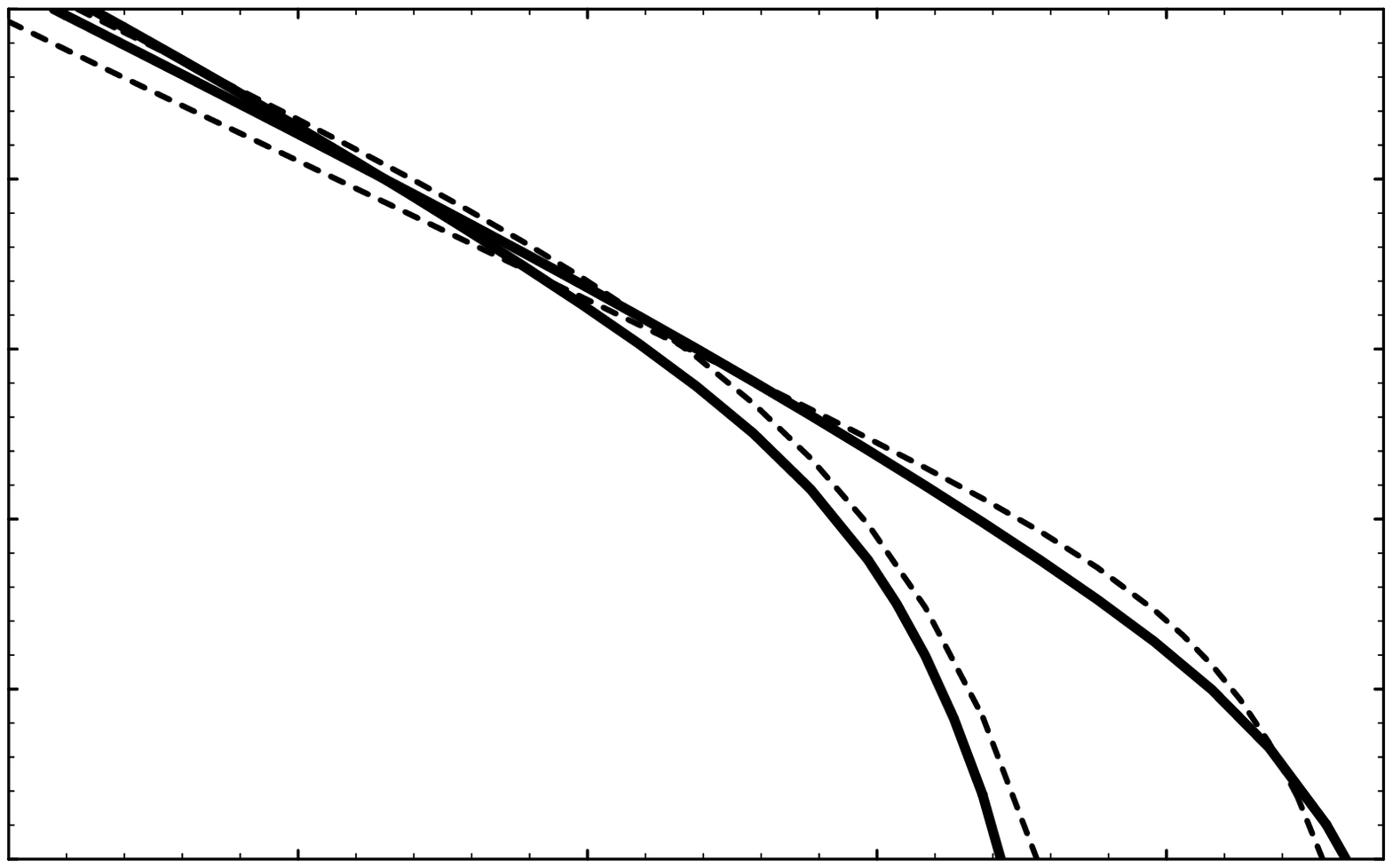}}
\put(-11,132){-2}
\put(-9,184){2}
\put(-9,234){6}
\put(116,135){$z=3$}
\put(193,140){$z=0$}
\put(15,145){CDM}
\put(15,130){$\Omega=1$}
\put(160,220){$\log[V_0 \; \eta(M)]$}
\put(105,114){$\log(M/M_0)$}

\epsfxsize=8.22 cm
\epsfysize=10.5 cm
\put(-14.5,-110){\epsfbox{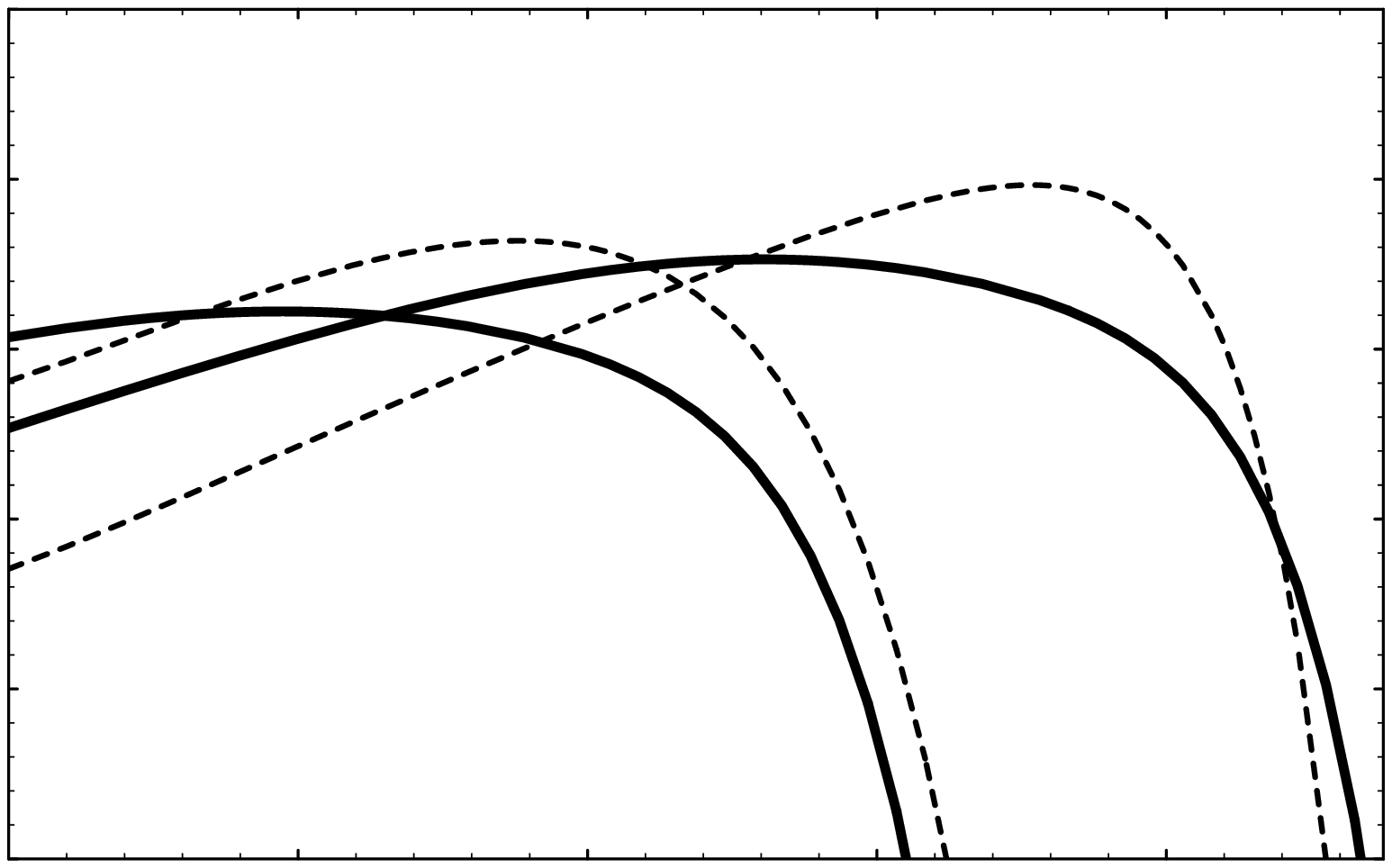}}
\put(40,-28){-6}
\put(88,-28){-4}
\put(134,-28){-2}
\put(183,-28){0}
\put(-11,-22){-3}
\put(-11,27){-2}
\put(-11,77){-1}
\put(142,15){$z=3$}
\put(170,35){$z=0$}
\put(60,7){CDM}
\put(60,-8){$\Omega=1$}
\put(7,85){$\log[\mu(M)]$}
\put(155,-15){$\log(M/M_0)$}

\end{picture}

\caption{ {\it Upper figure:} The evolved correlation function $\xia(R,z)$ (solid line), and the linear extrapolation $\Sigma(R)^2$ (dot-dashed line), in the case $\Omega=1$ for a CDM power-spectrum $P(k)$. The short dashed line is the power-law $(R/R_0)^{-\gamma}$ with $\gamma=1.8$.  {\it Intermediate figure:} Comoving mass function $\eta(M) \; dM/M$. The graph shows the quantity $V_0 \; \eta(M)$, for the non-linear prescription (solid line), with $\omega=0.4 \; , \; x_*=12.5 \; , \; \omega_s=-1.4$, and for the PS prescription (dashed line). The vertical solid lines show the position of $M_*$, while the vertical dashed lines show the position of $M_{*PS}$, for both redshifts. {\it Lower figure:} Mass fraction $\mu(M) \; dM/M$ for the same case.}
\label{figetaCDM}

\end{figure}

\begin{figure}[htb]

\begin{picture}(230,400)(-18,-15)

\epsfxsize=8 cm
\epsfysize=8.5 cm
\put(-5,180){\epsfbox{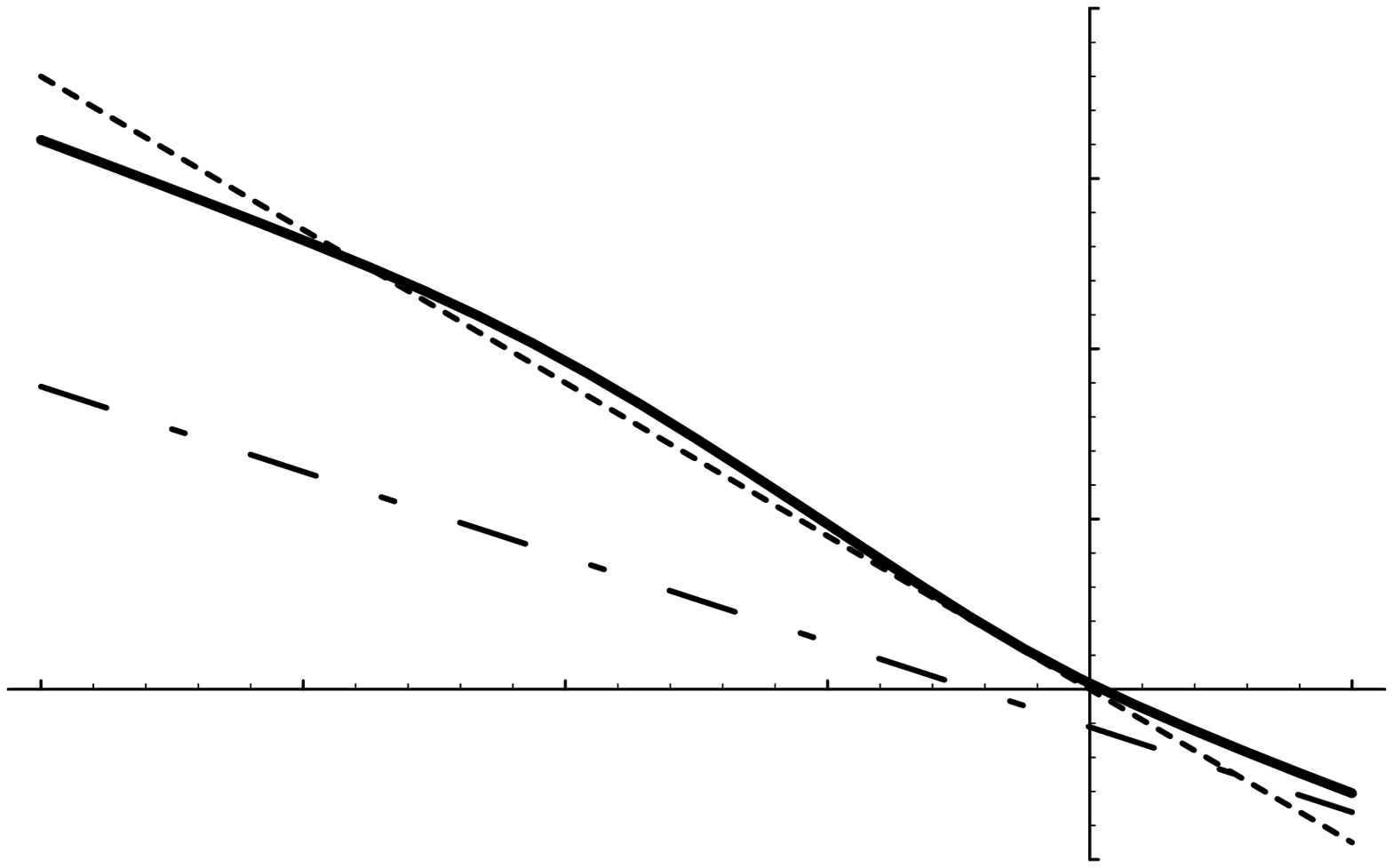}}
\put(-4,261){-2}
\put(83,261){-1}
\put(165,261){0}
\put(163,245){-1}
\put(165,290){1}
\put(165,330){3}
\put(165,360){$\log(\xia)$}
\put(188,273){$\log(R/R_0)$}
\put(20,301){$n=-2$}
\put(20,286){$\Omega_0=0.3 \; , \; \Lambda=0$}

\epsfxsize=8 cm
\epsfysize=10.5 cm
\put(-8,21){\epsfbox{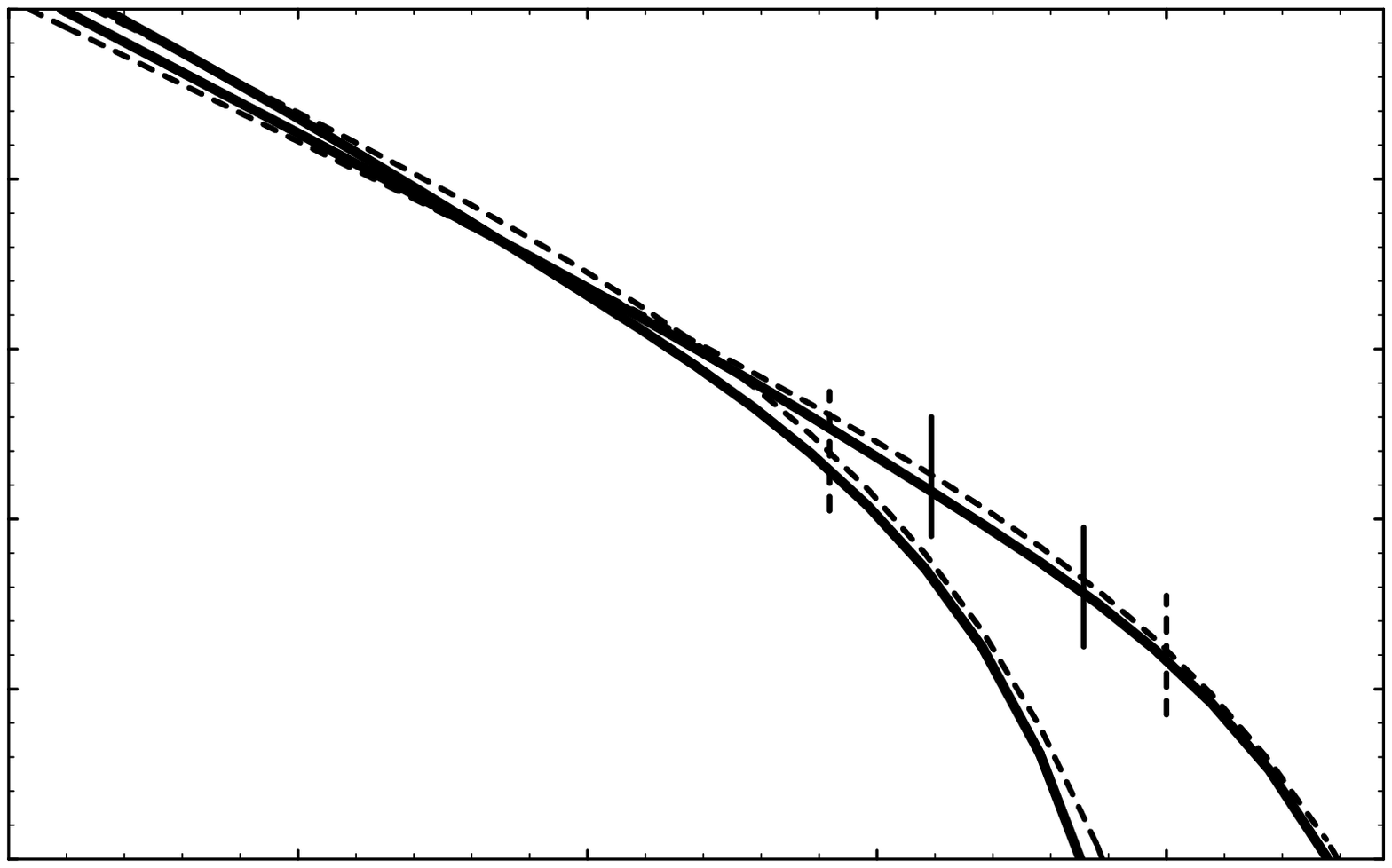}}
\put(-11,132){-2}
\put(-9,184){2}
\put(-9,234){6}
\put(180,153){$M_{*PS}$}
\put(165,163){$M_*$}
\put(130,135){$z=3$}
\put(193,140){$z=0$}
\put(15,145){$n=-2$}
\put(15,130){$\Omega_0=0.3 \; , \; \Lambda=0$}
\put(160,220){$\log[V_0 \; \eta(M)]$}
\put(115,114){$\log(M/M_0)$}

\epsfxsize=8.22 cm
\epsfysize=10.5 cm
\put(-14.5,-110){\epsfbox{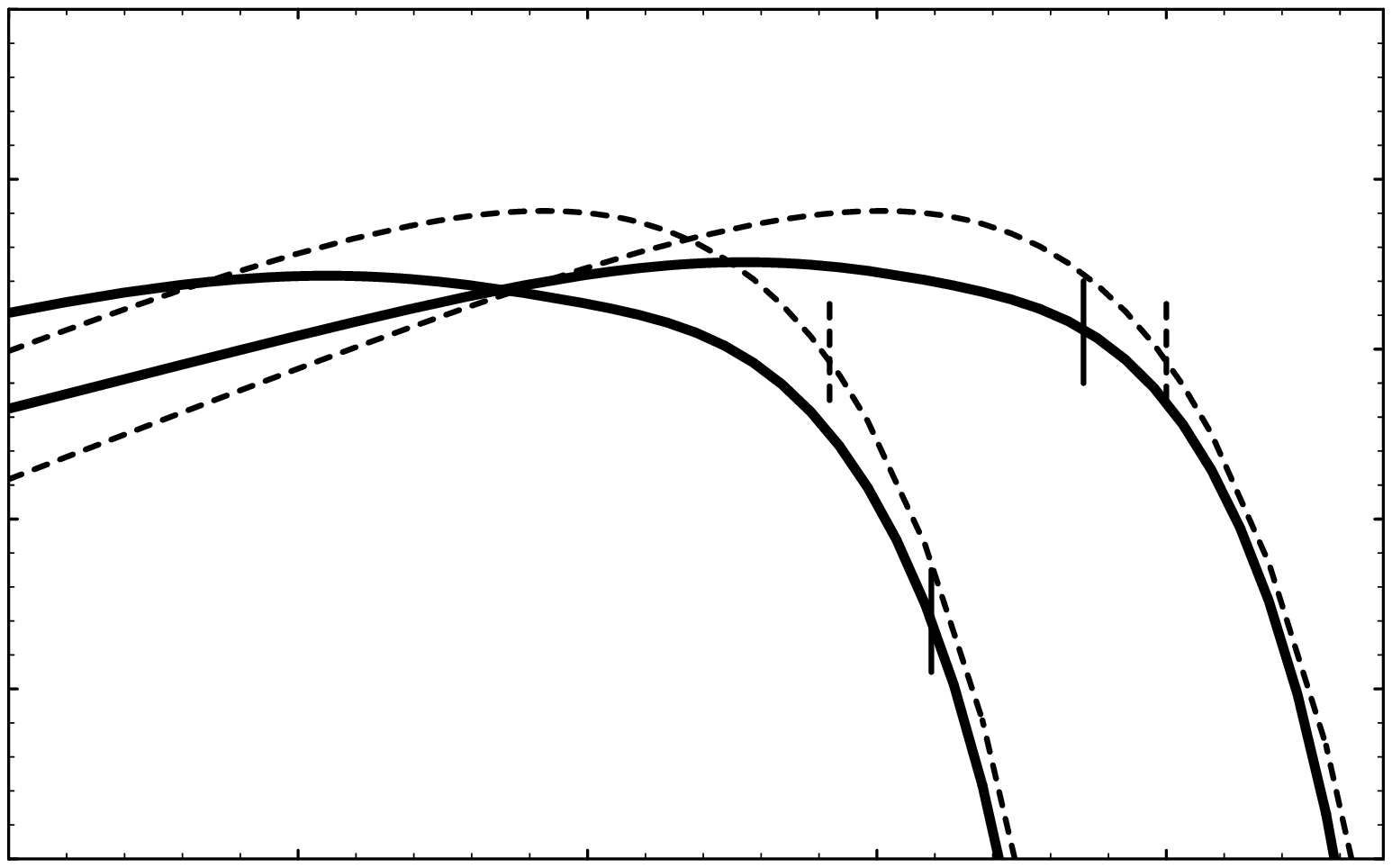}}
\put(40,-28){-6}
\put(88,-28){-4}
\put(134,-28){-2}
\put(183,-28){0}
\put(-11,-22){-3}
\put(-11,27){-2}
\put(-11,77){-1}
\put(180,66){$M_{*PS}$}
\put(165,70){$M_*$}
\put(117,18){$z=3$}
\put(167,30){$z=0$}
\put(20,10){$n=-2$}
\put(20,-5){$\Omega_0=0.3 \; , \; \Lambda=0$}
\put(7,85){$\log[\mu(M)]$}
\put(105,-15){$\log(M/M_0)$}

\end{picture}

\caption{ {\it Upper figure:} The evolved correlation function $\xia(R,z)$ (solid line), and the linear extrapolation $\Sigma(R)^2$ (dot-dashed line), in the case $\Omega_0=0.3 \; , \; \Lambda=0 \; ,\; P(k) \propto k^{-2} \; (n=-2)$. The short dashed line is the power-law $(R/R_0)^{-\gamma}$ with $\gamma=1.8$.  {\it Intermediate figure:} Comoving mass function $\eta(M) \; dM/M$. The graph shows the quantity $V_0 \; \eta(M)$, for the non-linear prescription (solid line), with $\omega=0.3 \; , \; x_*=18 \; , \; \omega_s=-1.6$, and for the PS prescription (dashed line). The vertical solid lines show the position of $M_*$, while the vertical dashed lines show the position of $M_{*PS}$, for both redshifts. {\it Lower figure:} Mass fraction $\mu(M) \; dM/M$ for the same case.}
\label{figeta2OM03}

\end{figure}

	The upper graph of Fig.\ref{figeta1} shows the non-linear correlation function $\xia$ in the case $n=-1, \; \Omega=1$. We can see that it is very close to the observed power-law $\xia(R) \propto R^{-1.8}$ in the range we consider presently. Fig.\ref{figeta1} also displays both mass fractions $\mu(M) dM/M$ (for the PS and non-linear prescriptions) in the same case $n=-1$, $\Omega=1$, at the redshifts $z=0$ and $z=3$. As we explained in the previous section, both mass fractions follow the same evolution with redshift, and their relative position remains constant. Contrary to what could be inferred from section 4.1. and Fig.\ref{fighPSNL} the non-linear prescription does not lead to many more halos than the linear formulation after the cutoff (at the large mass end). This is due to the fact that the analysis we developped above breaks down in this region where $\xia < \Delta_c$. Indeed, for these small values of $\xia$ the clustering is not stable: we do not have $\xia \propto \xia_L^{\;3/2}$ and $\xia$ is smaller than the value it would have if this relation were still valid (as can be seen on Fig.\ref{figXiO1} which clearly shows the sudden falloff in the relation $\tilde{\xi}(\tilde{\xi}_L)$ which translates into a similar break for $\xia(\xia_L)$). This decrease in $\xia$ implies an increase in $x$ which leads to a stronger cutoff than the one implied by our previous analysis. The net result is that both mass fractions are very close in this domain, especially in the case $n=-1$. This implies for instance that the multiplicity functions of clusters of galaxies given by both prescriptions will be very close. 
Naturally, although the relation $\nu-x$ we obtained in the strongly non-linear regime does not hold any longer for these weakly non-linear scales, the evolution with redshift of both mass fractions is still similar in this domain. Indeed, since the power spectrum is a power-law the shape of the mass fraction does not depend on redshift (for any formulation). As both mass fractions evolve in parallel in the highly non-linear end (which corresponds to small scales, and in the present case to small masses), all of their features evolve in parallel. This also means that there is a relation $\nu-x$, independent of $z$, for all values of $\nu$ and $x$. In fact, as we noticed in 2.1., the non-linear scaling $h(x)$ breaks down for small $\xia$ where we enter the quasi-gaussian regime. We can note that for other astrophysical objects defined by larger density contrasts, the analysis we developped in section 4.1. for the highly non-linear regime holds for large $x$ and $\nu$. Thus, the shape of the mutiplicity functions given by both approaches follows the comparison shown on Fig.\ref{fighPSNL}, and the non-linear formulation will lead to many more extreme objects located beyond the cutoff than the PS extension. This will be the case for instance for galaxies, since the cooling constraint will imply a high density contrast for bright galaxies at low redshifts. This point is developped in detail in Valageas \& Schaeffer (1997). Hence, although both prescriptions may lead to very close results for some objects like clusters, they can differ widely for others.

	In the case $n=-2$, $\Omega=1$, the mass functions are different for nearly all masses. Fig.\ref{figeta2} shows the evolved correlation function, which can still be approximated by a power-law, but with a shallower slope $\gamma = 1.4$. It also displays both mass fractions $\mu(M) dM/M$ at the same redshifts $z=0$ and $z=3$ as previously. As for the case $n=-1$, both multiplicity functions are very close at the high mass end.

	Fig.\ref{figeta0} shows similar results in the case $n=0$.

	Fig.\ref{figetaCDM} shows the evolution of the correlation function from $\Sigma(r)^2$ to $\xia(r)$ for a CDM power-spectrum in a critical universe. One can see that it is somewhere between the cases $n=-1$ and $n=-2$ at the scales we are interested in. Since the power-spectrum is no longer a power-law in this case, following Jain et al.(1995) we define an effective index $n_{eff}(z)$ to get the correction $\alpha(n_{eff})$, such that:
\[
n_{eff} = \left( \frac{ d\mbox{ln} P(k)}{d\mbox{ln} k} \right)_{k_e=1/r_e} \;\;\; \mbox{with} \;\;\; \xia_L(r_e)=1
\]
Fig.\ref{figetaCDM} also shows the mass functions $\eta(M) dM/M$ and the mass fractions $\mu(M) dM/M$ in this case of a CDM power-spectrum with $\Omega=1$. As the power-spectrum and the correlation functions $\Sigma(r)$ are no longer power-laws the shape of the curves $\mu(M) dM/M$ changes slightly with the redshift, as well as the relative position of both multiplicity functions.

	Finally, we consider the case $\Omega_0=0.3 \; , \; \Lambda=0$, which is shown on Fig.\ref{figeta2OM03}. We take the case $n=-2$ as an example, which shows the qualitative similarity with the case $\Omega=1$. As is well-known, the evolution with redshift of the multiplicity functions is slower than for a critical universe (compare to Fig.\ref{figeta2}). As we explained above, both mass fractions do not follow exactly the same evolution, as was the case for $\Omega=1$, but they remain very close to each other around their maxima and their cutoff. Other initial spectra would show analoguous features.

As a summary, the PS mass functions, which result from counting the overdensities in the linear regime with a prejudice for which ones are going to end up as clusters, and the non-linear mass functions, which directly count the overdensities at the epoch of interest, differ significantly, but somewhat less at the large mass end where occasionnally the two are identical. All examples given here were for overdensities of $\simeq 200$  (and $\simeq 400$ for $\Omega_0=0.3$), as is customary when describing ``virialized'' objects. To describe the mass function of galaxies (Valageas and Schaeffer (1997)), overdensities of $\sim 10^5$ are required. In the latter case the difference between the PS and the non-linear mass function is much larger, especially at the large mass end.

\section{Other astrophysical objects}

	So far, we have only considered the mass functions of dark matter halos defined by a constant density contrast threshold $\Delta$. We mainly focused on $\Delta = \Delta_c(z)$ which corresponds to just virialized halos (according to the usual spherical collapse criteria), but our analysis performed for the non-linear density field (characterized by $\varphi(y)$ or $h(x)$) applies to any constant density contrast threshold, which allows one to study underdense regions too, as in section 2.4. However, astrophysical objects are not always defined by a constant density threshold. For instance, galactic halos also need to satisfy a cooling constraint (together with the usual virialization condition) which leads to a specific relation $\Delta(R)$ (or $\Delta(M)$) which defines the external boundary of these objects of a given mass (through the relation $R(M)$ it implies), as is studied in detail in Valageas \& Schaeffer (1997). The relation $\Delta(R)$ is usually different from a constant: for instance in the case of galaxies one gets a constant density contrast at small masses (the usual virialization condition), which transforms into a constant radius at large masses (asymptotic cooling constraint at high temperature), which can be written as $(1+\Delta) \propto M$. In a similar fashion, the modellization of Lyman-$\alpha$ clouds leads to a non-constant relation $\Delta(R)$. As a consequence, it is important to look in detail what mass functions can be obtained when objects are no longer defined by a constant density threshold but by a more general relation $\Delta(R)$ or $\Delta(M)$. This is the aim of this chapter, where we only consider the non-linear prescription based on the scale invariance of the function $\varphi(y)$. Indeed, this is beyond the reach of the ``linear'' formulation (which cannot model underdense areas) except in the domain where it can be recast in the ``non-linear'' terms with a scaling function $2 \; h_{PS}(x)$, as was developped in Chapter 4.

\subsection{Counts-in-cells}

	We consider astrophysical objects defined by a function $(1+\Delta)(R)$, and we note:
\beq
\beta(R) = - \frac{dln(1+\Delta)}{dlnR}
\eeq
Hence the usual virialization condition corresponds to $\beta=0$, and a constant $\beta$ means:
\beq
\mbox{if} \; \beta = \mbox{constant} \; : \;\; (1+\Delta) \propto R^{-\beta} \; , \; x \propto R^{-\beta+\gam}
\eeq
where we used $\xia(R) \propto R^{-\gam}$ by definition of $\gam$, and $x=(1+\Delta)/\xia$ is the usual non-linear parameter. 

	We first consider the case $\beta < \gam$. As we did in section 2.1, we can define from the counts-in-cells the fraction of matter with an overdensity larger than $(1+\Delta)(R)$ at scale $R$:
\beq		      			                 
F_m (>\Delta(R),R) =  \int_{x(R)}^{\infty} x h(x) dx 
\eeq
and use for the fraction of matter $\mu(R) dR/R$ embedded in objects of scale $R$ - $R+dR$:
\beq
\mu(R) \frac{dR}{R} = - \frac{d}{dR} F_m (>\Delta(R),R) \; dR = x h(x) \frac{dx}{dR} \; dR    \label{dFm1}
\eeq
while the mass function is:
\beq
\eta(M) \frac{dM}{M}  = \frac{\rho_0}{M} x h(x) dx  \;\;\;\;\; \mbox{with} \;\;\;\;\; M = (1+\Delta) \rho_0 V    
\eeq
Whatever the relation $(1+\Delta)(R)$ this mass function is normalized to unity: $\int \mu(R) dR/R = 1$. We can note that the case $R$=constant for the definition of the objects, which corresponds here to $\beta = -\infty$, is not different from other cases in the final expression for the mass fonction. It simply means that $R$ is not a good variable, but one can still calculate the mass fraction $d{\cal M} = x h(x) dx$ and the multiplicity function $d{\cal N} = \rho_0/M \; x h(x) dx$, which can be expressed in terms of the mass of the objects $M$. Note that we can always use the variable $x$ to count objects, since $\beta < \gam$.

	Now we look at the case $\beta > \gam$. Since by construction we have $dR>0$ in (\ref{dFm1}), and $dx/dR<0$, we get a mass fraction which is negative if we apply directly (\ref{dFm1}). This is due to the fact that what we have to consider now is no longer $F_m (>\Delta(R),R)$ but $F_m (<\Delta(R),R)$. This is obvious if the density profile of typical halos (or more exactly of the cells) is $\rho \propto R^{-\gam}$. Indeed, a halo has a radius between $R$ and $R+dR$ if the curve $(1+\Delta)(r)$ intersects the actual halo profile $(1+\delta)(r)$ at a scale $r$ between $R$ and $R+dR$. If the halo profile is $\rho \propto r^{-\gam}$, and $\beta > \gam$, the curve $(1+\Delta)(r)$ is steeper than the halo density profile and the intersection only occurs between $R$ and $R+dR$ if $(1+\delta)(R) < (1+\Delta)(R)$ and $(1+\delta)(R+dR) > (1+\Delta)(R+dR)$. Hence, if a halo has a size larger than $R$ it satisfies $(1+\delta)(R) < (1+\Delta)(R)$. Note that the curve $(1+\Delta)(R)$ which defines the boundary of the dark matter halos is not the physical constraint which applies all along a given halo. For instance, in the case of galaxies (see Valageas \& Schaeffer 1997), the cooling constraint translates into a cooling radius $R_{cool}(T)$ defined by a temperature - density relation $\rho_{cool}(T)$. That is, if we consider to simplify that halos are isothermal ($\gam = 2$), for a given halo defined by its virial temperature $T$ all the matter inside the radius $R_{cool}(T)$ where the density is larger than $\rho_{cool}(T)$ will be considered as part of the galaxy (because the cooling time in these regions is small). If we express the relation $\rho_{cool}(T)$ as a relation $(1+\Delta)(R)$, we obtain a curve which is very steep: $\beta > 3$, and as a consequence in the inner regions of the halo $R < R_{cool}$ we have $(1+\delta)(R) < (1+\Delta)(R)$ while in the outer regions $R > R_{cool}$ we get $(1+\delta)(R) > (1+\Delta)(R)$. However, contrary to what the sign of these inequalities may suggest the density within the radius $R_{cool}$ is larger than what is required by the cooling constraint: one has to compare the local density to the fixed number $\rho_{cool}(T)$, and not to the local value of the function $(1+\Delta)(R)$ which has a different meaning. Hence we define the mass fraction by: 
\beq
\mu(R) \frac{dR}{R} = - \frac{d}{dR} F_m (<\Delta,R) \; dR = - x h(x) dx   \label{dFm2}
\eeq
while the mass function is:
\beq
\eta(M) \frac{dM}{M}  = - \frac{\rho_0}{M} x h(x) dx    
\eeq
As in the previous case, $\beta < \gam$, the particuliar cases $R=$constant ($\beta=+\infty$), or $M=$constant ($\beta=3$), present no difficulties, and one can always use $x$ as a variable.

	Thus, in all cases we can use for the mass fraction $d{\cal M}$ and the mass function $d{\cal N}$:
\beq
d{\cal M} = x h(x) |dx| \;\; \mbox{and} \;\; d{\cal N} = \frac{\rho_0}{M} x h(x) |dx|  \label{dMN1}
\eeq
Within this framework, we cannot define a mass function in the case $\beta = \gam$, where $x=$constant. This can also be understood if halos have a mean profile in $\rho \propto R^{-\gam}$, since in this case it is obvious that the criterium $(1+\Delta)(R) \propto R^{-\gam}$ cannot be used to define objects. However, such a picture may be misleading, and a more correct statement is simply that the mass fraction embedded in cells which lie above a certain density threshold $\Delta$ depends only on the parameter $x$. Then, if $\beta=\gam$ the derivative of the mass fraction is zero: $dF_m(>\Delta,R)/dR = dF_m(<\Delta,R)/dR =0$. This also suggests that the mass functions obtained in the case $\beta \simeq \gam$ may not be very accurate, but fortunately the cases of practical interest do not cross this domain: $\beta=0$ for usual virialized halos and $\beta>3$ in the case of galaxies or Lyman-$\alpha$ clouds.

	We can note however that the previous approximation for the mass function of objects defined by a curve $(1+\Delta)(R)$ is not entirely satisfactory in the cases $\beta<0$ and $\beta>\gam$. Indeed, contrary to the usual case $\Delta=$constant, there is no guarantee that the density of the ``objects'' picked out in this way decreases at larger radius, even locally at the considered scale $R$. One may even fear that such a mass function counts mainly ``valleys'' (cells with a density which increases when averaged over a larger radius) in the domain of small $x$. This should not be the case for large $x$ where most cells have a density profile in $\rho \propto R^{-\gam}$ (contrary to what happens for small $x$), as is recalled in App.C. This leads us to consider the immediate surroundings of cells, as we did in section 2.2, which will allow us to specify explicitely that the density of the cells we count is at least a decreasing function of radius locally at scale $R$.

\subsection{Surroundings of a cell}

	In a fashion similar to what we did in section 2.2, and from the considerations developped in the previous paragraph, we can define the mass fraction $d{\cal M}$ formed by objects of scale $R$ - $R+dR$ by: 
\beq
\left\{ \begin{array}{rrcl} \beta < \gam : & d{\cal M} & = & (1+\Delta) P(>\Delta,<\Delta') \\ \\  \beta > \gam : & d{\cal M} & = & (1+\Delta) P(<\Delta,>\Delta') \end{array} \right.   
\eeq
where $\Delta=\Delta(R)$, $\Delta'=\Delta(R')$ and $R'=R+dR$, with $dR>0$. These two expressions can be recast into one formulation:
\beq
d{\cal M} = (1+\Delta) P(>\Delta,<\Delta') \;\;\; \mbox{with} \;\; dx>0
\eeq
Then, for $\beta<\gam$ we have $dR>0$, $R'>R$, while for $\beta>\gam$ we have $dR<0$, $R'<R$. The mass function is simply $d{\cal N} = \rho_0/M d{\cal M}$. As shown in App.C, this definition leads to a mass function which is similar to the one encountered in the previous section (\ref{dMN1}):
\beq
d{\cal M} = x H(x) |dx| \;\; \mbox{and} \;\; d{\cal N} = \frac{\rho_0}{M} x H(x) |dx|  \label{dMN2}
\eeq
with a scaling function $H(x)$ which differs from $h(x)$, and depends on the slope $-\beta$ of the curve $(1+\Delta)(R)$. As was the case for a constant density threshold, we can obtain general bounds for $H(x)$, as compared to $h(x)$. They show that in all cases both scaling functions $h(x)$ and $H(x)$ have the same exponential cutoff at large $x \gg x_*$, apart from a possible normalization constant, and that they also have the same power-law behaviour at small $x \ll x_*$ if $\beta > \gam$. If $\beta>3$, we get $H(x) = h(x)$ in the limit of large $x \gg x_*$. In the case $|\beta| \rightarrow \infty$, which corresponds to objects defined by a constant radius $R$, we get $H(x)=h(x)$. This is natural, since in this case we obtain the usual counts-in-cells, at fixed radius $R$ and the behaviour of the density around the object is irrelevant (in this approach). As was the case in the previous section, if $\beta = \gam$ we cannot get a mass function within this framework. In general, $(1+\Delta)(R)$ is not a power-law, and the function $H(x)$ should depend on the scale $R$ through $\beta(R)$. However, in practice we shall simply replace $H(x)$ by $h(x)$ which is known from numerical simulations and should provide a fair approximation, according to the previous section and the relations between $h(x)$ and $H(x)$ described above.

	Finally, as we mentionned in the previous section, this formulation allows us to add explicitely the constraint that the density profile is locally decreasing at the external radius of the objects. Thus, if $\beta<0$ we can define $H(x)$ such that $d{\cal M} = x H(x) dx$ by:
\beq
x H(x) dx = (1+\Delta) \int_{\Delta}^{\infty} d\delta \int_{-1}^{\mbox{Min}(\Delta',\delta)} d\delta' P(\delta,\delta')
\eeq
while if $\beta>\gam$ we have:
\beq
x H(x) dx = - (1+\Delta) \int_{\Delta'}^{\Delta} d\delta \int_{\Delta'}^{\delta} d\delta' P(\delta,\delta')
\eeq
Hence, the mass functions we obtain still present the scaling in $x$ (if $\beta$ is constant), but no general bounds are available, except in the case $\beta>3$, which is the one which occurs in practical cases (for galaxies or Lyman-$\alpha$ clouds). Then we have an upper bound for $H(x)$ in both limits $x \ll x_*$ and $x \gg x_*$, which shows that the decrease of $x^2 H(x)$ in these limits cannot be slower than the decline of $x^2 h(x)$.  In the case $\beta \rightarrow \infty$, which corresponds to objects defined by a constant radius, and is of practical interest as it is the cooling constraint for galaxies at large temperature, we obtain:
\beq
x \geq 0 \; : \;\; H(x) \leq h(x) \;\;\; , \;\;\; x \gg x_* \; : \;\; H(x) \geq \frac{\gam}{3} h(x)   \label{HhRct}
\eeq
Thus, as we could expect, the large $x$ behaviour of both mass function is similar (apart from a possible normalization constant). This is natural since most cells have a density profile which decreases with radius in this regime, so that the additional constraint that the density is locally decreasing has not much influence as it was already fulfilled in most cases. This is not so obvious for low $x$ cells, where only a specific model for the density field (the functions $h(x)$ or $\varphi(y)$ do not contain all the needed information) can allow one to get a definitive answer. However, we showed that the mass function will in any case present the same scaling in $x$, with an upper bound if $\beta>3$. In practice, we shall replace $H(x)$ by $h(x)$, which is justified by (\ref{HhRct}) since we generally have $\beta \gg 3$ (nearly constant radius), or $\beta=0$ as usual.

\section{Conclusion}

	Our purpose in this paper was to examine various analytical forms of the mass function, to understand their relation, and to get some hint to judge their accuracy. The main motivation for this study is to get accurate enough a formulation so as to be able to use these analytical forms to get a robust tool for studying the {\it number evolution} of the structures, as a preparation for the oncoming data from the HST and KECK telescopes. A more theoretical motivation was to understand {\it how the universe gets non-linear} at galactic and extragalactic scales, and the relation of the presently observed structures with initial conditions.

	To achieve this purpose, we have compared in detail two approaches. The first -- that we call the linear approach -- describes, along the ideas  of Press and Schechter (1974), the clustering at the present epoch by recognizing in the primordial spectrum of fluctuations those overdensities that eventually form the structures and counting them there. It is in principle limited to objects with an overdensity $\Delta \sim 200$ at the epoch under consideration. The second -- called the non-linear approach -- , that we develop here  following earlier ideas (Schaeffer 1984, BS89, Bernardeau and Schaeffer 1991) on how the non-linear density field behaves (the probability to have an overdensity $(1+\Delta)$ in a cell of scale $R$ is given by $P(\Delta) = 1/\xia(R)^2 \; h [ (1+\Delta) / \xia(R) ]$ where $h$ is a universal function depending on initial conditions), describes the present-day clustering directly by means of the actual density fluctuations. This description holds for any overdensity $\Delta$ provided one is in the non-linear regime.

	A first comparison tackles the {\it cloud-in-cloud problem} raised by Bardeen et al (1986). The arguments invoqued in the calculation of these mass functions show that it stands as an approximation of a ``better'' calculation. The latter counts overdensities with a given contrast at their edge, defined as a future cluster in the primordial density field or as an actual cluster in the present density field. Such overdensities may actually be composite objects made of smaller size overdensities with the same density contrast surrounded by underdense regions, so that the same mass is counted several times, as a large object and also as several smaller objects. We have defined a measure of this overcounting by considering the probability for overdensities to be surrounded by underdensities, and for underdensities to be surrounded by overdensities. The problem turns out to be severe in the linear stage where fluctuations switch easily from above to below average, and quite mild in the non-linear stage where the overdensities are well separated. We also showed that the geometrical ``cloud-in-cloud'' problem does not provide a justification for the usual multiplicative parameter 2 generally affected to the PS mass function, and is only an ad-hoc correction to the dynamics.

	In the linear approach, the knowledge of the r.m.s. density fluctuation -- which are gaussian -- as a function of scale and time in the primordial universe is sufficient. In the non-linear approach, the behaviour of the actual correlation as a function of scale and time is needed, but turns out to be sufficient to describe the density field although the fluctuations are violently non-gaussian (see however the discussion sect 4.1 and in Colombi et al, 1996, and Bernardeau,1992, for a more precise statement). This illustrates an important difference of the two approaches. In the linear approach, clustering as a function of scale and time goes as the linear correlation function (whence a factor $(1+z)^2$), whereas in the non-linear approach, clustering goes as the non-linear scaling of the correlation function (whence a factor $(1+z)^{3-\gamma}$). Both scalings cannot be simultaneously true, and will provide for a way to distinguish between the two approaches.

This shows also that a more detailed comparison of the latter can be done provided one can understand {\it the non-linear evolution of the correlation function}. We have presented a model based on the spherical collapse plus virialization that we believe is the explanation of the non-linear behaviour of the correlation function, and that is seen to reproduce quantitatively the scaling observed by Hamilton et al (1988) of the non-linear correlation function computed in an $ \Omega = 1 $ universe as a function of the extrapolated linear correlation function. This scaling is seen not to exist for $\Omega < 1$ but our method allows to calculate the evolution of the correlation function in the latter case: $\Omega<1$ and $\Lambda=0$.

Specific examples of the predicted mass distribution and its evolution as a function of redshift are given using power-law initial spectra as well as CDM initial conditions. The linear and non-linear mass functions show an intriguing qualitatively similar behaviour, as a function of mass as well as a function of redshift. Their quantitative differences however are serious. The error made by using one of the mass functions instead of the other are as large as the difference between the cases with different initial spectra. One should thus {\it  be extremely careful when deriving by means of analytical approximations to the mass function conclusions on the adequacy or inadequacy of an initial spectrum} from an observed distribution of objects (galaxies, clusters, Ly$_\alpha$ clouds).

The understanding of the evolution of the non-linear correlation function allows one to phrase the difference between the linear and the non-linear approach in a way that clearly shows their difference, but also explains their apparent similarity in actual calculations. The basic assumption in the PS approach is that the seed of a cluster in the present-day universe may be recognized in the primordial universe. These seeds then evolve according to the spherical collapse model until they enter the stable clustering regime, but {\it the evolution conserves their number}. Using this assumed property, together with the evolution of the correlation function according to the spherical collapse, which agrees with numerical simulations, we derive  the non-linear density field implied by the PS approximation. For a power-law primordial spectrum taken as an illustrative case, this non-linear density field implied by the PS assumption has a similar scaling as the one seen in numerical simulations, but differs quantitatively, the difference being of the order of the range spanned by realistic changes of the initial conditions.

Finally, we present the mass functions one can obtain when objects are defined by a density threshold which is no longer constant but may vary with the mass of the objects. This allows one to consider other astrophysical objects (like galaxies, see Valageas \& Schaeffer, 1997, for a detailed study) than the usual just-virialized objects, and is of great practical importance. Since this is beyond the reach of usual approximations based on an extrapolation from the initial gaussian field like the PS prescription (which cannot model low density regions anyway), it provides a clear illustration of the advantages of our approach.

We relate these various multiplicity functions to the properties of the counts-in-cells (characterized by the scaling function $h(x)$). Thus, one single measure of $h(x)$ allows one to obtain all these mass functions, at any time, within the highly non-linear regime. Moreover, we show that they should obey the same specific scaling in $x$ (through a scaling function $H(x)$ close to $h(x)$) as the counts-in-cells. Numerical simulations are needed to confirm this scaling and precise the relation between $H(x)$ and $h(x)$. We can note that the PS approximation leads to the same scaling and provides a simple estimate of $h(x)$, so that we could still apply in this case the method described in the non-linear approach. However, as we argued above this approximate $h_{PS}(x)$ is probably too crude to give accurate results.

Numerical tests have been done to check the PS approximation. They involve for instance (Efstathiou et al 1988) to test the scaling of the mass function as a function of the scaling parameter $\nu=\delta_L/\Sigma$. But we have seen that the typical scaling of the non-linear mass function that depends on the variable $x$ can be reexpressed as a function of the former parameter $\nu$ for power-law initial conditions. This may be expected to be approximatively the case even for more complex initial conditions. The scaling test thus may not be  sufficient to distinguish among the two approaches. The shape of the mass function is to be explicitely considered. The non-linear predictions have been well checked against observations (Alimi et al 1990, Maurogordato et al 1992, Benoist et al 1996) as well as against simulations (Bouchet et al 1991, Bouchet and Hernquist 1992, Colombi et al 1994,1995,1996,1997). Clearly, in the light of the previous discussion, testing the scaling laws of the counts-in-cells may not be sufficient. But the galaxy void probability functions typically show ( Sharp 1980, Bouchet and Lachieze-Ray 1985, Maurogordato and Lachieze-Ray 1987,  Maurogordato et al 1992) the non-linear scaling predicted (White 1979, Schaeffer 1984) by the non-linear theory. Also, numerical simulations are accurate enough to give the shape of $h(x)$ that  is not the one that the PS approximation implies for the density field. We believe nevertheless that the numerical tests should be redone in the light of the present findings. Testing the PS and the non-linear predictions has been done by different groups. One now has two well defined prescriptions that can be written down using exactly the same choice of parameters and directly compared. Also, there may also be some dependence of the cluster mass function on the specific way in which clusters are defined. Clearly, would we define the number of clusters of scale $R$ as the number of overdense cells of that scale minus the number of overdense ones at scale $R + dR$, the non-linear mass function would be exact. Our explicit consideration of the surroundings of a given cell, and the probability for the latter to be underdense, shows that one may expect a friend-of-friends algorithm to give a similar answer. But this has to be examined numerically. Specific comparisons with numerical results (Lacey, Valageas, Schaeffer) are quite promising, but will be discussed elsewhere.

The issue is to have robust predictions of the {\it number evolution} of matter condensations, so as to be able to concentrate on the evolution of the internal properties of the various objects that may be considered. The explicit consideration of the galaxy luminosity function evolution, as well as the evolution of X-ray clusters and Lyman-$\alpha$ lines is in preparation. The present, still phenomenological, description of the dynamics of non-linearity may also be a guide to more rigorous calculations that would be done starting from first principles, by the use of the equations of motions induced by gravity.
\\

\vspace{1cm}	

\appendix

{\bf APPENDIX}

\section {The mass function calculated from the early gaussian fluctuations}

\subsection{Gaussian approximation}

In the case of a gaussian distribution, the probability for a cell of radius $R$ to have a density contrast $\delta$ is:
\beq
P(\delta,R) d\delta = \frac{1}{\sqrt{2 \pi}\Sigma(R)} \; e^{-\delta^2/(2\Sigma^2(R))} \; d\delta
\eeq
The probability distribution $P(\delta,R;\delta',R')$, or simply $P(\delta,\delta')$, to have a contrast $\delta$ in the radius $R$ and a contrast $\delta'$ in the larger radius $R'$ is
\beq
\begin{array}{rcl} P(\delta,\delta') & = & {\displaystyle \frac{1}{2\pi \sqrt{\Sigma^2 \Sigma'^2 - \Sigma''^4}} }  \\  &  & \hspace{0.5cm} {\displaystyle \times \exp \left[ -\frac{\delta^2}{2\Sigma^2} - \frac{(\delta \Sigma''^2 - \delta' \Sigma^2)^2}{2\Sigma^2 (\Sigma^2 \Sigma'^2 - \Sigma''^4)} \right] } \end{array}  \label{Pgaus}
\eeq
with
\[
\Sigma^2=\int d^3k \; W(kR)^2 P(k) \;\;\;\;\; , \;\;\;\;\; \Sigma'=\Sigma(R') 
\]
\[
\Sigma''^2=\int d^3k \; W(kR) W(kR') P(k)
\]
where $W(x)$ is the window function (for a top-hat filter in real space $W(x) = 3(\sin x - x \cos x)/x^3$) and $P(k)$ is the power-spectrum of the initial density fluctuations. If the window function is a top-hat in $R$, or a gaussian, we have when $R' = R+dR$:
\beq
\Sigma^2 \Sigma'^2 - \Sigma''^4 = \left( a \Sigma^2 \frac{dM}{M} \right)^2  
\eeq
where $a$ depends on the slope of $P(k)$ at scale $R$ (it is a constant if $P(k)$ is a power-law) and the distribution in $\delta'$, at fixed $\delta$, is a gaussian centered on $\delta_0$ and width $\sigma_0$ with:
\beq
\delta_0 = \delta \; \left( 1+ \frac{dln\Sigma}{dlnM} \frac{dM}{M} \right) \; \mbox{and} \; \sigma_0 = a \Sigma \frac{dM}{M}
\eeq
The case of a top-hat in $k$ is very peculiar, since $\Sigma''=\Sigma'$ and we obtain:
\beq
\delta_0 = \delta \; \left( 1+ 2 \frac{dln\Sigma}{dlnM} \frac{dM}{M} \right) \; \mbox{and} \; \sigma_0 = \Sigma \sqrt{2 \frac{dln\Sigma}{dlnM} \frac{dM}{M}}
\eeq
which is fundamentally different.

\subsection{The multiplicity function: Press-Schechter formulation}

	Much has been said about this way to calculate the number of objects with mass $M$ (Press and Schechter 1974 , Schaeffer and Silk 1985,1988, see White 1993 for a review). We recall here a few relevant features of this approach for the sake of comparison with our new  derivation of the mass function.

	The idea, here, in order to avoid entering into the description of the universe at the non-linear scales, is to try to recognize in the early universe those very small density fluctuations that are eventually going to become clusters or voids, to count them and to attach to them non-linear properties deduced from their properties in the linear stage. The mass of these objects is given as an initial condition. Their radius or density contrast then is assumed to be the result of their internal evolution. The main difficulty is to be granted that these proto-objects remain isolated during and after their formation, so that their number or their mass do not evolve in the non-linear stage.

	 An overdense cell containing the mass M initially following the general expansion, whose overdensity grows according to the linear theory, slows down and finally collapses. One often labels the overdensity by the value $\delta_0$ it would have presently would his evolution be given all the way by linear theory
\beq
\delta_L(t) = \delta_0  \; D(t) / D(t_0)  
\eeq
The statistics of $\delta_L$ in the very early universe, where it is simply equal to $\delta$, is thus the statistics of $\delta_0$ at the present epoch, using linear theory
\beq
P(\delta) d\delta = \frac{1}{\sqrt{2 \pi} \Sigma_0} \; e^{-\delta_0^2 / (2 \Sigma_0^2) }  \;  d\delta_0
\eeq
The mass of the object remains constant, equal to $M$, and $R_M$, see (\ref{R_M}), which plays the role of a mass scale, is the radius the object would have at the present epoch would it simply follow the general expansion. The actual radius $R$ has a maximum $R_{max}$ at a time $t=t_{max}(\delta_0)$. It then starts to shrink. With no kinetic energy at the maximum radius, energy conservation and virial equilibrium imply that the collapsed object has reached a stationary state with a radius $R_{vir} = R_{max} / 2$ at a time $t = t_{vir}(\delta_0)$, usually chosen as the time of collapse given by the spherical model, that is when $\delta = \delta_c$ with $\delta_c = 1.69$ for $\Omega = 1$ and $\delta_c = 1.62$ today for $\Omega_0 = 0.1$. The overdensity at the time of virialization is thus $\Delta_c = 178$ for $\Omega = 1$ and $\Delta_c = 978$ today for $\Omega_0 = 0.1$. All objects having reached this stage are supposed to give rise to a massive collapsed halo. The mass fraction within such objects of mass larger than $M$ is then given by the requirement that $\delta$ in the very early universe, where it is negligibly small, is such that, at time $t$, it will be above the required threshold $\delta_c$, at the scale $M$. Therefore
\beq
F_v(>\delta_c) = F_m(>\delta_c) =  \int_{\delta_c}^{\infty} \frac{d\delta}{\sqrt{2 \pi} \Sigma(M)} e^{-\delta^2 / (2 \Sigma^2(M)) } \label{Frac}
\eeq   
with
\[
\Sigma(M) = \Sigma_0(M) \; D(t) / D(t_0)   
\]
the mass and volume fractions relative to the considered cells being the same. The operation can be repeated for larger cells. Only the mass fraction above threshold at scale $M$, and not at scale $M+dM$, that is
\[
F_m(>\delta_c,M) - F_m(>\delta_c,M+dM) 
\]
\[
\;\;\;\;\;\;\;\;\;\;\;\;\;\;\;\;\;\;\;\; = - \frac{d}{dM} F_m(>\delta_c,M) dM 
\]
produces objects of mass between $M$ and $M+dM$. It is then possible to recognize in the very early universe those features that will, at a given much later epoch $t$, have gone above the given threshold. This leads to the mass function:
\beq
\eta(M) \frac{dM}{M} = \frac{1}{\sqrt{2\pi}} \frac{\rho_0}{M} \frac{\delta_c}{\Sigma} \left| \frac{\mbox{dln} \Sigma}{\mbox{dln}M} \right| e^{-\delta_c^2/(2\Sigma^2)}  \; \frac{dM}{M}  \label{etaM}
\eeq
a formula in principle valid for any mass $M$.

	The advantage of this approach is its extreme simplicity, if one does not consider the cloud-in-cloud problem, together with a direct connection with the primordial fluctuation spectrum. A rather serious drawback is that the statistics is done in the very early universe, trying to recognize protoclusters among small overdensities while it is not clear that there is no evolution of the number of objects after non-linearity has occurred. An obvious consequence of (\ref{etaM}) is that the total mass in the above halos
\beq
\int_0^{\infty}  M \eta(M) \frac{dM}{M} =  \rho_0  \int_0^{\infty} \frac{d\nu} {\sqrt{2\pi}} e^{-\nu^2/2}   =  \frac{\rho_0}{2}
\eeq
represents only half of the matter. From our experience in modelling the present, obviously non-linear universe, this is not true: most of the mass must be within the condensed objects. For this reason alone, the PS approach cannot be an exact theory as has been claimed at various places. This problem can be cured using the excursion sets formalism (Cole 1989, Bond et al. 1991) for the very specific case of a top-hat window function in $k$ (which should not be confused with the usual top-hat in $R$, that we use thougout this paper), which gives the PS mass function multiplied by the factor 2. However, as we argue in chapter 2 this cannot be extended to other window functions (the fact that a top-hat in $k$ is a very peculiar case was also apparant in the previous paragraph). Also,  despite the increased complexity, one  is still working within the linear universe (with no control of any evolution once in the non-linear stage). For similar reasons (cloud-in-cloud problem, and, although to a lesser extent, dependence on the window function, complexity even in the linear approximation), the attempt by Bardeen et al. (1986) to define the mass function by counting the peaks of the density distribution in the linear stage is not entirely satisfactory, as already stated by the authors.

\subsection{Surroundings of a cell}

	An alternative formulation starts from the remark that, in the prescription described above, one substracts the fraction of matter potentially non-linear at scale $M+dM$ from the fraction at scale $M$. This provides no guarantee that the immediate surroundings of a cell above threshold are below that same threshold, as one could expect for a true proto-cluster. It is, on the other hand, possible to calculate the probability for this to occur. Divide the universe into cells of size $R_M$ and consider, for each of these cells a cell centered at the same place but of size $R_M+dR_M$. The probability $P(>\delta_c,R_M ; <\delta_c,R_M+dR_M )$, which we shall note simply $P(>\delta_c,<\delta_c)$, for having an overdensity $\delta$ above a given threshold $\delta_c$ in the original cell and below $\delta_c$ in the larger cell is (using (\ref{Pgaus})), still in the ``linear'' gaussian regime,
\[
P(>\delta_c,<\delta_c) = \frac{1}{\sqrt{2\pi}} \frac{\delta_c}{\Sigma} \left| \frac{\mbox{dln} \Sigma}{\mbox{dln}M} \right| e^{-\delta_c^2/(2\Sigma^2)}  \; \frac{dM}{M}
\]
\beq
\hspace{2.6cm} \times  \int_{-\infty}^{\beta \nu} \left[1-\frac{u}{\beta \nu} \right] e^{-u^2/2} \frac{du}{\sqrt{2 \pi}}   \label{P><gau}
\eeq
with $\nu = \delta_c /\Sigma(M)$ and $\beta(M)$ is a parameter which depends on the form of the initial power-spectrum $P(k)$. If $P(k)$ is a power-law $\beta$ is a constant. Note that for a top-hat in $k$, one cannot define a mass function from $P(>\delta_c,R;<\delta_c,R+dR)$ as this latter quantity behaves as $\sqrt{dR/R}$ for small $dR/R$. The formula (\ref{P><gau}) looks very similar to (\ref{etaM}), and indeed this prescription is closely related to that previous formulation: 
\beq
\begin{array}{ll} P(>\delta_c,<\delta_c) =  & F_m(>\delta_c,M) - F_m(>\delta_c,M+dM)  \\  &  + P(<\delta_c,>\delta_c)  \end{array} \label{Psi}
\eeq
It is indeed staightforward to see that when substracting $F_m(>\delta_c,M+dM)$ from $F_m(>\delta_c,M)$, one substracts a little too much, the case where the density is low inside but high in the larger cell being included in $F_m(>\delta_c,M+dM)$.
\\

\begin{figure}[htb]

\begin{picture}(230,150)(-18,-15)

\epsfxsize=8 cm
\epsfysize=12 cm
\put(-8,-118){\epsfbox{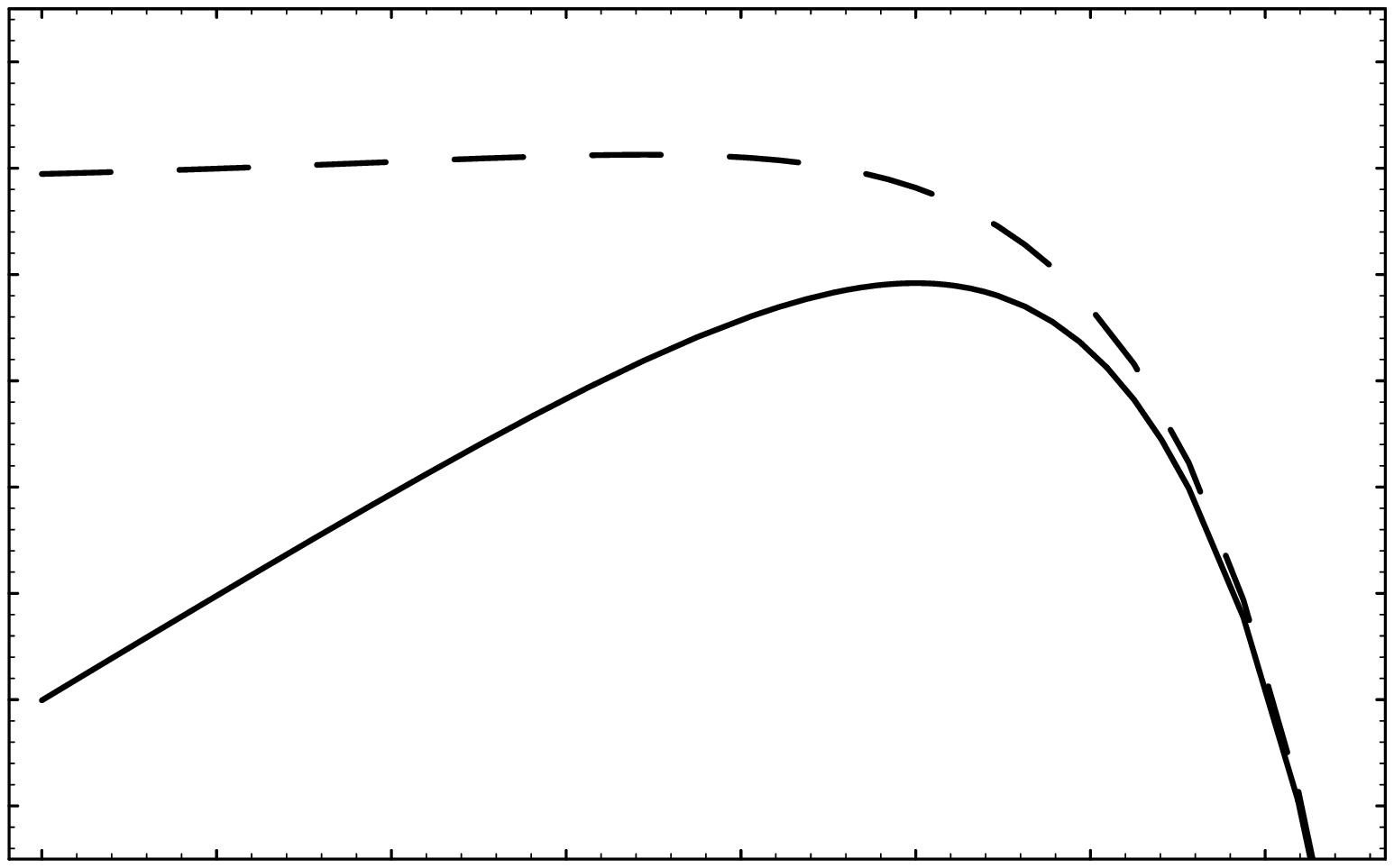}}
\put(28,-26){-0.8}
\put(82,-26){-0.4}
\put(144,-26){0}
\put(194,-26){0.4}
\put(-14,-10){-1.6}
\put(-6,45){-1}
\put(-14,98){-0.4}
\put(40,90){$\mu_{><}$}
\put(110,60){$\mu_{PS}$}
\put(173,113){$\log[\mu(\nu)]$}
\put(160,-9){$\log(\nu)$}

\end{picture}

\caption{Behaviour of the mass fraction at the present epoch $\mu(M) dM/M = \mu(\nu) d\nu/\nu$ , obtained by counting the overdensities in the early universe where the fluctuations are sufficiently small to be described by linear theory, with the assumption that their number does not evolve once they become non-linear. The graph shows the mass fraction per unit logarithmic interval of $\nu$ : $\mu_{><}(\nu)$ and $\mu_{PS}(\nu)$, in the case $n=-1.5 \; , \; \beta = 0.43$ }
\label{figmugaus}

\end{figure}

If the evaluation of the mass function by substraction of the mass fractions $F_m$  at different scales has any meaning, it should not be too different from $P(>\delta_c,<\delta_c)$, the difference between the two measuring the error due to the fact that some of the protoclusters, that is density patches in the linear regime with a given density threshold, may contain smaller patches with the same density theshold that are then also counted as protoclusters at a smaller mass. Similarly to Chapter 2.2., the use of $P(>\delta_c,<\delta_c)$ to define a mass function leads to the mass fraction in objects of mass $M$ :
\[
\mu_{><}(M) \frac{dM}{M} = \frac{1}{\sqrt{2\pi}} \; \nu \; e^{-\nu^2/2} \; \frac{d\nu}{\nu} 
\]
\beq
\hspace{2.5cm} \times  \int_{-\infty}^{\beta \nu} \left[1-\frac{u}{\beta \nu} \right] e^{-u^2/2} \frac{du}{\sqrt{2 \pi}}
\eeq
while the PS prescription gives
\beq
\mu_{PS}(M) \frac{dM}{M} = \frac{1}{\sqrt{2\pi}} \; \nu \; e^{-\nu^2/2} \; \frac{d\nu}{\nu}
\eeq
Here, $\mu(M) dM/M$ is the mass fraction in objects of mass between $M$ and $M+dM$, and $\eta(M) = \rho_0/M \; \mu(M)$. These two mass functions are similar for large masses (that is $\nu \gg 1$), as can be checked on Fig.\ref{figmugaus}, but at small masses $\mu_{><}(\nu)/\nu$ diverges as $1/\nu$ while $\mu_{PS}(\nu)/\nu$ remains finite. As a consequence, $\int_0^{\infty} \mu_{><}(M) dM/M = +\infty$ while $\int_0^{\infty} \mu_{PS}(M) dM/M = 1/2 \;$. This is due to the fact that when $\Sigma \rightarrow \infty$, ($M \rightarrow 0$) the density around a given point shows wide oscillations ($\sim \Sigma$) and crosses the threshold $\delta_c$ many times, which leads to a large over-counting. This is the well-known cloud-in-cloud problem raised by Bardeen et al.(1986), that renders this approach uncertain.

\section{Counts in cells and matter distribution}

	A convenient way to describe the density field above the galactic scales
is to divide the universe into cells of a given size $R$ and to consider the distribution of matter within these cells. It is then of interest to define the probability distribution of the  density $\rho$ within each cell or equivalently of the deviation $\delta$ of $\rho$ from the average density $\rho_0$ of matter in the Universe
\[
\delta   =    ( \rho - \rho_0 ) / \rho_0    
\]
and to study it as a function of the cell size $R$, that plays the role of a
smoothing scale. In the following, we shall, as a rule, omit to precise explicitely the radius of the cells we consider, for instance we shall note $P(\delta,R)$ simply as $P(\delta)$, except when we consider different lengths and there is a risk of confusion.

\subsection{Gaussian limit}

	In the regime where $\delta$ and $\xia$ are small, say $\delta \; , \; \xia \; \ll 1$, that is at very large scales, this probability takes the well-known gaussian form 
\beq
P(\delta) d\delta = \frac{1}{\sqrt{2 \pi \xia}} \; e^{-\delta^2/(2\xia)} \; d\delta
\eeq
and is determined once $\xia(R)$ is known. The latter is the average of the two-body correlation function $\xi_2 (r_1,r_2)$ within the volume of the cell
\[
\xia(R) =   \int_V \frac{d^3r_1 \; d^3r_2}{V^2} \; \xi_2 (r_1,r_2)  \;\;\;\;\; \mbox{with} \;\;\;\;\; V= \frac{4}{3} \pi R^3
\]
For a gaussian distribution it is usually denoted $\Sigma^2$ in the litterature. We use the notation $\xia(R)$ in use for the actual mass fluctuation at scale $R$ in the non-linear regime. We keep the notation $\Sigma(M)$ for the r.m.s. mass fluctuation obtained  in the {\em linear} approximation, calculated  at scale $M$ in the very early universe (where it is assumed to be infinitely small) and extrapolated to any epoch with the proper linear growth factor. We will also use for the linear correlation function the notation $\xia_L(R_M) = \Sigma^2(M)$ , where $R_M$ is a comoving length that plays the role of a mass scale, defined by
\beq
M = 4 \pi /3 \; \rho_0 \; R_M^3 \label{R_M}
\eeq

\subsection{Deviations from gaussian behaviour}

	Deviations from the gaussian behaviour, however, occur as soon as $\delta$ is not very small. They can be described as follows (White 1979, Schaeffer 1984, 1985, Hamilton 1988b, BS89, Bernardeau and Schaeffer 1992, Bernardeau 1992). One introduces the auxiliary function
\beq           
\varphi(y)   =   \sum_{n=1}^{\infty} \frac{(-)^{n-1}}{n!} \; S_n \; y^n 
\eeq
where the coefficients $S_n$ , which are all positive, describe, for $n>2$, the deviations from the gaussian behaviour. These coefficients are related to the many-body  correlation functions $\xi_n(r_1,...,r_n)$ by
\[
S_n = \frac{\xia_n}{\xia^{\; n-1}} \;\;\;\; \mbox{with} \;\;\;\; \xia_n =   \int_V \frac{d^3r_1 ... d^3r_n}{V^n} \xi_n (r_1,...,r_n)
\] 
They depend in principle on the scale $R$ and on the shape of the considered
cells. The function $\varphi(y)$ was introduced by the above authors in the case where the many-body correlation functions exhibit a {\em scaling} behaviour, precisely such as to make $S_n$ {\em scale independent} and argue that, to 
a rather good approximation, these parameters are also {\em shape independent}, the latter being not compulsory for the purpose we aim to in this paper. The probability distribution of the density within the cell is then given by
\beq                           
P(\delta) = \inta \frac{dy}{2 \pi i \xia} \;\; \exp \left[ \frac{(1+\delta)y - \varphi(y)}{\xia} \right]   \label{P(d)} 
\eeq	
which can be inverted as
\beq
e^{-\varphi(y)/\xia} = \int_{-1}^{\infty} e^{-(1+\delta)y/\xia} \;\; P(\delta) d\delta   \label{phiP(d)}
\eeq
In the limit where $\varphi(y) = y - y^2 / 2$, that is when all coefficients $S_n$ for $n>2$ vanish, the gaussian behaviour is recovered. Schaeffer (1985) by comparison with data, and BS89 for very general reasons due to the behaviour of $\varphi(y)$ however argue that $S_n$ should grow at least as $n!$ and show that, as soon as $\delta$ is not much smaller than $\xia$, the values of $y$ contributing to the integral are of order unity and the deviations from a gaussian behaviour are important. For $\delta$ much larger than $\xia$, large tails develop, the gaussian behaviour being replaced by a slow exponential decrease.

	In the limit $\xia \ll 1$, but for any $\delta$, i.e. in what he called the {\em quasi-linear} regime, expected to hold at large scales or early times, Bernardeau (1992, 1994) has been able to calculate explicitely the function $\varphi(y)$, that arises purely from the gravitational instability, and its relation with the primordial spectrum of fluctuations.

	In the limit $\xia \gg 1$, that is relevant at small scales and late times, under the sole assumption of scale-invariance of the many-body correlation functions expected (Peebles 1980) in the {\em deeply non-linear} regime, BS89 give the properties of the function $\varphi(y)$ independently of any model and whence of $P(\delta)$. They discuss possible analytical models for $\varphi(y)$. Further models closer to what is required by the data are given in Bernardeau and Schaeffer (1992). The scale-invariance of the correlations implies that the coefficients $S_n$ depend only on $n$, being independent of the cell size and time. Although the qualitative properties of these coefficients, thus, are similar in the non-linear and in the quasi-linear regimes, their values are expected, from theoretical reasons, to be different. Numerical simulations (Colombi et al 1996) indeed show they are constant in both regimes $\xia \ll 1$ and $\xia \gg 1$, with a strong transition at scales/times for which $\xia$ increases from $\simeq 1$ to $\simeq 100$, in agreement with the spherical collapse model originally derived only for the two-body correlation function (Gott and Rees 1975, Peebles 1980). Astonishingly, the functions $\varphi(y)$ mentioned above, that reproduce the data in the deeply non-linear regime and the ones calculated exactly in the quasi-linear regime do not look quantitatively very different. Indeed, their properties originate from general model-independent considerations such as the special scaling due to the fact the coefficients $S_n$ depend only on $n$, their rapid growth with $n$, the positivity of the probabilities and of the reduced moments $<\delta^{2p}>$.

\subsection{Properties of the density distribution as a function of smoothing scale}

	We summarize here the properties of the probability distribution of the
density within a cell, as derived by BS89 in both the $\xia \gg 1$ and the $\xia \ll 1$ regimes. These authors showed these properties describe the large-scale matter distribution in the Universe in the deeply non-linear regime, whereas Bernardeau (1992, 1994) showed they are also relevant -- with in principle different parameters -- to the quasi-linear regime.

\subsubsection{Model-independant properties of $\varphi(y)$}

	The function $\varphi(y)$ behaves as   
\beq
\left\{ \begin{array}{rcl}  y \rightarrow 0 & : & \varphi(y) = y-y^2/2+... \\ y \rightarrow \infty & : & \varphi(y) \sim a \; y^{1-\omega}  \;\;\;\;\; \mbox{with} \;\; 0 \leq \omega \leq 1 \end{array} \right.
\eeq
The rapid growth of $S_n$ as a function of $n$ implies a singular behaviour of $\varphi(y)$ at small negative values of $y$, say $y_s = -1 / x_*$ with $x_*$ being large ($x_* \sim 10$):
\beq
y \rightarrow y_s^{+}  \; : \;\; \varphi(y) = - a_s \Gamma(\omega_s) \; (y-y_s)^{-\omega_s}
\eeq
where we neglected less singular terms. Moreover, $\varphi(y)$ and all its derivatives $(-)^n \; \varphi^{(n)}(y)$ must be non-negative for all values of $y \geq 0$.

\subsubsection{Large scales / early times}

	For $\xia \ll 1$ the probability $P(\delta)$ can be evaluated using the saddle-point method. Defining
\[
\psi(\delta) = \varphi(y) - (1+\delta) y
\]
with $y$ given as a function $y(\delta)$ of $\delta$ by
\[
\varphi'(y) = 1+\delta
\]
we get
\beq
P(\delta) = \sqrt{ \frac{\psi''(\delta)}{2 \pi \xia}} \; e^{-\psi(\delta)/\xia}   \eeq
For $|\delta| \ll 1$ only, $\varphi(y)$ can be expanded around $y=0$ to yield $y(\delta)=-\delta$ and $\psi(\delta) = \delta^2 /2 \;$, which is the gaussian behaviour, more precisely valid when $|\delta| \ll 1/x_*$ so as to have the series 
giving $\varphi(y)$ converge, indeed an extremely small value (of the order of $1/x_* \simeq 0.1$).

\subsubsection{Small scales / late times}

	For $\xia \gg 1$, the many-body correlations totally dominate the density distribution. The probability $P(\delta)$ takes two different forms.
\\

	a) For values of $(1+\delta)$ that are not too small, $\varphi(y)/\xia$ in the exponent of (\ref{P(d)}) can be expanded, to give as the first non-vanishing contribution
\beq
P(\delta) = - \inta \frac{dy}{2 \pi i \xia^2} \; \varphi(y) \;\; e^{(1+\delta) y / \xia}         
\eeq
In this regime, $P(\delta)$ exhibits a {\em scaling law}, in the sense that $\xia^2 \; P(\delta)$ is a function of the unique parameter $x = (1+\delta)/ \xia \;$ :
\beq
1+\delta \gg  \xia^{\; -\omega/(1-\omega)} \;\; : \;\;\; P(\delta) = \frac{1}{\xia^{\;2}} \; h(x) \label{cond1}
\eeq
with 
\beq
h(x) = -\inta \frac{dy}{2 \pi i} \; \varphi(y) \; e^{xy} \\  = \inta \frac{dy}{2 \pi i} \; \frac{ \varphi'(y)}{x} \; e^{xy}
\label{Pdelhx}
\eeq
The behaviour of $\varphi(y)$ near $y=0$ implies the sum rule
\beq
\int_0^{\infty}  x h(x) \; dx  = 1
\eeq
which states that all the matter is within cells of density contrast satisfying (\ref{cond1}). In fact we have:
\beq
n \geq 1 \; : \;\; S_n = \int_0^{\infty}  x^n h(x) \; dx  \;\;\; , \;\;\; S_1=S_2=1  \label{Snhx}
\eeq

	At small values of $(1+\delta)/\xia$ that still obey the condition (\ref{cond1}), $P(\delta)$ is a power-law given by the behaviour of $\varphi(y)$ at large $y$
\beq
\xia^{\; \frac{-\omega}{1-\omega}} \ll 1+\delta \ll \xia \; : \;\;\;   P(\delta)  \simeq  \frac{a (1-\omega)}{\Gamma(\omega) \xia^2} \left[ \frac{1+\delta}{\xia} \right]^{\omega-2}   \label{hs}
\eeq
and for large $(1+\delta)/\xia$, the above saddle-point evaluation holds, to show that the singular behaviour of $\varphi(y)$ at $y=-1/x_*$ imposes
\beq
1+\delta \gg \xia \; : \;\;  P(\delta)  \simeq  \frac{a_s}{\xia^2} \left( \frac{1+\delta}{\xia} \right)^{\omega_s-1}  \exp \left[ - \frac{1+\delta}{x_* \xia} \right]  \label{hl}
\eeq

In the limit $\xia \rightarrow \infty$ we get:
\[
\varphi(y)/\xia \simeq  \int_{-1}^{\infty}(1- e^{-(1+\delta)y/\xia}) \;\;P(\delta) d\delta
\]
and
\beq
\varphi'(y) = \lim_{\xia \rightarrow \infty}  \int_{-1}^{\infty} e^{-(1+\delta)y/\xia} \;\; (1+\delta) P(\delta) d\delta   \label{hphip}
\eeq

	b)  For  values of $(1+\delta)$ that are not too large, only large values of $y$ contribute to the integral defining $P(\delta)$ so that $\varphi(y)$ can be replaced by its asymptotic form $a \; y^{1-\omega}$ to yield
\beq
P(\delta) = \inta \frac{dy}{2 \pi i \xia}  \; \exp \left[ \frac{(1+\delta)y - a y^{1-\omega} }{\xia} \right]    
\eeq		
that  implies another scaling of $P(\delta)$ as a function of the variable $z = (1+\delta) a^{-1/(1-\omega)}  \xia^{\; \omega/(1-\omega)}$
\beq
1+\delta \ll \xia \; : \;\;  P(\delta) = a^{-1/(1-\omega)} \; \xia^{\; \omega/(1-\omega)}  \; g(z)    \label{cond2}
\eeq
with
\beq
g(z) = \inta \frac{dt}{2 \pi i}  \exp(zt - t^{1-\omega} )
\eeq

	At small values of $z$, $P(\delta)$ takes the form 
\[
P(\delta)  \simeq   a^{\frac{-1}{1-\omega}} \; \xia^{\; \frac{\omega}{1-\omega}} \; \sqrt{ \frac{(1-\omega)^{1/\omega}}{2 \pi \omega z^{(1+\omega)/\omega} } } \; \exp \left[ - \omega \left( \frac{z}{1-\omega} \right)^{-\frac{1-\omega}{\omega}} \right]
\]
\beq
\mbox{if} \;\;\;\;\;  1+\delta \ll \xia^{\; -\omega/(1-\omega)}
\eeq
which vanishes exponentially when $(1+\delta)$ and whence $z$ goes to zero. For
large values of $z$, still under the condition (\ref{cond2}), $P(\delta)$ is the power-law
\beq
P(\delta)  \simeq  a^{-1/(1-\omega)} \; \xia^{\; \omega/(1-\omega)} \; \frac{1-\omega}{\Gamma(\omega)} \; z^{\omega-2}   \label{gl}
\eeq
\[
\mbox{when} \;\;\;\; \xia^{\; -\omega/(1-\omega)}  \ll  1+\delta  \ll  \xia
\]
This expression is identical to (\ref{hs}) since in the region $\xia^{\; -\omega/(1-\omega)}  \ll  1+\delta  \ll  \xia$ the conditions of validity of both expressions (\ref{hs}) and (\ref{gl}) overlap.

	The $h(x)$ and $g(z)$ scalings have been seen to hold in the deeply
non-linear regime in galaxy surveys -- the same scaling is expected for the galaxy distribution as advocated by BS89 but with a different function $h_{gal}(x)$ (since the distribution of galaxies is biased with respect to the underlying matter distribution) as shown by Bernardeau and Schaeffer 1992 -- for the CfA (Alimi et al 1990) as well as the SSRS (Maurogordato et al 1992, Benoist et al 1996). For galaxies one has $\omega$ measured to be $\simeq 0.6$ to 1, $a \simeq 1$, and a poorly determined value of $1/x_*$, that is definitely smaller than 0.3 . Much more precise tests can be done using numerical simulations of the matter distribution (Bouchet et al. 1991, Colombi et al. 1992,1994,1995) for various initial conditions. The density probability distribution is found to obey extremely well the above scaling, with functions $h(x)$ that depend on the initial conditions, with values of $\omega$ around 0.5 and a well defined exponential decrease with $1/x_*$ measured to be between 0.07 and 0.2 . In the quasi-linear regime, too, the theoretical predictions are practically indistinguishable from the results of the simulations at the times and scales corresponding to this regime (Bernardeau 1994).

\subsection{Picture of the density fluctuations implied by hierarchical clustering}

\begin{figure}[htb]

\begin{picture}(230,180)(-18,-5)

\epsfxsize=8 cm
\epsfysize=14 cm
\put(-8,-118){\epsfbox{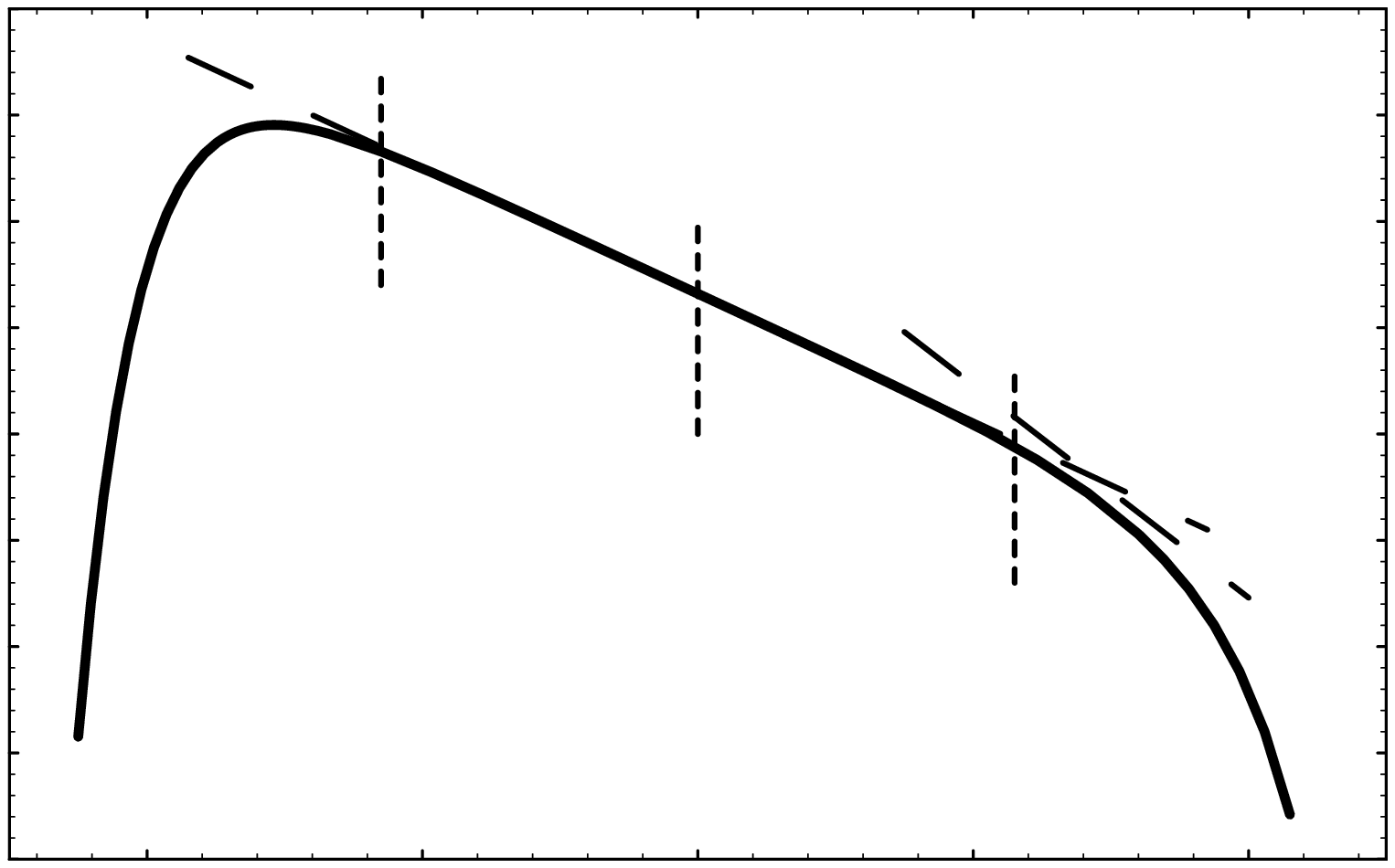}}
\put(25,-12){-4}
\put(111,-12){0}
\put(195,-12){4}
\put(-10,41){-10}
\put(-2,123){0}
\put(59,103){$\rho_v$}
\put(109,74){$\rho_0$}
\put(158,45){$\rho_c$}
\put(170,150){$\log[P(\rho)]$}
\put(160,5){$\log(\rho/\rho_0)$}
\put(60,35){$\xia = 200$}

\end{picture}

\caption{Probability $P(\rho)$ for finding a density $\rho$ in a cell within which the average correlation function is $\xia=200$ in the case $\omega=1/2, \; \omega_s=-3/2 \;$. The density distribution is a power-law with a cutoff at small ($\rho_v$) and large ($\rho_c$) densities. Both $\rho_v$ and $\rho_c$ are scale-dependent: for large $\xia$, $\rho_v$ goes to 0 while $\rho_c$ goes to infinity. The curve for $\rho \gg \rho_v$ corresponds to the part of $P(\rho)$ which  scales as $h(x)$, with a power-law decrease followed by an exponential  cut and the curve for  $\rho \ll \rho_c$ corresponds to the part of $P(\rho)$ which scales as $g(z)$. Note that both regions overlap. The dashed line on the left is the approximation (\ref{hs}), $P(\rho) \propto \rho^{\omega -2}$, and the dashed line on the right is the power-law $P(\rho) \propto \rho^{\omega_s -1}$ corresponding to (\ref{hl}) without the exponential falloff.}
\label{figPrho}

\end{figure}

	In the non-linear regime, the density distribution is thus a power-law of the density  $\rho = (1+\delta) \rho_0$ with an exponential falloff at $\rho > \rho_c = \rho_0 \xia$ at the high densitiy end, and another cut at $\rho < \rho_v = \rho_0 \; a^{1/(1-\omega)} \; \xia^{\; -\omega/(1-\omega)}$, as we can see on Fig.\ref{figPrho}. For a smaller scale $R$, $\xia$ gets extremely large, being infinite as $R \rightarrow 0$. So the range where $P(\delta)$ behaves as a power-law has a wider and wider extention. There is a non-vanishing probability to have arbitrary large densities, which means that sampling with small cells provides for rare  peaks whose hight increases without limit when $R$ goes to zero. The probability to find cells of arbitrary low density gets arbitrarily large in the same limit: there are numerous very underdense ``holes'' in the density distribution.

	The fraction of volume occupied by cells of density larger than $\rho = (1+\Delta) \rho_0$ is
\beq                          
F_v(>\Delta) = \int_{\Delta}^{\infty} P(\delta) \; d\delta  \simeq   \frac{a}{\Gamma(\omega)}  \frac{(1+\Delta)^{\omega-1}}{\xia^{\; \omega}}   \label{FvNL}
\eeq
for $1+\Delta  \gg  a^{1/(1-\omega)} \xia^{\; -\omega/(1-\omega)}$. The latter limit vanishes for small $R$: even for moderately small positive values of the density contrast $\Delta$, $F_v(>\Delta)$ is given by the above expression, and is small for large $\xia$: {\em the volume occupied by the overdense cells is vanishingly small}.

	The fraction of matter within cells of overdensity larger than $\Delta$ is, for $1+\Delta \ll \xia$,
\[                                         
F_m(>\Delta) =  \int_{\Delta}^{\infty} (1+\delta) P(\delta) \; d\delta
\]
\beq
\hspace{1.6cm}  \simeq  1 - \frac{a (1-\omega)}{\Gamma(\omega+1)} \left( \frac{1+\Delta}{\xia} \right)^{\omega}   \label{FmsupD}
\eeq
In the non-linear regime where $\xia$ is large, even for moderately large $\Delta$, {\em all the mass is contained in overdense cells}.

	At the larger scales, or the earlier times when $\xia$ gets smaller, the
power-law region in-between the two limits $\rho_v$ and $\rho_c$ where the probability falls down to zero shrinks and eventually disappears for $\xia \ll 1$, but whatever small $\xia$, {\em the gaussian behaviour appears only for densities very close to the average} $(|\delta| < 0.1)$. For $|\Delta| \ll \xia^{\; 1/2} \ll 1$, the volume in overdense $(\delta > \Delta)$ cells is
\beq
F_v(>\Delta) =  \int_{\Delta}^{\infty}  P(\delta) \; d\delta  \simeq  \frac{1}{2} - \frac{\Delta}{\sqrt{2 \pi \xia}}   \label{Fvgau}
\eeq            
while the fraction of mass is, since $\delta$ is small
\beq                                                               
\begin{array}{rl} F_m(>\Delta) & = {\displaystyle \int_{\Delta}^{\infty} (1+\delta) P(\delta) \; d\delta \simeq \int_{\Delta}^{\infty}  P(\delta) \; d\delta   } \\  \\   & {\displaystyle \simeq   \frac{1}{2} - \frac{\Delta}{\sqrt{2 \pi \xia}}  }  \end{array}
\eeq   
Thus, these overdense cells occupy half of the volume and contain half of the mass. This is genuinely due to the smallness of the density fluctuations in this regime.

\section{Non-linear models: surroundings of a cell}

\subsection{General results}

As we noticed in the main text, an alternative definition for the mass function that takes the surroundings of an overdensity  explicitely into account is $\eta (M) dM/M = \rho_0 /M f_m(>\Delta,<\Delta;R) dR$ where $f_m(>\Delta,<\Delta;R) dR = F_m(>\Delta,R;<\Delta,R+dR)$ is the mass fraction in cells with an overdensity larger than $\Delta$ in $R$ but smaller than $\Delta$ in $R+dR$. The mass fraction $F_m(>\Delta,R;<\Delta,R')$ in cells with overdensity larger than $\Delta$ in $R$ but smaller than $\Delta$ in $R'$ verifies $F_m(>\Delta,R;<\Delta,R') = (1+\Delta) P(>\Delta,R;<\Delta,R')$ so that
\beq
\eta (M) \frac{dM}{M} = \rho_0 /M  \; (1+\Delta) P(>\Delta,R;<\Delta,R')
\eeq
In the same way  we introduce a second definition for  the multiplicity function of underdense regions $\eta_u(R) \; dR/R$ that  can be formulated as:
\beq
\eta_u(R) \; \frac{dR}{R} = \frac{1}{V} \; P(<-\Delta ,R; >-\Delta,R')
\eeq
There is an important relation among these probabilities that may  be written
\beq
\begin{array}{rcl} P(>\Delta,R;<\Delta,R') &  =  & P(>\Delta,R) - P(>\Delta,R') \\ & & + P(<\Delta,R;>\Delta,R') \end{array}
\label{P(><)}
\eeq
or
\[
\begin{array}{rcl} P(<-\Delta,R;>-\Delta,R') & = & P(<-\Delta,R) - P(<-\Delta,R') \\ & & + P(>-\Delta,R;<-\Delta,R') \end{array}
\]
both being valid for either sign of $\Delta$ . The difference in the r.h.s is our standard definition of the mass function. The alternative definition of the mass function, the standard one and the distribution of underdense regions are thus related

\subsubsection{Distribution within two volumes}

We are thus led to consider two volumes: a central sphere of radius $R$, volume $V$, and 
a sphere
of radius $R'$, volume$V'$, or the corona of volume $V''=V'-V$.
We need the probability distribution $P(\delta,R;\delta',R')$ for a density contrast $\delta$ over $R$ and $\delta'$ over $R'$. As one does for a simple cell, we define the generating function:
\beq
{\cal P}(\lambda,\lambda'')=\sum_{N,N''} \lambda^N \lambda''^{N''} P(N,N'')
\eeq
where $P(N,N'')$ is the probability that there are $N$ particles in the sphere $V$ and $N''$ particles in the corona $V''$. As is well-known, the generating function is related to the many-body correlation function through:
\beq
{\cal P}(\lambda,\lambda'') = \exp [\chi (\lambda,\lambda'')]
\eeq
with
\[
\chi (\lambda,\lambda'') = \sum_{n=1}^{\infty} \frac{(-\rho_0)^n}{n!} \int_{V'} \prod_n d^3r_i \prod_n \mu(r_i) \; \xi_n(r_1,...,r_n)
\]
and
\[
\mu(r)=\left\{ \begin{array}{ll} (1-\lambda) & \mbox{over} \; V \\ (1-\lambda'') & \mbox{over} \; V'' \end{array} \right.
\]
Then, as is usual for simple cells, we define (BS89)
\[	
\xia_{k,n-k} = \int_V \prod_k \frac{d^3r_i}{V} \int_{V'} \prod_{n-k} \frac{d^3r''_j}{V''} \; \xi_n(r_1,..,r_k,r''_1,..,r''_{n-k})
\]
\[
S_{k,n-k} = \frac{\xia_{k,n-k}}{\xia^{\;n-1}} \;\; , \;\; N_c = \rho_0 V \xia \;\; , \;\; \epsilon=V''/V
\]
\[
y = (1-\lambda)N_c \;\; , \;\; y'' = \epsilon (1-\lambda'')N_c  \;\; , \;\; \chi (\lambda,\lambda'') = - \psi(y,y'') /\xia
\]
Hence:
\beq
\psi(y,y'') = \sum_{n=1}^{\infty} \frac{(-)^{n-1}}{n!} \sum_{k=0}^n \left( \begin{array}{c} n \\ k \end{array} \right) \; S_{k,n-k} \; y^k \;  y''^{\;n-k}
\eeq
and the continuous limit in $N,N''$ ( denoted $\nu,\nu''$ in this case ), gives:
\[
P(\nu,\nu'') \! = \!\! \inta \!\!\! \frac{dydy''}{\epsilon(2 \pi i N_c)^2} \exp \left[ \frac{\nu}{N_c}y+\frac{\nu''}{\epsilon N_c}y''-\frac{\psi(y,y'')}{\xia} \right]
\]
The function $\psi(y,y'')$ depends on $\epsilon$ through the coefficients $S_{k,n-k}$. It is the analog of the function $\varphi(y)$ we used for simple cells. Using the fact that $\chi_{V,V''}(\lambda,\lambda) = \chi_{V+V''}(\lambda)$, we have:
\beq
\psi(y,y) = (1+\epsilon)^{\gamma/3} \; \varphi \left[ y(1+\epsilon)^{1-\gamma/3} \right] 
\eeq

\subsubsection{Limit of an infinitesimally thin corona}

Since $S_{n-1,1}=[1-\gamma/3 +\gamma/(3n) ] S_n$ to zeroth order in $\epsilon$, the expansion of $\psi(y,y'')$ to the first order in $\epsilon$ or $y''$ is:
\beq
\psi(y,y'') = \varphi(y) +  y'' \left[ \left( 1-\frac{\gamma}{3} \right) \varphi'(y) + \frac{\gamma}{3} \frac{\varphi(y)}{y} \right] + ...       \label{delPsi}
\eeq
If we now consider the density contrasts $\delta$ over $V$ and $\delta'$ over $V'$ we get:
\[
P(\delta,\delta') = \inta \frac{dydy''}{\epsilon(2 \pi i \xia)^2} (1+\epsilon) 
\]
\[
\times \exp \left[ \frac{1+\delta}{\xia} y + \frac{(1+\epsilon)(1+\delta')-(1+\delta)}{\epsilon \xia}y'' - \frac{\psi(y,y'')}{\xia} \right] 
\]
Then a simple integration gives to first order in $\epsilon \;$:
\beq
P(>\Delta,<\Delta) = \epsilon \inta \frac{dydy''}{(2 \pi i)^2} \frac{1}{y''^2} e^{ \{ (1+\Delta)(y+y'') - \tilde{\psi} \}/{\xia} } 
\eeq
with $\;\;$ Re $(y'')>0 \;\;$ along the integration path, and
\beq
\tilde{\psi}(y,y'') = \sum_{n=1}^{\infty} \frac{(-)^{n-1}}{n!} \sum_{k=0}^n \left( \begin{array}{c} n \\ k \end{array} \right) \; \tilde{S}_{k,n-k} \; y^k \;  y''^{n-k}  
\eeq
where the function $\tilde{\psi}(y,y'')$ is independent of $\epsilon$ and:
\[
\tilde{S}_{k,n-k} = \lim_{\epsilon \rightarrow 0} S_{k,n-k} \;\;\;\; , \;\;\;\; \tilde{\psi}(y,y'') = \lim_{\epsilon \rightarrow 0} \psi(y,y'')
\]

Since $\eta(M) dM/M = \rho_0 /M \; F_m(>\Delta,R;<\Delta,R')$ ,
we get writing $dM=\epsilon M$:
\[
\eta(M) = \frac{\rho_0}{M} \inta \frac{dydy''}{(2 \pi i)^2} \frac{1+\Delta}{y''^2
} e^{ \{ (1+\Delta)(y+y'') - \tilde{\psi}(y,y'') \}/{\xia} } 
\]
Using the same approximations as for simple cells we have in the limit of large  $\xia$:
\beq
\eta(M) \frac{dM}{M} = \frac{\rho_0}{M} x H(x) dx \;\;\; \mbox{with} \; x=\frac{M}{N_c}=\frac{1+\Delta}{\xia}
\eeq
and
\beq
H(x) = -\frac{3}{\gamma} \inta \; \frac{dydy''}{(2\pi i)^2} \; \frac{\tilde{\psi}(y,y'')}{xy''^2} \; e^{x(y+y'')}  
\eeq
with Re$(y'')>0$ along the integration path, which replaces the function $h(x)$ we had for simple cells:
Hence to get the behaviour of the new mass function as compared to the one we got from simple cells we only need to compare $h(x)$ and $H(x)$. It is also convenient to write $H(x)$ as:
\beq
H(x) = \inta \frac{dy}{2 \pi i} \frac{Q(y)}{x} e^{xy}
\eeq
with
\beq
Q(y) = -\frac{3}{\gamma} \inta \frac{dy''}{2 \pi i} \frac{\tilde{\psi}(y-y'',y'')}
{y''^2} \;\; , \; \mbox{Re} (y'')>0       \label{Qy}
\eeq
that can be usefully compared to the same expression (\ref{Pdelhx}) for $h(x)$
in which $\varphi'(y)$ replaces $Q(y)$.

\subsubsection{Properties of $ H(x)$ }

It is also possible to relate the normalization of the mass function to the function $Q(y)$. Indeed, if we note that $\eta(M) dM/M = \rho_0/M \; \mu(M) dM/M$ where $\mu(M)dM/M$ is the mass fraction in objects of mass between $M$ and $M+dM$, we have:
\beq
\int_0^{\infty} \mu(M) \frac{dM}{M} =  \int_0^{\infty} x H(x)dx = Q(0)
\eeq
Hence $Q(0)$ should be equal to unity if all the mass is distributed within the objects defined this way, and if the same overdensity has not been counted several times at different scales (as can be feared due to the cloud-in-cloud problem)
\\

It is readily seen from (\ref{P(><)}) that $H(x)$ is related to $h(x)$ by
\[
\begin{array}{rcl} xH(x) \; dx & = & {\displaystyle - \; \frac {d}{dx} \left( x \int_x^{\infty}h(x)dx \right) \; dx } \\ \\ & & + (1+\Delta) P(<\Delta,R;>\Delta,R') \end{array}
\]

We have from the expression of $P(\nu,\nu'')$, in the limit $\epsilon \rightarrow 0$ :
\beq
P(\delta,\delta'') =  \inta \frac{dydy''}{(2 \pi i \xia)^2} e^{ \{ (1+\delta) y + (1+\delta'') y'' - \tilde{\psi}(y,y'') \}/ \xia } 
\eeq
The corresponding Laplace transform is:
\beq
e^{-\tilde{\psi}(y,y'')/\xia} = \int_{-1}^{\infty} e^{ -\{ (1+\delta) y + (1+\delta'') y''\}/\xia } \;\; P(\delta,\delta'') d\delta d\delta''
\eeq
while for simple cells, we have(\ref{phiP(d)}). We can note that these relations imply:
\[
\left\{  \begin{array}
{l} \mbox{Re}(y) \geq 0 \; : \;\; \left| e^{-\varphi(y)/\xia} \right| \leq 1 \;\; , \;\; \mbox{Re}[\varphi(y)] \geq 0 \\
 \mbox{Re}(y) \geq 0  \; , \;  \mbox{Re}(y'') \geq 0  \; : \;\; \mbox{Re}[\tilde{\psi}(y,y'')] \geq 0  \end{array} \right.
\]
For double cells we get:
\[
\tilde{\psi}(y,y'')/\xia \simeq  \int_{-1}^{\infty} (1-e^{ -\{ (1+\delta) y + (1+\delta'') y''\}/\xia }) P(\delta,\delta'') d\delta d\delta''  
\]
If we report this expression into (\ref{Qy}) we obtain:
\[
Q(y) = \lim_{\xia \rightarrow \infty} \;\; \frac{3}{\gamma} \int_{-1}^{\infty} d\delta \; e^{-(1+\delta)y/\xia}
\]
\beq
\hspace{1.6cm} \times \int_{-1}^{\delta} d\delta'' \;\;[(1+\delta)-(1+\delta'')] P(\delta,\delta'')  
\eeq

Moreover, we get from the definition of $h(x)$ and $H(x)$:
\[
\left\{ \begin{array}{rcl} x h(x) & = & {\displaystyle \lim_{\xia \rightarrow \infty}} \;\; \xia \; (1+\delta) P(\delta) \\ \\  x H(x) & = & {\displaystyle \lim_{\xia \rightarrow \infty}} \;\; \xia \; \frac{3}{\gamma} {\displaystyle \int}_{-1}^{\delta}  [(1+\delta)-(1+\delta'')] P(\delta,\delta'') d\delta''  \end{array} \right.
\]
with $x = (1+\delta)/\xia$. Hence we obtain:
\beq
x \geq 0 \; : \;\; \frac{3}{\gamma} \left[ 1 - \frac{<1+\delta''>_{/\delta}}{1+\delta} \right]   \leq 
 \frac{H(x)}{h(x)} \leq \frac{3}{\gamma}   \label{Hhall}
\eeq
where $<1+\delta''>_{/\delta}$ is the mean value of $(1+\delta'')$ when there is a fixed density contrast $\delta$ in the internal sphere $V$. When $x \gg x_*$, one can show that $<1+\delta''>_{/\delta} = (1-\gamma/3) (1+\delta)$, which corresponds to a fractal dimension $3-\gamma$, 
so that:
\beq
x \gg x_* \; : \;\;  h(x)  \leq H(x) \leq \frac{3}{\gamma} h(x)
\label{H><}
\eeq
Finally, the normalization of the mass function is given by:
\beq
Q(0) = \frac{3}{\gamma} \int_{-1}^{\infty} d\delta \int_{-1}^{\delta} d\delta'' [(1+\delta)-(1+\delta'')] P(\delta,\delta'')    
\eeq
Thus we obtain:
\beq
\forall x \; : \;\; 0 \leq H(x) \leq \frac{3}{\gamma} h(x) \;\;\; \mbox{and} \;\;\; 0 \leq Q(0) \leq \frac{3}{\gamma}
\label {Q><}
\eeq
The mass function defined by means of $ P(>\Delta,R;<\Delta,R')$ due to possible density inversions measured by $ P(<\Delta,R;>\Delta,R')$ counts the same overdensities several times. This illustrates the overcounting inherent in a density threshold algorithm, due to the cloud-in-cloud problem. The bound (\ref{Q><}) however shows this problem is mild when the counts are done in the non-linear regime. The same counts of overdensities in the early universe lead to a dramatic overcounting since each overdensities are counted infinitely many times at different scales (App.A) in the limit $\xia \rightarrow \infty$. We can note on Fig.\ref{figPrho} and eq.(\ref{hs}) that integrals of the form $\int (1+\delta) P(\delta) d\delta$ are dominated by values of $x \sim x_*$, or $(1+\delta) \sim x_* \xia$. Hence, we can expect that $Q(0) \simeq 1$ in the general case. Thus, both mass functions, expressed with $h(x)$ and $H(x)$, are probably quite similar. We shall confirm this in the next section in the case of a specific model.
\\

We also obtain the general relations:
\beq
\left\{ \begin{array}{l} \frac{P(>\Delta,<\Delta)}{\epsilon} = {\displaystyle \int}_{-1}^{\Delta} [(1+\Delta)-(1+\delta'')] P(\Delta,\delta'') d\delta'' \\  \\  \frac{P(<\Delta,>\Delta)}{\epsilon} = {\displaystyle \int}_{\Delta}^{\infty} [(1+\delta'')-(1+\Delta)] P(\Delta,\delta'') d\delta'' \end{array} \right.  \label{Psupinf}
\eeq
in the limit $\epsilon \rightarrow 0$. We shall see in the next section in some particular cases that for large densities $P(>\Delta,<\Delta) \gg P(<\Delta,>\Delta)$ while for very small densities $P(>\Delta,<\Delta) \ll P(<\Delta,>\Delta)$.
\\

\subsubsection{Underdense regions}

In a fashion similar to the calculation of $P(>\Delta,<\Delta)$ one gets:
\[
P(<-\Delta,>-\Delta) = \inta \frac{dydy''}{(2 \pi i)^2} \frac{\epsilon}{y''^2} 
e^{ \{ (1-\Delta)(y+y'') - \tilde{\psi} \}/{\xia} } 
\]
with $\;\;$ Re $(y'')<0 \;\;$ along the integration path. We shall see in the next section, with a particular model for $\tilde{\psi}(y,y'')$, that this leads to
\beq
\eta_u(R) \; \frac{dR}{R} = - \; \frac{1}{V} \; G(z) \; dz
\eeq
where $G(z)$ replaces the function $g(z)$ we had for simple cells, and is defined by:
\beq
P(<-\Delta , >-\Delta) = - G(z) \; dz 
\eeq
and $z=(1-\Delta) \; a^{-1/(1-\omega)} \; \xia^{\;\omega/(1-\omega)}$ is the variable we already encountered in the case of simple cells. Using (\ref{PVoid}) we can see that 
\beq
-G(z) dz  = - g(z) dz + P(>-\Delta,<-\Delta)  \label{Gg}
\eeq
which implies that
\beq
\forall z \geq 0 \; : \;\; G(z) \geq g(z)
\eeq
We can note that this also means that $P(>-\Delta,<-\Delta) \leq P(<-\Delta,>-\Delta)$. This is consistent with the fact that, when $x \ll x_*$, hence when the scaling in $g(z)$ is valid, we have $<1+\delta''>_{/\delta} = (1+\gamma/3 \; \omega/(1-\omega)) (1+\delta)$, which corresponds to a fractal dimension $3+\gamma \; \omega/(1-\omega)$ larger than 3. From (\ref{Gg}) and (\ref{Psupinf}) we also see that
\[
G(z)  =  g(z) - \frac{\epsilon}{dz}  \int_{-1}^{-\Delta}
 [(1-\Delta)-(1+\delta'')] P(-\Delta,\delta'') d\delta''
\]
which, since $\epsilon = - \frac{3(1-\omega)}{\gamma\omega} \frac{dz}{z}$, implies
\[
G(z)  \leq  g(z) + \frac{3(1-\omega)}{\gamma\omega \; z}  \int_{-1}^{\infty}
 (1-\Delta) P(-\Delta,\delta'') d\delta''
\]
that is
\[
G(z)  \leq  \left[ 1 +\frac{3(1-\omega)}{\gamma\omega} \right] g(z)
\]
Whence, collecting all inequalities we obtain
\beq
g(z) \leq G(z)  \leq  \left[ 1 +\frac{3(1-\omega)}{\gamma\omega} \right] g(z) \label{gGg}
\eeq
that implies very strong constraints on the asymptotic forms of $G(z)$ which are the same as those of $g(z)$, except for a normalization constant, for both $z \rightarrow 0$ and for $z \rightarrow \infty$. In the latter case, writing $G(z) = k \; g(z) \propto z^{\omega-2}$, we then have $k \geq 1$. When $z \gg 1$ and $x \ll 1$ we are in the region where both scalings in $z$ and in $x$ hold: $H(x)$ thus behaves as $x^{\omega-2}$ as does $h(x)$, except in the special case where $k=1$ and where $H(x)$ then decreases faster, as $x^{\omega-1}$ in the generic case. This completes the proof that in general $H(x)$ and $h(x)$ have the same asymptotic forms.

	In a more specific way, one can show that if the singularity of $\tilde{\psi}(y,y'')$ near $(y_s,0)$ can be written as:
\beq
\tilde{\psi}(y,y'') = -a_s \Gamma(\omega_s) \; \left[ y-y_s + (1-\gamma/3) y'' \right]^{-\omega_s}
\eeq
(where the factor $(1-\gamma/3)$ is implied by (\ref{delPsi})) one gets for $x \gg x_*, \; \xia \gg 1$ :
\beq
\left\{ \begin{array}{l}  H(x) \simeq h(x) \\  P(>\Delta,<\Delta) \gg P(<\Delta,>\Delta) \end{array} \right.
\eeq
This will also be the case for the Hamilton model considered below. Similarly, if $\zeta(\tau) \propto \tau^{-\kappa}$ for large $\tau$ with $\kappa>0$, one obtains for $z \ll 1, \; \xia \gg 1$ :
\beq
\left\{ \begin{array}{l}  G(z) \simeq g(z) \\  P(<\Delta,>\Delta) \gg P(>\Delta,<\Delta) \end{array} \right.
\eeq

\subsection{Tree-model}

In the peculiar case of the tree-model (Schaeffer 1984, Bernardeau and Schaeffer 1992) , the many-body correlation functions verify:
\beq
\xi_n(r_1,...,r_n) = \sum_{(\alpha)} Q_n^{(\alpha)} \sum_{t_{\alpha}} \prod_{n-1} \xi(r_i,r_j)
\eeq
where $(\alpha)$ is a particular tree topology connecting the $(n-1)$ points without making any loop, $Q_n^{(\alpha)}$ is a parameter associated with the order of the correlation and the topology involved, $t_\alpha$ is a particular labelling of the given topology $(\alpha)$ and the last product is made over the $(n-1)$ links between the $n$ points with two-body correlation functions. Moreover, we assume that the parameters $Q_n^{(\alpha)}$ are given by
\beq
Q_n^{(\alpha)} =  \prod_{i=2}^{\infty} \nu_i^{d_i(\alpha)}
\eeq
$\nu_i$ being a weight associated with a vertex of $i$ lines and $d_i(\alpha)$ the number of such vertices in the topology $(\alpha)$, and we define the generating function 
\beq
\zeta(\tau) = \sum_{i=0}^{\infty} \frac{\nu_i}{i \: !}\;\;\;\;\;\; \mbox{with} \;\;\;\;\; \nu_0 = \nu_1 = 1
\eeq
In this case, one can show that $\chi(\lambda,\lambda')$, that is $-\psi(y,y'')/\xia$, is given by the system
\beq
\left\{ \begin{array}{l} \tau(r)  =  - {\displaystyle \int}_{V+V''} d^3r' \rho_0 \mu(r') \xi(r,r') \zeta'(\tau(r')) \\        \\
\chi  =  - {\displaystyle \int_{V+V''} d^3r \rho_0 \mu(r) \left[ \zeta(\tau(r)) - \frac{\tau(r) \zeta'(\tau(r))}{2} \right] }
\end{array} \right.       \label{chi}   
\eeq
The equations for $\tilde{\chi}=-\tilde{\psi}(y,y'')/\xia$ with $\mu(r) = 1-\lambda$ in volume $V$ and $1-\lambda''$ in $V''$ are then obtained from (\ref{chi}) by taking the limit $\epsilon \rightarrow 0$ at fixed, finite $y''$:
\beq
\left\{ \begin{array}{rcl} \tau(r) & = & - y { \displaystyle \int_V\frac{d^3r'}{V} \frac{\xi(r,r')}{\xia} \zeta'(\tau(r'))-y''\frac{\xi(r,R)}{\xia} \zeta'(\tau(R)) }
 \\     \\
\tilde{\psi} & = &  y {\displaystyle \int_V\frac{ d^3r}{V}  \left[ \zeta(\tau(r)) - \frac{\tau(r) \zeta'(\tau(r))}{2} \right] } \\ & & \\ & & + \; y'' {\displaystyle \left[ \zeta(\tau(R)) - \frac{\tau(R) \zeta'(\tau(R))}{2} \right] } \end{array} \right.       \label{chitilde}
\eeq
which are exact equations.

\subsubsection{Constant $ \tau $ approximation}

A very useful approximation, for a single cell ($V''=0$) in the case where $\mu (r)$ is constant in the volume $V$, which turns out to be astonishingly accurate in the tree-model, is to use for the solution $\tau(r)$ of (\ref{chi}) a constant $\tau$. This leads to the equations
\beq
\left\{ \begin{array}{rcl} \tau & = & - y \zeta'(\tau) \\  \\ \hat{\varphi}(y) & = & y [ \zeta(\tau) - \tau \zeta'(\tau) /2 ] \end{array} \right. \;\;\; \mbox{and} \;\; \hat{\varphi}'(y) = \zeta(\tau)
\eeq

\subsubsection{Hamilton model}

We shall now consider the more specific case of the Hamilton model:
\beq
\zeta(\tau) = 1-\tau+\frac{\tau^2}{4} 
\eeq
For a single cell, this leads to the exact solution
\beq
\varphi(y) = y \int_V \frac{ d^3r}{V} \frac{ d^3r'}{V} \left( 1+\frac{y}{2}\frac{\xi}{\xia} \right)^{-1}
\eeq
where the inverse is meant in the operator sense, within the cell $V$. In the constant $\tau$ approximation we get 
\beq
 \hat{\varphi}(y)= \frac{y}{1+y/2}
\eeq
and
\[
x>0 \; : \;\; \hat{h}(x)=- \inta \frac{dy}{2 \pi i} \; \frac{y}{1+y/2} \; e^{xy} = 4 e^{-2x}
\]

For a cell with an infinitesimally thin edge, we have the exact solution of (\ref{chitilde}), after the elimination of $\tau(R)$ and $\xi(r,R)$ using (\ref{delPsi}):   
\beq
\tilde{\psi}(y,y'') = \varphi(y) + \frac{y'' \; [ \; (1-\alpha) \frac{\varphi(y)}{y}+\alpha \varphi'(y) \; ]} {1+\beta\frac{y''}{2}-\alpha\frac{yy''}{4}
Tr[\frac{\xi}{\xia}(1+\frac{y}{2}\frac{\xi}{\xia})^{-1}\frac{\xi}{\xia}]}
\label{psiham}
\eeq
with
\[
\left\{ \begin{array}{l} \alpha = (1-\gamma/3) = 0.47 \\ \beta = (1-\gamma/3) (1-\gamma/4) (1-\gamma/6)/(1-\gamma/2) = 1.54 \end{array} \right.
\]
where we have used  $\gamma = 1.8$. 

In the constant $\tau$ approximation (\ref{psiham}) reduces to
\beq
\hat{\psi}(y,y'') = \frac{y+y''+yy'' (\beta-\alpha)/2}{1+y/2+y''\beta/2+yy'' (\beta-\alpha)/4}
\eeq

We have
\[
\mbox{Re}(y'') >0 \;\; , \;\;\;\; \hat{Q}(y) = -\frac{3}{\gamma} \inta \frac{dy''}{2 \pi i} \frac{1}{y''^2}  
\]
\[
\;\;\;\;\;\; \times \; \frac{y+(y-y'')y'' (\beta-\alpha)/2}{1+(y-y'')/2+y''\beta/2+(y-y'')y'' (\beta-\alpha)/4} 
\]
A simple calculation shows that:
\beq
\left\{ \begin{array}{rcl} y \rightarrow -2 & : & \hat{Q}(y) \sim \ \frac{3}{\gamma} \frac{1}{(1+\frac{y}{2})^2} \\ & & \\ y \rightarrow \infty & : & \hat{Q}(y) \sim \frac{3}{\gamma} \left( \frac{2\beta-\alpha-1}{\beta-\alpha} \right)^2 \frac{1}{y^2} \end{array} \right.
\eeq
and
\beq
\left\{ \begin{array}{rcll} x \rightarrow \infty & : & \hat{H}(x) \sim \hat{h}(x) \\ & & & \\ x \rightarrow 0 & : &  \hat{H}(x) \sim \frac{3}{4\gamma} \left( \frac{2\beta-\alpha-1}{\beta-\alpha} \right)^2 \hat{h}(x) & \simeq 1.07 \hat{h}(x) \end{array} \right.
\eeq
The normalization of the mass function is now:
\beq
\int_0^{\infty} \hat{\mu}(M)dM = \hat{Q}(0) 
\eeq
which leads to:
\beq
\int_0^{\infty} \hat{\mu}(M)dM = \frac{3}{\gamma} \frac{\beta-\alpha}{\sqrt{4(\beta-\alpha)+(\beta-1)^2}} \simeq 1.09
\eeq
which is close to unity. Thus, the scaling functions $h(x)$ and $H(x)$ are very close.

\subsection{Non-constant density contrast $(1+\Delta)(R)$}

	As explained in the main text, Chapter 5, one is led to consider objects defined by a density threshold which is not necessarily constant, and given by a curve $(1+\Delta)(R)$. If we note $-\beta$ the slope of the curve $(1+\Delta)(R)$, we have:
\beq
\beta(R) = - \frac{dln(1+\Delta)}{dlnR} \;\;\; , \;\;\; \frac{dlnx}{dlnR} = \gam -\beta
\eeq
Then, we define the mass fraction $d{\cal M}$ formed by objects of scale $R$ - $R+dR$ by: 
\beq
\left\{ \begin{array}{rrcl} \beta < \gam : & d{\cal M} & = & (1+\Delta) P(>\Delta,<\Delta') \\ \\  \beta > \gam : & d{\cal M} & = & (1+\Delta) P(<\Delta,>\Delta') \end{array} \right.   
\eeq
where $\Delta=\Delta(R)$, $\Delta'=\Delta(R')$ and $R'=R+dR$, with $dR>0$. 

	We first consider the case $\beta<\gam$. Then, in a fashion similar to what we did in App.C.1.2, we obtain:
\[
P(>\Delta,<\Delta') = \int \frac{dydy''}{(2 \pi i)^2} \frac{\epsilon}{y''^2} e^{ \{ (1+\Delta)[y+(1-\beta/3)y''] - \tilde{\psi} \}/{\xia} } 
\]
with Re$(y'')>0$ along the integration path. As was the case for a constant density threshold, we can write in the regime where the scaling in $x$ is valid: $(1+\Delta) P(>\Delta,<\Delta') = x H(x) dx$ with:
\beq
x H(x) = \frac{-3}{\gam-\beta} \inta \frac{dydy''}{(2 \pi i)^2} \;\; \frac{\tilde{\psi}}{y''^2} \;\; e^{ x [y+(1-\beta/3)y'']  } 
\eeq
with Re$(y'')>0$. Hence, the mass fraction can be written $d{\cal M} = x H(x) dx$ as usual. In terms of the probability distribution $P(\Delta,\delta'')$ in the corona we obtain:
\[
x H(x) = \lim_{\xia \rightarrow \infty} \;\;  \frac{3\xia}{\gamma-\beta}  \int_{-1}^{(1+\Delta)(1-\beta/3)-1}  d\delta'' \]
\beq
\hspace{1cm} \times \; [(1+\Delta)(1-\beta/3)-(1+\delta'')] P(\Delta,\delta'') 
\eeq
Hence we obtain the general bounds:
\beq
\frac{3}{\gamma-\beta} \left[ 1-\frac{\beta}{3} - \frac{<1+\delta''>_{/\Delta}}{1+\Delta} \right]   \leq 
 \frac{H(x)}{h(x)} \leq \frac{3-\beta}{\gamma-\beta}  
\eeq
where $<1+\delta''>_{/\Delta}$ is the mean value of $(1+\delta'')$ when there is a fixed density contrast $\Delta$ in the internal sphere $V$. For large $x$ we have:
\beq
x \gg x_* \; : \;\;  h(x)  \leq H(x) \leq \frac{3-\beta}{\gamma-\beta} h(x)
\eeq

	In the regime where the scaling in $x$ for $P(\Delta)$ is valid, we obtained $d{\cal M} = (1+\Delta) P(>\Delta,<\Delta') = \epsilon f(x)$ where $f(x)$ is a function of the sole variable $x$, and $\epsilon=3 dR/R$. If $\beta=\gam$, objects are defined by a constant $x$, which leads within this framework to a mass fraction $\mu(R) dR/R$ in objects of scale $R$ to $R+dR$: $\mu(R) = 3 f(x)$=constant. Hence we cannot define a mass function in the case $\beta = \gam$. The fact this value of $\beta$ leads to some problems was already apparant in section 5.1.

	In the case $\beta>\gam$, we obtain in a similar fashion:
\[
P(<\Delta,>\Delta') = \int \frac{dydy''}{(2 \pi i)^2} \frac{\epsilon}{y''^2} e^{ \{ (1+\Delta)[y+(1-\beta/3)y''] - \tilde{\psi} \}/{\xia} } 
\]
with Re$(y'')<0$, and we can write $(1+\Delta) P(<\Delta,>\Delta') = - x H(x) dx$, with $dx<0$. We can also express $H(x)$ as:
\[
x H(x) = \lim_{\xia \rightarrow \infty} \;\;  \frac{3 \xia}{\beta-\gamma}  \int_{(1+\Delta)(1-\beta/3)-1}^{\infty}  d\delta''
\]
\beq
\hspace{1cm} \times \; [(1+\delta'') - (1+\Delta)(1-\beta/3)] P(\Delta,\delta'')    \label{H<>2}
\eeq
If $\gam < \beta <3$, we obtain the bounds:
\[
\left[ \frac{<1+\delta''>_{/\Delta}}{1+\Delta} -1 + \frac{\beta}{3} \right]   \leq \frac{\beta-\gamma}{3} \frac{H(x)}{h(x)} \leq \frac{<1+\delta''>_{/\Delta}}{1+\Delta}
\]
which lead to:
\beq
x \gg x_* \; : \;\; 1 \leq \frac{H(x)}{h(x)} \leq \frac{3-\gam}{\beta-\gamma}
\eeq
\[
x \ll x_* \; : \;\;  \left[ \beta + \frac{\gam \omega}{1-\omega} \right] \leq (\beta-\gam) \frac{H(x)}{h(x)} \leq \left[ 3 + \frac{\gam \omega}{1-\omega} \right]
\]
In the case $\beta > 3$, a simplification occurs because the lower bound of the integral in (\ref{H<>2}) is smaller than $-1$. This is linked to the fact that for such a steep constraint the other probability $P(>\Delta,<\Delta')$ is zero, because the average density of the halo cannot decrease faster than $R^{-3}$ by conservation of the mass (in other words there are no negative densities). Then, (\ref{H<>2}) reads:
\[
x H(x) = \lim_{\xia \rightarrow \infty} \;\; \frac{3 \xia P(\Delta)} {\beta-\gamma}    \left[ <1+\delta''>_{/\Delta} - (1+\Delta)(1-\frac{\beta}{3}) \right] 
\]
so that 
\beq 
H(x) = h(x) \; \frac{3}{\beta-\gamma} \left[ \frac{<1+\delta''>_{/\Delta}}{1+\Delta} -1 +\beta/3 \right]
\eeq
which leads to
\beq
x \gg x_* \; : \;\; H(x) = h(x) 
\eeq
\beq
x \ll x_* \; : \;\; H(x) = h(x) \; \frac{1}{\beta-\gamma} \left[ \beta +  \frac{\gam \omega}{1-\omega} \right] 
\eeq

	We can note that for $|\beta| \rightarrow \infty$, which corresponds to objects defined by a constant radius $R$, we get $H(x)=h(x)$. 
 
	As we noticed in the main text, Chapter 5, the mass functions obtained above are not entirely satisfactory as there is no guarantee that the density of the ``objects'' picked out in this way decreases at larger radius, even locally at the considered scale $R$, if $\beta<0$ or $\beta>\gam$. Hence we are led to include explicitely the additional constraint that the density profile is locally decreasing. 

	Thus, in the case $\beta<0$ we can define the function $H(x)$ such that the mass fraction of the objects of interest is $d{\cal M} = x H(x) dx$ by:
\beq
x H(x) dx = (1+\Delta) \int_{\Delta}^{\infty} d\delta \int_{-1}^{Min(\Delta',\delta)} d\delta' P(\delta,\delta')
\eeq
However, no general bounds can be obtained.

	In the case $\beta>\gam$, we define in a similar fashion:
\beq
x H(x) dx = - (1+\Delta) \int_{\Delta'}^{\Delta} d\delta \int_{\Delta'}^{\delta} d\delta' P(\delta,\delta')
\eeq
which leads to 
\beq
\begin{array}{rl} x H(x) = & {\displaystyle \lim_{\xia \rightarrow \infty} \;\; \frac{3 \xia}{\beta-\gamma}  \int_{(1+\Delta)(1-\beta/3)-1}^{\Delta} d\delta'' } \\  \\  & \times \; [(1+\delta'') - (1+\Delta)(1-\beta/3)] P(\Delta,\delta'') \end{array}
\eeq
If $\beta>3$ we obtain the upper bound:
\beq 
\frac{H(x)}{h(x)} \leq \frac{3}{\beta-\gamma} \left[ \frac{<1+\delta''>_{/\Delta}}{1+\Delta} -1 +\frac{\beta}{3} \right]
\eeq
which gives
\beq
x \gg x_* \; : \;\; H(x) \leq h(x) 
\eeq
\beq
x \ll x_* \; : \;\; H(x) \leq h(x) \; \frac{1}{\beta-\gamma} \left[ \beta +  \frac{\gam \omega}{1-\omega} \right] 
\eeq
If $\beta \rightarrow \infty$, we obtain $H(x)/h(x) = P(\Delta, \delta''<\Delta)/P(\Delta)$. Hence $H(x) \leq h(x)$. Moreover, for $x \gg x_*$ we have $<1+\delta''>_{/\Delta} = (1-\gam/3) (1+\Delta)$ which implies that $P(\Delta, \delta''<\Delta) \geq \gam/3 \; P(\Delta)$. Hence we get: 
\beq
\beta \rightarrow \infty \; : \;\; H(x) \leq h(x) \; , \; x \gg x_* : \; H(x) \geq \gam/3 \; h(x)
\eeq

	Hence, when we add the constraint that the density profile is locally decreasing we usually cannot get general bounds, but the mass functions still present the scaling in $x$.

\section{Evolution of an over(under)-density (Peebles 1980)}

\subsection{Spherical collapse for $\Omega = 1$}

We consider the collapse of a sphere of mass $M$, radius $r$. The motion of the external shell is given by
\beq
\frac{d^2r}{dt^2} = -\frac{{\cal G}M}{r^2} 
\eeq
and a parametric solution of the dynamics is:
\beq
\left\{ \begin{array}{ll} r = A (1-\cos \theta) \\ & \;\;\; \mbox{with} \;\;A^3={\cal G} MB^2 \\ t = B(\theta-\sin \theta) \end{array} \right. \label{dyncoll}
\eeq
The sphere will collapse at $\theta_{coll}=2 \pi$, $t_{coll}=2 \pi B$. The density of the background universe is $\rho_b = 1/(6 \pi {\cal G} t^2)$, which translates into a non-perturbed radius $r_b^3 = 9/2 \; {\cal G} Mt^2$. The density contrast of the overdensity is
\beq
1+\delta = \left( \frac{r_b}{r} \right)^3 = \frac{9}{2} \frac{1}{(1-\cos \theta)^3} \left( \frac{t}{B} \right)^2   \label{A3}
\eeq
corresponds to the contrast evaluated using the linear theory
\beq
\delta_L(t) = \frac{3}{20} \left( \frac{6 t}{B} \right)^{2/3}  \label{A4}
\eeq
At the time of collapse the latter is 
\beq
\delta_{c0} = 3/20 \; (12 \pi)^{2/3} \simeq 1.69
\eeq
If there is no kinetic energy at the maximum radius $R_m = 2 A$, the energy conservation and the virial theorem imply that the equilibrium radius is $R_v = R_m /2 = A$, and if the object is virialized at the time $t_{coll}$ it has a density contrast $\Delta_c$ at this instant given by:
\beq
1 + \Delta_c = \left( \frac{r_b(t_{coll})}{R_v} \right)^3 = 18 \pi^2 \simeq 178
\eeq
If the overdensity collapsed at the redshift $z$, its present ``linear'' density contrast is thus:
\beq
\delta_c(z) = \delta_{c0} \; (1+z)  
\eeq

The dynamics of an underdense region is given by:
\beq
\left\{ \begin{array}{lll} r & = A (\cosh \eta -1) \\ & & \;\;\; \mbox{with} \;\; A^3={\cal G} MB^2 \\ t & = B(\sinh \eta -\eta) \end{array} \right. \label{AA2}
\eeq
We have:
\beq
1+\delta = \left( \frac{r_b}{r} \right)^3 = \frac{9}{2} \frac{1}{(\cosh \eta -1)^3} \left( \frac{t}{B} \right)^2   \label{AA3}
\eeq
and
\beq
\delta_L(t) = - \frac{3}{20} \left( \frac{6 t}{B} \right)^{2/3}   \label{AA4}
\eeq

\subsection{Spherical collapse for $\Omega < 1 \; , \; \Lambda = 0$}

 The dynamics of a spherical overdensity of mass $M$ which will eventually collapse is the same as previously, but the evolution of the background universe is now given by:
\beq
\left\{ \begin{array}{lll} r_b & = A_b (\cosh \eta_b -1) \\ & & \;\;\; \mbox{with} \;\; A_b^3={\cal G} MB_b^2 \\ t & = B_b(\sinh \eta_b -\eta_b) \end{array} \right.
\eeq
One also has:
\[
\eta_{b0} = \cosh^{-1} \left( \frac{2}{\Omega_0} - 1 \right) \;\; \mbox{and} \;\; B_b = \frac{1}{2} H_0^{-1} \Omega_0 (1-\Omega_0)^{-3/2} 
\]
The density contrast is then
\beq
1+\delta = \left( \frac{r_b}{r} \right)^3 = \left( \frac{\cosh \eta_b -1}{1-\cos \theta} \right)^3 \left( \frac{B_b}{B} \right)^2   \label{A9}
\eeq
The growing mode of the linear approximation is
\beq
D(t) = \frac{3 \sinh \eta_b (\sinh \eta_b - \eta_b)}{(\cosh \eta_b-1)^2} -2
\eeq
which is normalized so that $D(t \rightarrow \infty)=1$. At small times we have
\beq
\delta_L(t) = \frac{3}{2} D(t) \left[ 1+\left( \frac{B_b}{B} \right)^{2/3} \right]   \label{A11}
\eeq
with
\[
D(t) \simeq \frac{1}{10} \left( \frac{6t}{B_b} \right)^{2/3}  \hspace{1cm} \mbox{for} \; t \ll B_b
\]
At the time of collapse $t_{coll}=2\pi B$ the linear density contrast is
\beq
\delta_c = \frac{3}{2} D(t_{coll}) \left[ 1+\left( \frac{2\pi B_b}{t_{coll}} \right)^{2/3} \right]
\eeq
and the density contrast of the virialized object is
\beq
1+\Delta_c = (\cosh \eta_{bcoll} -1)^3 \left( \frac{2\pi}{\sinh \eta_{bcoll} -\eta_{bcoll}} \right)^2
\eeq
If $\Omega_0 = 0.1$, their present values are : $\delta_c \simeq 1.62$ and $\Delta_c \simeq 978$. If the overdensity collapsed at the redshift $z$, its present ``linear'' density contrast is thus:
\beq
\delta_c(z) =  \frac{3}{2} D(t_0) \left[ 1+\left( \frac{2\pi B_b}{t(z)} \right)^{2/3} \right]    
\eeq
where $t(z)$ is the age of the universe at the redshift $z$.
\\

However, in a low density universe some overdensities will never collapse. As can be seen from (\ref{A11}), they are characterized by $0<\delta_L(t)<3/2 \; D(t)$. In such a case, the dynamics of the overdensity is given by (\ref{AA2}). The density contrast is
\beq
1+\delta = \left( \frac{r_b}{r} \right)^3 = \left( \frac{\cosh \eta_b -1}{\cosh \eta -1} \right)^3 \left( \frac{B_b}{B} \right)^2  \label{resul1}
\eeq
and the linear density contrast is:
\beq
\delta_L(t) = \frac{3}{2} D(t) \left[ 1-\left( \frac{B_b}{B} \right)^{2/3} \right]   \label{resul2}
\eeq

The dynamics of underdense regions characterized by $\delta_L(t)<0$ is also given by (\ref{AA2}), hence the previous results (\ref{resul1}) and (\ref{resul2}) apply. Note that overdensities which will collapse satisfy $\delta_L(t) > 3/2 \; D(t)$ and $B>0$, while overdensities which will never collapse have $0 < \delta_L(t) < 3/2 \; D(t)$ and $B>B_b$, finally underdensities are characterized by $\delta_L(t)<0$ and $0<B<B_b$.

\section{Evolution of the correlation function}

\subsection{Relation of $\xia$ with $\delta$}

Initially we divide the early universe into cells of comoving size $x$, and we consider a simplified model where we have only two populations of cells: overdense ones with density $1+\delta_+$ and underdense ones with density $1+\delta_-$, such that when $a \rightarrow 0$ we have $\delta_+ = - \delta_- = \delta_L$,  and the density contrast of these cells evolves with time according to the spherical model. In the early universe there are as many cells $\delta_+$ as cells $\delta_-$. At any time, separating the universe in cells of the same size $r = ax$, we have a fraction $\eta_+$ of overdense cells and a fraction $\eta_-$ of underdense ones, with
\beq
\left\{ \begin{array}{rclcl} \eta_+ & + & \eta_- & = & 1  \\  \eta_+ \; \delta_+ & + & \eta_- \; \delta_- & = & 0  \\  \eta_+ \; \delta_+^2 & + & \eta_- \; \delta_-^2 & = & \xia  \end{array} \right.  \label{etapetam}
\eeq
From this system we obtain:
\beq
\xia =  - \; \delta_- \; \delta_+  \label{del1del2}
\eeq
Thus, in the linear stage we have
\beq
\xia \simeq \delta_L^2
\eeq
and $\xia \simeq \xia_L$, while in the non-linear regime we get 
\beq
\xia \simeq \delta_+
\eeq
since $\delta_- \simeq -1$. These two relations fit exactly into Peebles' qualitative picture. They show that the hierarchical clustering argument describes the evolution of the fluctuations in cells of a given size, which we have called $\xia$ throughout this paper (and is called $\xia_L$ or $\Sigma^2$ in the linear case). As we said above we take the time evolution of the {\it average} density contrasts $\delta_+$ and $\delta_-$ to be the one given by the spherical collapse model. Of course, this implies another simplification since the radii of the cells $\delta_+$ and $\delta_-$ evolve in a different way with time, while (\ref{del1del2}) is derived in the case where the universe is divided into cells of the same radius. Hence we consider that the density contrast within underdense regions $\delta_-$ is constant, so that we get the previous results with a comoving cell radius $x_{NL}$, entering $\xia$, equal to the radius of the cells $\delta_+$.

	It is interesting to note that, even in this very simple picture, we already find that, at late times, although overdense cells occupy a negligible fraction of the volume: $\eta_+ \simeq 0$ and $\eta_- \simeq 1$, they contain all the mass: $(1+\delta_+)\eta_+ \simeq 1 $ and $ (1+\delta_-)\eta_- \simeq 0 $.

\subsection{Evolution of $\xia$ if $\Omega=1$}

From (\ref{A3}) it can be seen that $\delta_+$ is a function of $t/B$, since $\theta$ depends only on $t/B$ from (\ref{dyncoll}). Hence it is a well-known function of $\delta_L = a \; \Sigma_0$ through (\ref{A4}). Similarly, through (\ref{AA2}) and (\ref{AA3}) $\delta_-$ is seen to be a function of the same ratio $t/B$, and using (\ref{AA4}) of $\delta_L$. We thus get
\beq
F(\xia_L) = - \; \delta_-(\xia_L) \; \delta_+(\xia_L)
\eeq
which defines a model for $F$, or at least explains the origin of the scaling for $\xia$ as a function of $\xia_L = a^2 \Sigma_0^2$. However, we cannot follow strictly the spherical collapse model for $\delta_+$ as we have to take into account the virialization of the overdensity within a finite radius. To do so, we write
\beq
1+\delta_+ = \frac{9}{2} \; \frac{1}{f(\theta)^3} \; \left( \frac{t}{B} \right)^2     \label{delp1}
\eeq
where $f(\theta) = r/A$ is equal to $(1-\cos \theta)$ as $\theta \rightarrow 0$, see (\ref{A3}), and $f(\theta) \simeq 1$ when $\theta \gg \pi$, with $R_v = A$ the final virialization radius, following the usual model with no kinetic energy at the time of maximum expansion. We shall get the precise shape of $f(\theta)$ from the numerical simulations performed by Jain et al.(1995). 

In a more explicit form, if we define the function $g(\tau)$ by
\[
\left\{ \begin{array}{rcrclcrcl} \tau >0 & : & g(\tau) & = & \tau^2/f(\theta)^3 & \mbox{with} & \tau & = & \theta -\sin \theta  \\   \\  \tau <0 & : & g(\tau) & = & \tau^2/(\cosh \eta-1)^3 & \mbox{with} & \tau & = & \sinh \eta-\eta  \end{array} \right. 
\]
we get
\beq
\xia(r,z) = - \; \left\{ \frac{9}{2}\; g\left(-\;\frac{t}{B}\right) -1 \right\} \; \left\{ \frac{9}{2}\; g\left(\frac{t}{B}\right) -1 \right\}   \label{Hamil1}
\eeq
where $r$ is the physical length at which we evaluate $\xia$, and
\beq
\frac{t}{B} = \frac{1}{6} \; \left( \frac{20 \;\delta_L}{3} \right)^{3/2}
\eeq
with
\beq
\delta_L = (1+z)^{-1} \; \Sigma_0\left[ (1+z) r [1+\xia(r,z)]^{1/3} \right]
\eeq
This defines the function $F$ such that $\xia = F(\xia_L)$, and gives an implicit equation for $\xia$ once $\Sigma_0(x)$, the initial fluctuation spectrum extrapolated up to $z=0$ by linear theory, is known. More precisely, we get for $F(x)$ the expression:
\[
F(x) =  - \; \left\{ \frac{9}{2}\; g\left[-\;\frac{1}{6} \left( \frac{20}{3} \right)^{3/2} x^{3/4}  \right] -1 \right\} 
\]
\beq
\;\;\;\;\;\;\;\;\;\;\;\;\;\; \times \left\{ \frac{9}{2}\; g\left[ \frac{1}{6} \left( \frac{20}{3} \right)^{3/2} x^{3/4}  \right] -1 \right\}  \label{FNL}
\eeq
and we recover the asymptotic behaviour of $F(x)$: for small $x$ we have $F(x) \simeq x$ while for large $x$ we get $F(x) \simeq (10/3)^3 \; x^{3/2}$. In fact, as was shown by Jain et al.(1995) from numerical simulations, the dynamics is not exactly given by the spherical model, and there is a correction which depends on the index $n$ of the power-spectrum. Thus we write, in a fashion similar to these authors:
\beq
\xia=F_n \left( \xia_L \right) \;\;\;\; \mbox{with} \;\;\;\; F_n(x) = \alpha^6 \; F \left( \frac{x}{\alpha^6} \right)    \label{FnO1}
\eeq
which ensures that we still have $\xia \simeq \xia_L$ if $\xia_L \ll 1$ and $\xia \propto \xia_L^{\;3/2}$ if $\xia_L \gg 1$. The parameter $\alpha$ depends on $n$, and is such that the virialization radius is in this picture $R_v=\alpha \; A$. Moreover, Jain et al.(1995) considered the quantity $\tilde{\xi}$, so we have to add a correction to get $\alpha$ for $\xia$ which is the quantity we are interested in. If the two-body correlation function is a power-law $\xi(r) \propto r^{-\gamma}$ we have:
\beq
\tilde{\xi} = 2^{\gamma} \; \left(1-\frac{\gamma}{4}\right) \; \left(1-\frac{\gamma}{6}\right) \; \xia
\eeq
In the linear regime $\gamma_L = 3+n$ while in the highly non-linear regime ($\xia > 177$) we have $\gamma = 3 (3+n)/(5+n)$. This allows us to get the correction $\alpha$ (for $\xia$) from the correction $\tilde{\alpha}$ (for $\tilde{\xi}$) obtained from the fits given by Jain et al.(1995). These parameters are shown on table 2.

\begin{table}
\begin{center}
\caption{Parameters $\tilde{\alpha}$ and $\alpha$ for various power-spectra.}
\begin{tabular}{ccccccc}\hline

$n$ & $\gamma_L$ & $\tilde{\xi}_L/\xia_L$ & $\gamma$ & $\tilde{\xi}/\xia$ & $\tilde{\alpha}$ & $\alpha$  \\ 
\hline\hline
\\ 

0 & 3 & 1 & 1.8 & 1.34 & 1.30 & 1.43 \\ 

-1 & 2 & 1.33 & 1.5 & 1.33 & 1.24 & 1.18 \\

-2 & 1 & 1.25 & 1 & 1.25 & 1.13 & 1.09 \\

\end{tabular}
\end{center}
\label{table2}
\end{table}

Note that we consider in this model the evolution of an overdensity which is initially equal to $\xia_L^{\;1/2}=\Sigma$, and we relate the evolution of the correlation function to the evolution of this particular density contrast. In fact, a more detailed treatment would be to consider the evolution of all initial density contrasts (distributed according to the usual gaussian) from the spherical model, and then take an average to get the fluctuations $\xia$ produced by these overdensities, as in Munshi and Padmanabhan (1996). However, with such a model only half the mass is in overdensities, as in the PS approximation, contrary to our  model. One usually cures this by multiplying the PS mass function, or in the present case the mass contained in any range of overdensities, by a factor of  2, so that all the mass is in overdense regions, and none in underdense ones. Nevertheless, this correction is rather artificial, in spite of the argument based on the result of the excursion sets formalism (that anyway holds only for the quite unrealistic top-hat filter in $k$), and as we note in section 4.1 it is indeed insufficient to give acceptable results. Therefore, a detailed treatment of all initial density fluctuations, with a final averaging, is probably unnecessary as it is based on a picture of the density field which is wrong at least by a factor of  2, and thus should not give a very reliable value of the normalization $\alpha$. Thus, we feel that although the spherical collapse model contains parts of the relevant physical processes, and can be used to get averaged quantities like $\xia$ within a correction factor close to unity, it should not be pushed too far to study the individual behaviour of all density fluctuations. 

From the explicit analytical fit for $\tilde{F}_0(x)$ given by Jain et al.(1995) we get directly $\tilde{f}_0(\theta)$. More precisely, we use the following form for $\tilde{f}_0(\theta)$ as a function of $\tau=t/B$, which in turn is related to $\theta$ through (\ref{dyncoll}):
\beq
\tilde{f}_0(\theta) = \frac{ (9/2)^{1/3} \tau^{2/3} - 3/10 \; (3/4)^{1/3} \tau^{4/3} + \tilde{\alpha}_0 \; (\tau/6)^4 } { 1+ (\tau/6)^4 }
\eeq
The terms in $\tau^{2/3}$ and $\tau^{4/3}$ correspond to the first and second order of $(1-\cos \theta)$ as $\theta \rightarrow 0$ and $\tau \rightarrow 0$, which we need to recover $\delta_+ \simeq \delta_L$ as $\delta_L \rightarrow 0$, and the term in $\tau^4$ ensures that $\tilde{f}_0(\tau) \rightarrow \tilde{\alpha}_0$ as $\tau \rightarrow \infty$. Then we can derive $F(x)$ and $F_n(x)$ from $\tilde{F}_0(x)$, since we have: $F(x) = \tilde{\alpha}_0^{-6} \; \tilde{F}_0 ( \tilde{\alpha}_0^6 \; x)$ and $F_n(x) = (\alpha/\tilde{\alpha}_0)^6 \; \tilde{F}_0 [ (\alpha/\tilde{\alpha}_0)^{-6} \; x]$, where the parameter $\alpha$ which depends on the power-spectrum is given in table 1.

\begin{figure}[htb]
 
\begin{picture}(230,110)(-10,0)

\epsfxsize=7 cm
\epsfysize=9 cm
\put(5,-82){\epsfbox{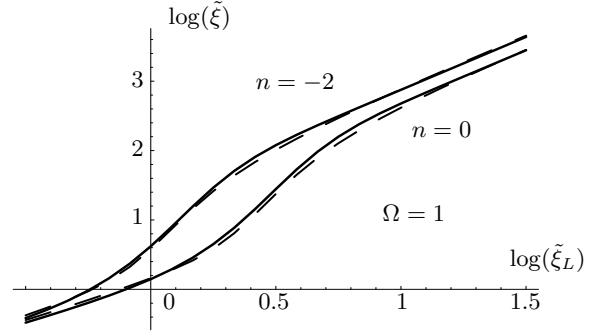}}
\put(62,-3){0}
\put(98,-3){0.5}
\put(150,-3){1}
\put(193,-3){1.5}
\put(50,29){1}
\put(50,55){2}
\put(50,81){3}
\put(64,105){$\log(\tilde{\xi})$}
\put(193,13){$\log(\tilde{\xi}_L)$}
\put(156,62){$n=0$}
\put(97,80){$n=-2$}
\put(145,30){$\Omega=1$}

\end{picture}

\caption{The correlation function $\tilde{\xi}$ as a function of $\tilde{\xi}_L$, for $n=-2$ and $n=0$. The solid line corresponds to our formulation (\ref{Hamil1}) or (\ref{FNL}) while the dashed-line is the fit given by Jain et al.(1995).}
\label{figXiO1}

\end{figure}

Fig.\ref{figXiO1} shows the comparison for $\tilde{\xi}$ between this analytical formulation and the fits given by Jain et al.(1995) for the power-spectra $n=-2$ and $n=0$. It is interesting to note on this figure that while the highly non-linear regimes are quite close for all spectra -- indeed we have $\xia = (10/3\alpha)^3 \; \xia_L^{\;3/2}$ -- the intermediate regime for $1<\xia<100$ shows wider variations. Note that the parameterization (\ref{FnO1}) implies that the ``virialized'' regime is reached at a constant value of $\xia_L/\alpha^6$, that is at $\delta_L \propto \alpha^3$, or at a time $t \propto \alpha^{9/2}$, for $\xia \propto \alpha^6$.

The advantage of our calculation, in addition to the fact that it enlightens the origin of the function $F$, is that it can readily be extended to the case $\Omega < 1 \; , \; \Lambda = 0,$ as we shall see now.

\subsection{$\Omega < 1 \; , \; \Lambda = 0$}

In the case of a low-density universe we still use (\ref{del1del2}) but the time evolution of $\delta_+$ and $\delta_-$ and their dependence on $\delta_L$ are different. Moreover, as we note in App.D.2 , some overdensities $\delta_+$ will never collapse. We first consider the case of an overdensity which will eventually collapse (large $\xia_L\;: \;\; \delta_L(t) > 3/2 \; D(t)$). Using the relation
\beq
(\cosh \eta_b-1)^3 \; B_b^2 = 2 \; H_0^{-2} \Omega_0^{-1} \; (1+z)^{-3}
\eeq
we get from (\ref{A9}) 
\beq
1+\delta_+ = \frac{2}{\Omega_0 (H_0 t)^2 (1+z)^3} \;  \frac{1}{f(\theta)^3} \; \left( \frac{t}{B} \right)^2
\eeq
where $f(\theta)$ is the function we introduced above, see (\ref{delp1}), in the case $\Omega=1$. We can note that $\delta_+$ is no longer a function of the sole ratio $t/B$. Indeed a simplification occurs for $\Omega=1$, where $\Omega_0 (H_0 t)^2 (1+z)^3 = 4/9$ is a constant, which leads to (\ref{delp1}). Next, to get the relation $\xia_L \leftrightarrow \xia$ at a given redshift $z$, we have to infer $B$ from $\delta_L = \sqrt{\xia_L}$. This is done from (\ref{A11}), which leads to
\beq
B = B_b \; \left[ \frac{\delta_L(t)}{\frac{3}{2} D(t)} -1 \right]^{-3/2}  \label{Bp}
\eeq
The dynamics of overdensities which will never collapse, and of underdensities (corresponding to $\delta_-$), is described by (\ref{AA2}). Hence $\delta_+$ or $\delta_-$ is given by (\ref{resul1}), which leads to
\beq
1+\delta_{+,-} = \frac{2}{\Omega_0 (H_0 t)^2 (1+z)^3} \;  \frac{1}{(\cosh \eta -1)^3} \; \left( \frac{t}{B} \right)^2
\eeq
where $\eta$ is defined by (\ref{AA2}). Finally, $B$ is given by (\ref{resul2}) which leads to
\beq
B = B_b \; \left[1- \frac{\delta_L(t)}{\frac{3}{2} D(t)} \right]^{-3/2}  \label{Bm}
\eeq
Note that contrary to the case $\Omega=1$ the cells $\delta_+$ and $\delta_-$ have a different parameter $B$. All these relations define a function $F$ analogous to the case $\Omega=1$ such that
\beq
\xia = F(\xia_L,a)
\eeq
but $F$ is now a function of two variables: $\xia_L$ and the redshift $z$ or expansion factor $a$.

More explicitely, we can write
\[
\xia(r,z) = - \; \left\{ \frac{2 \; g(t/B_-)}{\Omega_0 (H_0 t)^2 (1+z)^3} -1 \right\} 
\]
\beq
\;\;\;\;\;\;\;\;\;\;\;\;\;\;\; \times \left\{ \frac{2 \; g(t/B_+)}{\Omega_0 (H_0 t)^2 (1+z)^3} -1 \right\}   \label{Hamil03}
\eeq
where $r$ is the physical length at which we evaluate $\xia$, $g$ is the function we defined in the previous section in the case $\Omega=1$, $B_+$ and $B_-$ are given by
\beq
\left\{ \begin{array}{rcrcl}  \delta_L > 3/2\;D(t) & : & B_+ & = & B_b \; [ 2/3 \; \delta_L /D(t) -1]^{-3/2} \\   \\  \delta_L < 3/2\;D(t) & : & B_+ & = & -\; B_b \; [1 - 2/3 \; \delta_L /D(t)]^{-3/2}  \end{array} \right.
\eeq
and
\beq
B_-  =  -\; B_b \; [1 + 2/3 \; \delta_L /D(t)]^{-3/2}
\eeq
Finally, $\delta_L$ is obtained from
\beq
\delta_L = \frac{D(t)}{D(0)} \; \Sigma_0\left[ (1+z) r [1+\xia(r,z)]^{1/3} \right]
\eeq
This defines the function $F$ such that $\xia = F(\xia_L,a)$, and gives an implicit equation for $\xia$ once $\Sigma_0(x)$ is known. 

As was the case for $\Omega=1$, we must add a correction which depends on the index $n$ of the power-spectrum. Thus, in a fashion similar to what we did in the previous section we write:
\beq
\xia=F_n \left( \xia_L,a \right) \;\;\;\; \mbox{with} \;\;\;\; F_n(x,a) = \alpha^6 \; F \left( \frac{x}{\alpha^6} ,a \right)    \label{FnO03}
\eeq
The parameter $\alpha$, which measures the virialization radius of overdensities, is taken as independent of time. Since $\Omega \rightarrow 1$ when $a \rightarrow 0$, it must take the values given in table 1 obtained for a critical universe.

\begin{figure}[htb]
 
\begin{picture}(230,120)(-8,0)

\epsfxsize=7 cm
\epsfysize=9 cm
\put(5,-80){\epsfbox{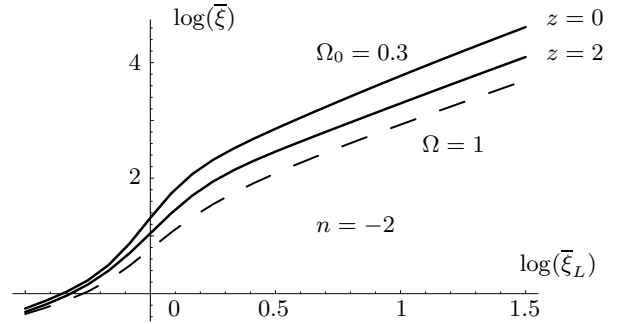}}
\put(64,-7){0}
\put(98,-7){0.5}
\put(150,-7){1}
\put(193,-7){1.5}
\put(49,43){2}
\put(49,86){4}
\put(66,103){$\log(\xia)$}
\put(197,10){$\log(\xia_L)$}
\put(207,103){$z=0$}
\put(207,90){$z=2$}
\put(160,55){$\Omega=1$}
\put(120,90){$\Omega_0=0.3$}
\put(120,25){$n=-2$}

\end{picture}

\caption{Evolution of the correlation function $\xia$ as a function of $\xia_L$ with the redshift for $n=-2$. The dashed-line corresponds to the case $\Omega=1$ (or $z \rightarrow \infty$), while the solid lines correspond to $\Omega_0=0.3$ for $z=0$ and $z=2$.}
\label{figXiO03}

\end{figure}

When $t \rightarrow 0$, we have $\Omega \rightarrow 1$, and relations (\ref{Hamil03}) and (\ref{FnO03}) tend to the relations (\ref{Hamil1}) and (\ref{FnO1}) we got in the previous section in the case of a critical universe, as we can see on Fig.\ref{figXiO03}. Moreover, in all cases we have $\xia \simeq \xia_L$ as $\xia_L \rightarrow 0$, and when $\xia \gg 200$, $\xia(r,z) \propto (1+z)^{-3}$. We can notice on the figure that $\xia \propto \xia_L^{\;3/2}$ for $\xia_L \gg 10$ whatever the redshift even for $\Omega_0 = 0.3$. This is natural, since the overdensities corresponding to these large $\xia_L$ became non-linear in the early universe, when $\Omega(t) \simeq 1$, hence they follow the same behaviour $\xia \propto \xia_L^{\;3/2}$ as the one we got for a critical universe. The value of $\xia_L$ which marks the start of this dependence does not vary very much with redshift because $D(t \rightarrow \infty) = 1$, hence the $\xia_L$ corresponding to a given scale which turned non-linear in the early universe will not increase up to infinity but will remain finite as $t \rightarrow \infty$. This also implies that for a given overdensity, or initial scale, we can see on the figure that while $\xia_L$ ``saturates'' and remains constant after some time, the corresponding value of $\xia$ keeps increasing with time as $\xia \propto (1+z)^{-3}$ and moves further away from the $\Omega=1$ curve. More precisely, for a given overdensity characterized by its fixed parameter $B_+ \ll B_b$, we have in the virialized regime: $\xia \simeq \delta_+ \propto a^3 \; B_+^{-2}$ while $\delta_L \propto D(t) \; B_+^{-2/3}$, hence we get:
\beq
\xia \propto (a/D(t))^3 \;\; \xia_L^{\;3/2}
\eeq
This may be written as a function of the density parameter $\Omega(t)$:
\beq
\xia \propto d(\Omega)^{-3} \;\; \xia_L^{\;3/2}
\eeq
with
\beq
d(\Omega) \simeq \frac{5 \; \Omega}{2 \left[ 1 + \Omega^{4/7} + \Omega/2 \right]}
\eeq
using the fitting formula of Carroll et al. (1992). Thus, by construction we recover the behaviour noticed by Peacock and Dodds (1996) in numerical simulations. The fact that the latter show that all the redshift dependence is accounted for by the factor $d[\Omega(t)]$ confirms that the parameter $\alpha$ should be constant, otherwise it would produce an additional redshift dependence.

\end{document}